\pdfoutput=1
\documentclass[aps, floatfix, a4paper, nofootinbib,
superscriptaddress, 11pt]{revtex4}
\usepackage{amsthm}
\usepackage[english]{babel} 
\usepackage{graphicx,float,tikz}
\usepackage[all]{xy}
\usepackage{amsmath,upgreek}
\usepackage{amsfonts}
\usepackage{amsthm}   
\usepackage{bbold}
\usepackage{bm}
\usepackage{amssymb}
\usepackage{color}
\usepackage{epsfig}		
\usepackage{graphicx,epstopdf}
\usepackage{subfigure}
\usepackage{pdfpages}
\usepackage{dsfont}
\usepackage{float}
\usepackage{hyperref}
\usepackage{nameref}
\usepackage{bookmark}
\usepackage{xcolor}

\usepackage{multirow}
\definecolor{green1}{RGB}{0,128,0} 
\hypersetup{backref=true,pagebackref=true}
\hypersetup{%
  colorlinks = true,
  linkcolor  = blue,
  citecolor = blue,
}
\usepackage{bookmark,textgreek}
\hypersetup{hidelinks,hyperindex=true,colorlinks=true,breaklinks=true,urlcolor=blue}
\hypersetup{%
  colorlinks = true,
  linkcolor  = blue
}

\begin{document}

\title{Pseudospectra in Holographic QCD Models}

\author{Luiz  F. Ferreira}
\email{lffaulhaber@gmail.com}
\affiliation{Instituto de Física y Astronomía, Universidad de Valparaíso, A. Gran Bretana 1111, Valparaíso, Chile}

\begin{abstract}

We investigate the spectral stability of quasinormal modes in nonconformal holographic QCD models using pseudospectral analysis. We first consider the soft-wall model, focusing on scalar and vector fields, and then extend the analysis to the analytic Improved Holographic QCD (IHQCD) model, where we study tensor glueball fluctuations. In both cases, the equations of motion are formulated as  linear eigenvalue problems, allowing the instability of the quasinormal spectrum to small perturbations to be quantified through the pseudospectrum. We find that the spectral stability of the quasinormal modes exhibits a nontrivial dependence on temperature, reflecting the presence of intrinsic dimensionful scales in holographic QCD models. As the temperature increases, the magnitude of the imaginary part of the quasinormal frequencies grows, signaling an increasing decay rate and the progressive loss of coherence of the corresponding hadronic excitations. At sufficiently high temperatures, the quasinormal frequencies approach an approximately linear dependence on temperature, although this regime is reached only after the associated hadronic states have melted. We further compare the pseudospectral instability with the behavior of the corresponding spectral functions. Our results indicate that the growth of pseudospectral instability provides a complementary characterization of the loss of spectral stability and can serve as a useful diagnostic of hadronic dissociation and melting in holographic QCD.

 \end{abstract}


\maketitle

\newpage
\medbreak

\section{Introduction}

Recently, pseudospectral methods~\cite{Trefethen:2005,Sjöstrand:2019,Davies:2007} have been widely applied to the study of spectral instabilities in black-hole quasinormal modes (QNMs)~\cite{Jaramillo2021,Destounis:2021lum,Cheung:2021bol,Jaramillo:2022kuv,Konoplya:2022pbc,Boyanov:2022ark,Sarkar:2023rhp,Courty:2023rxk,Destounis:2023ruj,Destounis:2023nmb,Cao:2024oud}. The presence of an event horizon, which acts as an absorbing boundary, renders the dynamics of linearized perturbations intrinsically non-conservative. As a result, the associated quasinormal-mode problem is governed by a non-self-adjoint operator, reflecting the non-conservative nature of the system as an open system. Consequently, its spectrum can exhibit strong spectral instability under small perturbations. In closed and conservative systems, by contrast, the relevant operators are typically self-adjoint and therefore normal, ensuring that their spectra are comparatively stable under small perturbations~\cite{Trefethen:2005}. For non-self-adjoint and, more generally, non-normal operators, however, arbitrarily small perturbations can produce substantial shifts in the eigenvalues. This phenomenon, known as spectral instability, is particularly relevant for quasinormal modes, since small uncertainties or approximations in the underlying physical system can lead to significant changes in the computed spectrum.

Within the framework of gauge/gravity duality \cite{Maldacena:1997re,Witten:1998qj,Gubser:1998bc}, the spectral instability \footnote{The spectral instability considered here has the same meaning as that discussed in Ref.~\cite{Arean:2023ejh}: it indicates that the frequencies and damping rates obtained from an idealized model may differ from those realized in nature. In the holographic context, this suggests that the transport properties of the underlying quantum many-body system may deviate significantly from those predicted by the holographic model.} of quasinormal modes has recently been investigated using pseudospectral analysis~\cite{Arean:2023ejh,Cownden:2023dam,Arean:2024afl,Garcia-Farina:2024pdd,Garcia-Farina:2026qug}. These studies, however, have so far focused on conformal AdS/CFT settings, where temperature is the only dimensionful scale and the quasinormal frequencies scale linearly with $T$. In this context, spectral instability acquires additional physical significance, since black-hole quasinormal frequencies are identified with the poles of retarded Green's functions in the dual strongly coupled quantum field theory.

Nonconformal holographic QCD models provide a natural framework in which to extend this analysis. In contrast to conformal AdS/CFT, these models introduce additional dimensionful scales and therefore exhibit a nontrivial temperature dependence of the quasinormal spectrum. A wide variety of holographic QCD models~\cite{Karch:2006pv,Gursoy:2007cb,Gursoy:2007er,Colangelo:2007pt,Colangelo:2008us,Gubser:2008ny,Gubser:2008sz,Gursoy:2009jd,Colangelo:2009ra,Gursoy:2010fj,Branz:2010ub,Li:2011hp,Kajantie:2011nx,Gutsche:2011vb,Alho:2012mh,Cai:2012xh,Li:2013oda,Dudal:2014jfa,Finazzo:2014zga,Braga:2016wkm,Braga:2017bml,Braga:2018zlu,Bohra:2020qom,Ballon-Bayona:2021ibm,Ballon-Bayona:2023zal,Arefeva:2024xmg,Ballon-Bayona:2024twa,Toniato:2025gts,Ferreira:2025iqe} incorporate nonconformal effects and exhibit a mass gap~\cite{Gursoy:2007cb,Gursoy:2007er}. Despite their relevance for describing the finite-temperature dynamics of QCD, their quasinormal spectra have not yet been investigated from the perspective of pseudospectral stability.

At zero temperature, confining holographic QCD models describe a discrete spectrum of stable hadronic states. In this regime, the corresponding fluctuation operator is effectively self-adjoint, and the spectrum is consequently insensitive to small perturbations. At finite temperature, however, the emergence of a black-hole horizon changes the character of the spectral problem: normal modes are replaced by quasinormal modes, whose complex frequencies encode the decay of excitations into the thermal medium.

This provides a natural setting in which to investigate the evolution of spectral stability from the low-temperature regime, characterized by well-defined hadronic states, to the high-temperature regime, where these states progressively dissociate in the thermal medium. As the temperature increases, the magnitude of the imaginary part of the quasinormal frequency grows, reflecting an increasing decay rate and a progressive loss of coherence of the corresponding quasiparticle state. At sufficiently high temperatures, the quasinormal frequencies recover an approximately linear scaling with temperature, similar to that observed in conformal theories. However, this regime is reached only after the hadronic states have already melted, and consequently no well-defined peaks are expected to remain in the corresponding spectral functions. The pseudospectrum therefore provides a complementary criterion for characterizing the development of hadronic melting, with its increasing instability offering a quantitative measure of the spectral instability associated with the progressive dissociation of the hadronic state.

In this work, we investigate the spectral instability of quasinormal modes in holographic QCD models using pseudospectral analysis. We first consider the soft-wall model \cite{Karch:2006pv}, one of the simplest realizations of holographic QCD, focusing on the scalar and vector fields. We then extend our analysis to the analytic Improved Holographic QCD (IHQCD) model~\cite{Gursoy:2007er,Kajantie:2011nx}, where we study the quasinormal modes associated with tensor glueball fluctuations of the background metric. In addition, we compare the results obtained from the pseudospectrum with those derived from the corresponding spectral functions. This comparison allows us to investigate whether the spectral instability of the quasinormal modes is reflected in the spectral functions and, consequently, whether pseudospectral instability can provide a useful diagnostic of the melting of hadronic states.

The paper is organized as follows. In Sec.~\ref{Pseudo}, we introduce the definition of the pseudospectrum. In Sec.~\ref{HQCD}, we present the holographic QCD models considered in this work and formulate the equations of motion for the corresponding fields as eigenvalue problems. In Sec.~\ref{NU}, we describe the numerical procedure used to compute the pseudospectra of the quasinormal modes of these models. In Sec.~\ref{Re}, we present our numerical results. Finally, in Sec.~\ref{Con}, we discuss our findings and present our conclusions.

\section{Pseudospectrum and Stability}\label{Pseudo}

In this section, we review the basic concepts of pseudospectra and condition numbers that are relevant for the analysis of the spectral stability of non-normal operators. We begin by recalling some standard definitions from the theory of linear operators on Hilbert spaces \cite{Trefethen:2005}.

Let
\[
\mathfrak{L}:\mathcal{D}(\mathfrak{L})\subset\mathcal{H}\rightarrow\mathcal{H}
\]
be a closed linear operator acting on a Hilbert space $\mathcal{H}$, with domain $\mathcal{D}(\mathfrak{L})$. The spectrum of $\mathfrak{L}$, denoted by $\sigma(\mathfrak{L})$, is defined as the set of complex numbers $\lambda$ for which the operator $(\mathfrak{L}-\lambda I)$ is not invertible, namely
\begin{equation}
\sigma(\mathfrak{L})
=
\left\{
\lambda\in\mathbb{C}:
(\mathfrak{L}-\lambda I)^{-1}
\text{ does not exist}
\right\},
\end{equation}
where $I$ denotes the identity operator. The spectrum of the operator $\mathfrak{L}$ is therefore defined as the set of points in the complex plane for which the resolvent operator is not well defined,
\begin{equation}
    \sigma(\mathfrak{L})
    =
    \left\{
    \lambda\in\mathbb{C}:
    R(\mathfrak{L};\lambda)=(\mathfrak{L}-\lambda I)^{-1}
    \ \text{does not exist}
    \right\}.
\end{equation}

An eigenvalue is a particular element of the spectrum for which there exists a non-trivial solution of the eigenvalue equation
\begin{equation}
    (\mathfrak{L}-\lambda)u_{\lambda}=0,
\end{equation}
where the corresponding eigenfunction satisfies
\begin{equation}
    u_{\lambda}\in \mathcal{D}(\mathfrak{L}).
\end{equation}

For self-adjoint operators, which are relevant to conservative systems in physical applications, the spectral theorem ensures that the spectrum is stable under small bounded perturbations. More precisely, a perturbation of size $\varepsilon$ can displace the eigenvalues by no more than an amount proportional to $\varepsilon$~\cite{Courant1989,Kato2013}. This result is not restricted to self-adjoint operators but applies to the entire class of normal operators satisfying
\begin{equation}
    AA^\dagger=A^\dagger A .
\end{equation}
Hence, for normal operators, the eigenvalues provide a reliable characterization of spectral stability.

In contrast, for non-conservative systems the associated operators are not necessarily normal. As a consequence, even perturbations of arbitrarily small magnitude may lead to significant displacements of the eigenvalues, making the spectrum highly sensitive to perturbations. Therefore, in such systems, the eigenvalue spectrum alone does not provide a complete characterization of the sensitivity of the eigenvalues to perturbations.

Therefore, to properly characterize the sensitivity and stability of eigenvalues under perturbations, it is necessary to go beyond the conventional spectral analysis. This motivates the introduction of the $\varepsilon$-pseudospectrum, which admits three mathematically equivalent formulations~\cite{Trefethen:2005}.

Following Ref.~\cite{Trefethen:2005}, the $\varepsilon$-pseudospectrum
can be characterized in terms of the norm of the resolvent. For a
closed linear operator $\mathfrak{L}$ acting on a Hilbert space
$\mathcal{H}$ with domain $\mathcal{D}(\mathfrak{L})$, and for
$\varepsilon>0$, the $\varepsilon$-pseudospectrum is given by
\begin{equation}
    \sigma_\varepsilon(\mathfrak{L})
    =
    \left\{
    z\in\mathbb{C}:
    \|R(\mathfrak{L};z)\|>\frac{1}{\varepsilon}
    \right\},
\end{equation}
where the resolvent is defined as
\begin{equation}
    R(\mathfrak{L};z)
    =
    (\mathfrak{L}-z)^{-1},
\end{equation}
and, by convention,
$\|R(\mathfrak{L};z)\|=\infty$ for
$z\in\sigma(\mathfrak{L})$.

An equivalent characterization of the pseudospectrum is obtained by
considering perturbations of the operator $\mathfrak{L}$. This
perturbative perspective provides a useful interpretation of the
pseudospectrum in terms of the spectral response to small changes in
the operator. Specifically, for a closed linear operator $\mathfrak{L}$
acting on a Hilbert space $\mathcal{H}$ with domain
$\mathcal{D}(\mathfrak{L})$, and for $\varepsilon>0$, the
$\varepsilon$-pseudospectrum can be written as
\begin{equation}
    \sigma_{\varepsilon}(\mathfrak{L})
    =
    \left\{
    z\in\mathbb{C}:
    \exists\,\mathcal{V},\ 
    \|\mathcal{V}\|<\varepsilon,\ 
    z\in\sigma(\mathfrak{L}+\mathcal{V})
    \right\}.
\end{equation}
This characterization has a simple physical interpretation: the
$\varepsilon$-pseudospectrum consists of those points in the complex
plane that can become eigenvalues under a perturbation of the operator
whose norm is smaller than $\varepsilon$. Thus, unlike the spectrum,
the pseudospectrum depends explicitly on the choice of norm used to
measure the size of the perturbation.

A third equivalent characterization of the pseudospectrum is obtained
in terms of approximate eigenvectors. For a closed linear operator
$\mathfrak{L}$ acting on a Hilbert space $\mathcal{H}$ with domain
$\mathcal{D}(\mathfrak{L})$, and for $\varepsilon>0$, the
$\varepsilon$-pseudospectrum can be expressed as
\begin{equation}
    \sigma_{\varepsilon}(\mathfrak{L})
    =
    \left\{
    z\in\mathbb{C}:
    \exists\,u_{\varepsilon}\in\mathcal{D}(\mathfrak{L}),
    \quad
    \|(\mathfrak{L}-z)u_{\varepsilon}\|
    <
    \varepsilon\|u_{\varepsilon}\|
    \right\},
\end{equation}
where $u_{\varepsilon}$ is an $\varepsilon$-pseudoeigenvector
associated with the pseudoeigenvalue $z$.

These equivalent characterizations highlight an important distinction
between the spectrum and the pseudospectrum. While the spectrum is
determined by the operator and its domain, the pseudospectrum also
depends on the choice of norm used to measure perturbations. In
particular, the operator norm quantifies the magnitude of a perturbation
$\mathcal{V}$ according to
\begin{equation}
    \|\mathcal{V}\|
    =
    \sup_{u\in\mathcal{H},\,u\neq0}
    \frac{\|\mathcal{V}u\|}{\|u\|},
\end{equation}
thereby specifying which perturbations of the operator are regarded as
small.

Among the equivalent characterizations of the pseudospectrum, the
perturbative formulation provides the most direct physical
interpretation. It identifies the $\varepsilon$-pseudospectrum as the
region of the complex plane in which eigenvalues can appear under
perturbations of the operator whose norm is smaller than
$\varepsilon$. This interpretation naturally motivates the
visualization of the pseudospectrum through contour plots, with each
contour corresponding to a different perturbation strength.

Despite its intuitive interpretation, a direct implementation of the
perturbative formulation is computationally impractical, since it
would require evaluating the spectrum of
$\mathfrak{L}+\mathcal{V}$ over the space of all perturbations satisfying
$\|\mathcal{V}\|<\varepsilon$. For the study of general spectral
instability, it is therefore more convenient to use the
resolvent-based formulation, which can be efficiently evaluated for
the finite-dimensional matrices obtained from the numerical
discretization of the differential operator.

Another useful measure of spectral stability is provided by the
condition number associated with each eigenvalue. Following
Ref.~\cite{Trefethen:2005},
the condition number $\kappa_i$ associated with an isolated eigenvalue
$\lambda_i$ of a closed linear operator $\mathfrak{L}$ can be written as
\begin{equation}
    \kappa_i
    =
    \frac{\|v_i\|\,\|u_i\|}
    {\left|\langle v_i,u_i\rangle\right|},
\end{equation}
where $\langle\cdot,\cdot\rangle$ denotes the inner product associated
with the norm $\|\cdot\|$. Here, $u_i$ and $v_i$ are the corresponding
right and left eigenvectors, satisfying
\begin{equation}
    \mathfrak{L}u_i=\lambda_i u_i,
    \qquad
    \mathfrak{L}^\dagger v_i=\lambda_i^* v_i,
\end{equation}
respectively.

The condition number quantifies the stability of an eigenvalue to
small perturbations of the operator. In particular, consider a
perturbation
\begin{equation}
    \mathfrak{L}\rightarrow
    \mathfrak{L}+\mathcal{V},
    \qquad
    \|\mathcal{V}\|=\varepsilon.
\end{equation}
For an isolated simple eigenvalue, its displacement under a sufficiently
small perturbation is, to leading order, bounded by
\begin{equation}
    |\delta\lambda_i|
    \lesssim
    \varepsilon\,\kappa_i.
\end{equation}
Thus, eigenvalues with large condition numbers are more sensitive to
small perturbations, whereas eigenvalues with $\kappa_i$ close to unity
are comparatively stable.

For a normal operator, the left and right eigenvectors can be chosen to
coincide up to normalization. Consequently,
\begin{equation}
    \kappa_i=1,
\end{equation}
for all eigenvalues. Therefore, condition numbers larger than unity provide a direct indication of non-normality and increasing spectral instability.

Since the differential operators considered in this work are discretized
before the numerical computation of the quasinormal modes, the spectral
stability analysis is ultimately performed on finite-dimensional
matrices. It is therefore useful to formulate the pseudospectrum and
condition numbers directly in terms of matrix quantities. A convenient
way to incorporate the physical inner product is through a similarity
transformation that maps the corresponding weighted norm to the
usual norm. Following
Refs.~\cite{Trefethen:2005,Warnick2015}, consider the standard
$\ell_2$ inner product on $\mathbb{C}^n$,
\begin{equation}
    \langle v,u\rangle_{2}
    =
    \sum_{i=1}^{n}\bar{v}_i u_i
    =
    v^\dagger u,
\end{equation}
and a general inner product of the form
\begin{equation}
    \langle v,u\rangle_{G}
    =
    v^\dagger G u,
\end{equation}
where $G$ is a positive-definite Hermitian matrix. Writing
\begin{equation}
    G=F^\dagger F,
\end{equation}
with $F$ invertible, the $G$-inner product can be mapped to the
 inner product through the transformation
$u\mapsto Fu$.

For a matrix $M\in M_n(\mathbb{C})$, the pseudospectrum computed with
respect to the $G$-norm is consequently related to the pseudospectrum by
\begin{equation}
    \sigma_{\varepsilon}^{G}(M)
    =
    \sigma_{\varepsilon}^{\ell_2}
    \!\left(FMF^{-1}\right).
\end{equation}
This relation allows the pseudospectrum associated with the physical
inner product to be computed using standard  matrix norms
after transforming the operator by $F$.

The same transformation can be used to evaluate the condition numbers.
If $\lambda_i$ is an eigenvalue of $M$, and $\tilde{u}_i$ and
$\tilde{v}_i$ denote the right and left eigenvectors of the transformed
matrix $FMF^{-1}$, respectively, then
\begin{equation}
    FMF^{-1}\tilde{u}_i
    =
    \lambda_i\tilde{u}_i,
    \qquad
    \left(FMF^{-1}\right)^\dagger\tilde{v}_i
    =
    \bar{\lambda}_i\tilde{v}_i,
\end{equation}
and the corresponding condition number in the $G$-norm is
\begin{equation}
    \kappa_i^{G}
    =
    \frac{
        \|\tilde{v}_i\|_{2}\,
        \|\tilde{u}_i\|_{2}
    }{
        \left|
        \langle\tilde{v}_i,\tilde{u}_i\rangle_{2}
        \right|
    }.
\end{equation}
Thus, both the pseudospectrum and the condition numbers associated with
the physical inner product can be evaluated using 
linear-algebra routines applied to the similarity-transformed matrix.

In the present analysis, we restrict our attention to sufficiently small perturbations satisfying $\varepsilon \ll d_{\mathrm{min}}$, where $d_{\mathrm{min}}$ denotes the minimum separation between distinct spectral components. In this regime, the pseudospectrum primarily probes the local stability of individual eigenvalues, while overlap between different spectral components remains negligible.

\section{Holographic QCD Models}\label{HQCD}

This section presents the holographic QCD models considered in this work and formulates the corresponding quasinormal-mode  equations as eigenvalue problems. These formulations provide the basis for the subsequent analysis of quasinormal mode spectral stability. We consider two holographic QCD models: the soft-wall model~\cite{Karch:2006pv}, which provides a simple phenomenological framework incorporating nonconformal effects through a prescribed dilaton background, and the Improved Holographic QCD (IHQCD) model~\cite{Gursoy:2007er,Kajantie:2011nx}, in which the dilaton is dynamically determined by the bulk gravitational dynamics. We begin with the soft-wall model.

\subsection{Soft-Wall Model}

\subsubsection{Soft-Wall Model and Eigenvalue Formulation for Quasinormal Modes}

In the soft-wall model, where the backreaction of the dilaton field on the geometry is neglected, the $AdS$-Schwarzschild black brane is described in the Poincar\'e patch by the following metric:
\begin{equation}\label{eqsec31}
    ds^2=\frac{R^2}{z^2}\left(
    -f(z)dt^2+d\vec{x}\cdot d\vec{x}
    +\frac{dz^2}{f(z)}
    \right),
\end{equation}
where
\begin{equation}
    f(z)=1-\frac{z^4}{z_h^4},
\end{equation}
$z_h$ denotes the position of the horizon, and $R$ represents the curvature radius of $AdS_5$. The black-hole temperature is given by
\begin{equation}\label{eqsec32}
    T=\frac{1}{4\pi}\left|f'(z_h)\right|
    =\frac{1}{\pi z_h}.
\end{equation}

In order to cast the problem into a standard eigenvalue form for the quasinormal modes, we follow Refs.~\cite{Arean:2023ejh,Arean:2024afl,Garcia-Farina:2024pdd,Garcia-Farina:2026qug} and introduce a set of regular coordinates that resemble the ingoing Eddington-Finkelstein (EF) coordinates near the horizon. This is achieved through the following change of variables:
\begin{eqnarray}\label{eqsec33}
    && \mathfrak{t}=t-(z-z_h)+\int\frac{dz}{f(z)}, \cr
    && z=z, \cr
    && x_i=x_i.
\end{eqnarray}
Furthermore, unlike the Poincar\'e coordinates used in the metric~\eqref{eqsec31}, the coordinates introduced above are regular at the horizon \footnote{In Ref.~\cite{Cownden:2023dam}, the authors show that the pseudospectral stability of quasinormal frequencies can be studied in Eddington--Finkelstein coordinates by constructing a generalized eigenvalue problem.}. Consequently, the quasinormal mode solutions are nonsingular there, allowing the corresponding energy norm to be well defined~\cite{Arean:2023ejh}. Under this transformation, the metric becomes
\begin{equation}\label{eqsec34}
    ds^2=\frac{R^2}{z^2}\left(
    -f(z)d\mathfrak{t}^2+d\vec{x}\cdot d\vec{x}
    -2\bigl(1-f(z)\bigr)d\mathfrak{t}dz
    +\bigl(2-f(z)\bigr)dz^2
    \right).
\end{equation}
To facilitate the numerical treatment of the problem, it is useful to introduce the dimensionless compactified radial coordinate
\begin{equation}\label{eqsec35}
    \rho=1-\frac{z}{z_h},
\end{equation}
such that the horizon is located at $\rho=0$, while the AdS boundary is mapped to $\rho=1$. The black brane metric~\eqref{eqsec34} then takes the form
\begin{equation}\label{eqsec36}
    ds^2=\frac{R^2}{z_h^2(1-\rho)^2}\left(
    -f(\rho)d\mathfrak{t}^2+d\vec{x}\cdot d\vec{x}
    -2\bigl(1-f(\rho)\bigr)d\mathfrak{t}d\rho
    +\bigl(2-f(\rho)\bigr)d\rho^2
    \right),
\end{equation}
where
\begin{equation}
    f(\rho)=1-(1-\rho)^4.
\end{equation}

Having presented the background geometry in terms of the compactified radial coordinate, we now consider the dynamics of the bulk scalar and vector fields in this background. Their respective actions, for the massless scalar field $\varphi$ and the vector field $V_M=(V_\mu,V_\rho)$, with $\mu=0,1,2,3$, are given by
\begin{eqnarray}\label{eqsec37}
    &&S=-\frac{1}{2}\int d^5x\,\sqrt{-g}\,e^{-\Phi}g^{MN}
    \partial_M\varphi\partial_N\varphi,\cr
    &&S=-\int d^5x\,\sqrt{-g}\,e^{-\Phi}
    \frac{1}{4}F_{MN}F^{MN},
\end{eqnarray}
where
\begin{equation}
    F_{MN}=\partial_MV_N-\partial_NV_M
\end{equation}
is the field-strength tensor associated with the vector field, and the dilaton profile is
\begin{equation}\label{eqsec38d}
    \Phi=c^2z^2=c^2z_h^2(1-\rho)^2.
\end{equation}
The parameter $c$ sets the infrared (IR) energy scale of the model and is associated with the hadronic mass scale  \cite{Karch:2006pv}.

Thus, by varying the actions in Eq.~\eqref{eqsec37} with respect to the corresponding fields in the background metric~\eqref{eqsec36}, we obtain the equations of motion for the scalar  and vector  fields. We then adopt the plane-wave ansatz
\begin{equation}\label{eqsec38}
    \varphi(\mathfrak{t},\rho)
    =e^{-i\omega\mathfrak{t}}\bar{\varphi}(\rho),
    \qquad
    V_x(\mathfrak{t},\rho)\equiv V(\mathfrak{t},\rho)
    =e^{-i\omega\mathfrak{t}}\bar{V}(\rho).
\end{equation}
For the vector field, we work in the radial gauge $V_\rho=0$. With this ansatz, the equations of motion reduce to
\begin{eqnarray}\label{eqsec39}
    \partial_\rho^2\bar{V}
    &+&\partial_\rho\bar{V}
    \left(
    \frac{1}{1-\rho}
    +\frac{\dot{f}}{f}
    -\dot{\Phi}
    +2i\omega z_h\frac{f-1}{f}
    \right)
    +\nonumber\\
    &&\bar{V}
    \left(
    \omega^2z_h^2\frac{2-f}{f}
    +i\omega z_h
    \left(
    \frac{1}{f}\frac{f-1}{1-\rho}
    +\frac{\dot{f}}{f}
    -\frac{\dot{\Phi}(f-1)}{f}
    \right)
    \right)=0,
\end{eqnarray}
and
\begin{eqnarray}\label{eqsec310}
    \partial_\rho^2\bar{\varphi}
    &+&\partial_\rho\bar{\varphi}
    \left(
    \frac{3}{1-\rho}
    +\frac{\dot{f}}{f}
    -\dot{\Phi}
    +2i\omega z_h\frac{f-1}{f}
    \right)
    +\nonumber\\
    &&\bar{\varphi}
    \left(
    \omega^2z_h^2\frac{2-f}{f}
    +i\omega z_h
    \left(
    \frac{3}{f}\frac{f-1}{1-\rho}
    +\frac{\dot{f}}{f}
    -\frac{\dot{\Phi}(f-1)}{f}
    \right)
    \right)=0.
\end{eqnarray}
Here and throughout this subsection, a dot denotes differentiation with respect to the dimensionless radial coordinate \(\rho\).

For the subsequent analysis of the associated eigenvalue problem, it is convenient to recast the equations of motion, Eqs.~\eqref{eqsec39} and~\eqref{eqsec310}, in the form
\begin{eqnarray}\label{eqx3}
    \omega^2z_h^2V(\mathfrak{t},\rho)
    &=&
    \frac{i\omega z_h}{f-2}
    \left[
    (1-\rho)e^\Phi
    \left(
    \frac{e^{-\Phi}(f-1)}{1-\rho}
    \right)'
    +2(f-1)\partial_\rho
    \right]V(\mathfrak{t},\rho)
    \nonumber\\
    &&-\frac{1}{f-2}
    \left[
    -(1-\rho)e^\Phi
    \left(
    \frac{fe^{-\Phi}}{1-\rho}
    \right)'\partial_\rho
    -f\partial_\rho^2
    \right]V(\mathfrak{t},\rho),
\end{eqnarray}
and
\begin{eqnarray}\label{eqx4}
    \omega^2z_h^2\varphi(\mathfrak{t},\rho)
    &=&
    \frac{i\omega z_h}{f-2}
    \left[
    (1-\rho)^3e^\Phi
    \left(
    \frac{e^{-\Phi}(f-1)}{(1-\rho)^3}
    \right)'
    +2(f-1)\partial_\rho
    \right]\varphi(\mathfrak{t},\rho)
    \nonumber\\
    &&-\frac{1}{f-2}
    \left[
    -(1-\rho)^3e^\Phi
    \left(
    \frac{fe^{-\Phi}}{(1-\rho)^3}
    \right)'\partial_\rho
    -f\partial_\rho^2
    \right]\varphi(\mathfrak{t},\rho).
\end{eqnarray}

As a final step, to recast the equations of motion into a first-order system in time, we introduce the auxiliary fields
\begin{eqnarray}\label{eqx4a}
    &&\xi(\mathfrak{t},\rho)\equiv z_h\partial_{\mathfrak{t}}V(\mathfrak{t},\rho),\cr
    &&\chi(\mathfrak{t},\rho)\equiv z_h\partial_{\mathfrak{t}}\varphi(\mathfrak{t},\rho),
\end{eqnarray}
and, for convenience, define the dimensionless frequency and dilaton parameter,
\begin{equation}
    \mathfrak{w}=\omega z_h=\frac{\omega}{\pi T},
    \qquad
    \mathfrak{c}=cz_h=\frac{c}{\pi T}.
\end{equation}
With these definitions, the equations of motion~\eqref{eqx3} and~\eqref{eqx4} can be recast as the following standard eigenvalue problems:
\begin{equation}\label{eqx5}
    \mathfrak{w}
    \begin{pmatrix}
        V(\mathfrak{t},\rho)\\
        \xi(\mathfrak{t},\rho)
    \end{pmatrix}
    =
    i
    \begin{pmatrix}
        0 & 1\\
        \Xi_1(\partial_\rho^2,\partial_\rho,\rho) &
        \Xi_2(\partial_\rho,\rho)
    \end{pmatrix}
    \begin{pmatrix}
        V(\mathfrak{t},\rho)\\
        \xi(\mathfrak{t},\rho)
    \end{pmatrix},
\end{equation}
and
\begin{equation}\label{eqx6}
    \mathfrak{w}
    \begin{pmatrix}
        \varphi(\mathfrak{t},\rho)\\
        \chi(\mathfrak{t},\rho)
    \end{pmatrix}
    =
    i
    \begin{pmatrix}
        0 & 1\\
        \mathfrak{L}_1(\partial_\rho^2,\partial_\rho,\rho) &
        \mathfrak{L}_2(\partial_\rho,\rho)
    \end{pmatrix}
    \begin{pmatrix}
        \varphi(\mathfrak{t},\rho)\\
        \chi(\mathfrak{t},\rho)
    \end{pmatrix}.
\end{equation}

The differential operators $\Xi_1$, $\Xi_2$, $\mathfrak{L}_1$, and $\mathfrak{L}_2$ are defined, respectively, as
\begin{align}\label{eqx7}
    \Xi_1 &=
    \frac{1}{f-2}
    \left[
    -(1-\rho)e^\Phi
    \left(
    \frac{fe^{-\Phi}}{1-\rho}
    \right)'\partial_\rho
    -f\partial_\rho^2
    \right],\\
    \Xi_2 &=
    \frac{1}{f-2}
    \left[
    (1-\rho)e^\Phi
    \left(
    \frac{e^{-\Phi}(f-1)}{1-\rho}
    \right)'
    +2(f-1)\partial_\rho
    \right],\\
    \mathfrak{L}_1 &=
    \frac{1}{f-2}
    \left[
    -(1-\rho)^3e^\Phi
    \left(
    \frac{fe^{-\Phi}}{(1-\rho)^3}
    \right)'\partial_\rho
    -f\partial_\rho^2
    \right],\\
    \mathfrak{L}_2 &=
    \frac{1}{f-2}
    \left[
    (1-\rho)^3e^\Phi
    \left(
    \frac{e^{-\Phi}(f-1)}{(1-\rho)^3}
    \right)'
    +2(f-1)\partial_\rho
    \right].
\end{align}
The operators whose spectral stability we aim to study within the framework of the soft-wall model are given by
\begin{equation}\label{opes}
    \mathfrak{L}
    =
    i
    \begin{pmatrix}
        0 & 1\\
        \mathfrak{L}_1(\partial_\rho^2,\partial_\rho,\rho) &
        \mathfrak{L}_2(\partial_\rho,\rho)
    \end{pmatrix},
\end{equation}
and
\begin{equation}\label{opev}
    \Xi
    =
    i
    \begin{pmatrix}
        0 & 1\\
        \Xi_1(\partial_\rho^2,\partial_\rho,\rho) &
        \Xi_2(\partial_\rho,\rho)
    \end{pmatrix}.
\end{equation}
Their corresponding adjoint operators are 
\begin{equation}
    \mathfrak{L}^{\dagger}
    =
    i
    \begin{pmatrix}
        0 & 1\\
        \mathfrak{L}_1(\partial_\rho^2,\partial_\rho,\rho) &
        \mathfrak{L}_2(\partial_\rho,\rho)
        -2i\delta(\rho)\dfrac{f(\rho)-1}{f(\rho)-2}
    \end{pmatrix},
\end{equation}
and
\begin{equation}
    \Xi^{\dagger}
    =
    i
    \begin{pmatrix}
        0 & 1\\
        \Xi_1(\partial_\rho^2,\partial_\rho,\rho) &
        \Xi_2(\partial_\rho,\rho)
        -2i\delta(\rho)\dfrac{f(\rho)-1}{f(\rho)-2}
    \end{pmatrix}.
\end{equation}
These adjoint operators differ from $\mathfrak{L}$ and $\Xi$ by a boundary term localized at the horizon ($\rho=0$), reflecting the non-self-adjoint character of the operators in the presence of a horizon.

\subsubsection{Boundary Conditions}

The boundary condition at the event horizon corresponds to a purely infalling solution, which is regular in ingoing Eddington–Finkelstein coordinates, ensuring the regularity of the fields. They are given by
\begin{eqnarray}\label{eqx7b}
    &&V(\mathfrak{t},\rho\rightarrow0)
    \approx \mathfrak{B}_1e^{-i\omega\mathfrak{t}},\cr
    &&\varphi(\mathfrak{t},\rho\rightarrow0)
    \approx \mathfrak{A}_1e^{-i\omega\mathfrak{t}}.
\end{eqnarray}

Near the AdS boundary, the asymptotic behavior of the fields is determined by their near-boundary expansions. Since quasinormal modes are required to be normalizable, we retain only the normalizable branch of the solution. Consequently, the fields exhibit the asymptotic behavior
\begin{eqnarray}\label{eqx8}
    &&V(\mathfrak{t},\rho\rightarrow1)
    =\mathfrak{C}(\mathfrak{t})(1-\rho)^2,
    \qquad \,\cr
    &&\varphi(\mathfrak{t},\rho\rightarrow1)
    =\mathfrak{D}(\mathfrak{t})(1-\rho)^4\,
    .
\end{eqnarray}
These boundary conditions ensure the correct asymptotic behavior of the quasinormal modes while guaranteeing that the associated energy norm is finite and positive definite.

To simplify the implementation of these boundary conditions, we introduce the rescaled fields
\begin{equation}\label{eqx9}
    \begin{pmatrix}
        \tilde{V}\\
        \tilde{\xi}
    \end{pmatrix}
    =
    \frac{1}{1-\rho}
    \begin{pmatrix}
        V\\
        \xi
    \end{pmatrix},
\end{equation}
and
\begin{equation}\label{eqx10}
    \begin{pmatrix}
        \hat{\varphi}\\
        \hat{\chi}
    \end{pmatrix}
    =
    \frac{1}{(1-\rho)^2}
    \begin{pmatrix}
        \varphi\\
        \chi
    \end{pmatrix}.
\end{equation}
With these definitions, imposing the asymptotic behavior in Eq.~\eqref{eqx8} is equivalent to imposing homogeneous Dirichlet boundary conditions on the rescaled fields at the AdS boundary.

\subsubsection{Energy Norm in the soft-wall model}

In the previous subsection, we formulated the equations of motion for the scalar and vector fields as  eigenvalue problems. To investigate the spectral stability of the corresponding quasinormal modes, it is necessary to equip the space of perturbations with an appropriate norm. Following Refs.~\cite{Jaramillo:2020tuu,Gasperin:2021kfv,Arean:2023ejh}, we adopt the energy norm.

This choice is naturally motivated by the probe approximation, in which the quasinormal mode perturbations are assumed to be sufficiently small so that their backreaction on the background geometry can be neglected. In this regime, the energy associated with a perturbation provides a physically meaningful measure of its amplitude. This energy is determined by the corresponding energy-momentum tensor and is directly related to the backreaction that the perturbation would induce on the background geometry. The energy norm therefore provides a physically motivated measure of the size of quasinormal mode perturbations and allows us to quantify their stability to small perturbations of the underlying spectral problem.

Accordingly, the norm of a quasinormal mode associated with a field $\psi$ is defined as the energy evaluated on a constant-time hypersurface,
\begin{equation}
    \|\psi\|_{E}^{2}
    =
    \int_{\Sigma_\mathfrak{t}}
    d^{d-1}x\sqrt{\gamma_{\Sigma}}\,
    n^{A}_{\Sigma}T_{AB}\mathfrak{t}^{B}
    =
    -\int d\rho\,dx^1dx^2dx^3\,
    \sqrt{-g}\,\mathfrak{t}^{B}T_{B}^{t},
\end{equation}
where $\mathfrak{t}=\partial_{\mathfrak{t}}$ is the Killing vector associated with time translations, $n_{\Sigma}^{A}$ is the unit normal vector to the hypersurface $\Sigma_\mathfrak{t}$, $\gamma_{\Sigma}$ denotes the induced metric on $\Sigma_\mathfrak{t}$, and $T_{AB}$ is the energy-momentum tensor associated with the field.

For a scalar field $\varphi$, the corresponding energy-momentum tensor is given by
\begin{equation}\label{eq:stress_tensor}
    T_{MN}
    =\, e^{-\Phi}\left(
    \nabla_{M}\varphi\,\nabla_{N}\varphi
    -\frac{1}{2}g_{MN}
    \left(
    \nabla_{S}\varphi\,\nabla^{S}\varphi
    \right) \right).
\end{equation}
For the Maxwell field, the energy-momentum tensor is instead given by
\begin{equation}\label{eq:stress_tensor_maxwell}
    T_{MN}
    =
   e^{-\Phi}\left( F_{MS}F_{N}^{\ S}
    -\frac{1}{4}g_{MN}F_{RS}F^{RS}. \right)
\end{equation}
Therefore, the corresponding energy norms for the scalar and Maxwell fields can be written as
\begin{equation}
    \|\varphi\|_{E}^{2}
    =
    \int d\rho\,dx^1dx^2dx^3 \, e^{-\Phi}
    \left[
    \frac{1}{(1-\rho)^3}
    \left(
    f(\partial_{\rho}\varphi)^2
    +z_h^2(\partial_t\varphi)^2
    \right)
    \right],
\end{equation}
and
\begin{equation}
    \|V\|_{E}^{2}
    =
    \int d\rho\,dx^1dx^2dx^3 \, e^{-\Phi}
    \left[
    \frac{1}{1-\rho}
    \left(
    f(\partial_{\rho}V)^2
    +z_h^2(\partial_tV)^2
    \right)
    \right].
\end{equation}

These expressions can be simplified by decomposing the fields into Fourier modes.\footnote{Since we consider a fixed momentum mode, the integral over momentum can be omitted.} Introducing the auxiliary fields, we obtain
\begin{equation}
    \|\varphi\|_{E}^{2}
    =
    \int_{\mathfrak{t}=\mathrm{const}}d\rho\, 
    \begin{pmatrix}
        \varphi^{*} & \chi^{*}
    \end{pmatrix}
    \mathcal{G}
    \left(
    \overleftarrow{\partial}_{\rho},
    \overrightarrow{\partial}_{\rho},\rho
    \right)
    \begin{pmatrix}
        \varphi\\[0.2cm]
        \chi
    \end{pmatrix},
\end{equation}
and
\begin{equation}
    \|V\|_{E}^{2}
    =
    \int_{\mathfrak{t}=\mathrm{const}}d\rho\,  
    \begin{pmatrix}
        V^{*} & \xi^{*}
    \end{pmatrix}
    \mathcal{G}
    \left(
    \overleftarrow{\partial}_{\rho},
    \overrightarrow{\partial}_{\rho},\rho
    \right)
    \begin{pmatrix}
        V\\[0.2cm]
        \xi
    \end{pmatrix}.
\end{equation}
Here, $\mathcal{G}$ is a matrix-valued differential operator whose explicit form depends on the field content. For the scalar field, $\mathcal{G}$ is given by
\begin{equation}\label{eq:Gmatrix_scalar}
    \mathcal{G}_{\varphi}
    =
   e^{-\Phi}\, \begin{pmatrix}
        \overleftarrow{\partial}_{\rho}
        \dfrac{f(\rho)}{(1-\rho)^3}
        \overrightarrow{\partial}_{\rho}
        & 0\\[0.4cm]
        0 &
        \dfrac{2-f(\rho)}{(1-\rho)^3}
    \end{pmatrix}.
\end{equation}

For the Maxwell field, the corresponding operator is
\begin{equation}\label{eq:Gmatrix_vector}
    \mathcal{G}_{V}
    =
   e^{-\Phi}\, \begin{pmatrix}
        \overleftarrow{\partial}_{\rho}
        \dfrac{f(\rho)}{1-\rho}
        \overrightarrow{\partial}_{\rho}
        & 0\\[0.4cm]
        0 &
        \dfrac{2-f(\rho)}{1-\rho}
    \end{pmatrix}.
\end{equation}

Thus, for both the scalar and vector fields in the soft-wall model, the corresponding inner product can be written in the common form
\begin{equation}\label{eq:inner_product}
    \langle u_1,u_2\rangle_E
    =
    \int_{t=\mathrm{const}}
    d\rho\,
    \Psi_1^{\dagger}(t,\rho)\,
    \mathcal{G}
    \left(
    \overleftarrow{\partial}_{\rho},
    \overrightarrow{\partial}_{\rho},\rho
    \right)
    \Psi_2(t,\rho),
\end{equation}
where
\begin{equation}
    \Psi_i=
    \begin{cases}
        (\varphi_i,\chi_i)^T, & \text{for the scalar sector},\\[0.1cm]
        (V_i,\xi_i)^T, & \text{for the vector sector}.
    \end{cases}
\end{equation}

\subsection{Improved holographic QCD model (IHQCD)}

\subsubsection{IHQCD Model and Eigenvalue Formulation for Quasinormal Modes}

We now turn to the formulation of the eigenvalue problem in a more sophisticated holographic QCD model. We begin by introducing the effective five-dimensional action in the Einstein frame,
\begin{equation}\label{eq:IHQCD1}
S=\frac{1}{16\pi G_5}\int d^5x\,\sqrt{-g}
\left[
R-\frac{4}{3}(\partial\Phi)^2+V(\Phi)
\right]+S_{GH},
\end{equation}
which is supplemented by the Gibbons--Hawking boundary term, ensuring a well-defined variational principle. Moreover, the asymptotically AdS$_5$ geometry introduces ultraviolet divergences in the on-shell action, which are removed through holographic renormalization by adding the appropriate counterterms. Since these boundary contributions do not affect the present analysis, they will not be discussed further. The dilaton potential $V(\Phi)$ effectively encodes the noncritical five-dimensional string-theory corrections.

In the Einstein frame, the metric is written as
\begin{equation}\label{eq:IHQCD2}
ds^2=b(z)^2
\left[
-f(z)dt^2+d\vec{x}^{\,2}
+\frac{dz^2}{f(z)}
\right],
\end{equation}
while the dilaton depends only on the holographic coordinate, $\Phi=\Phi(z)$. In this case, the background is obtained by solving the coupled Einstein-dilaton equations, so that the dilaton is fully backreacted on the geometry rather than treated as a probe field.

The Einstein equations are obtained by varying the action in Eq.~(\ref{eq:IHQCD1}) with respect to the metric in Eq.~(\ref{eq:IHQCD2}), yielding
\begin{align}  
6\frac{b'^2}{b^2}-3\frac{b''}{b} &=\frac{4}{3}\Phi'^2, \label{eq:IHQCD3a} \\ \frac{f''}{f'}+3\frac{b'}{b} &=0, \label{eq:IHQCD3b} \\ 6\frac{b'^2}{b^2}+3\frac{b''}{b} &=\frac{b^2}{f}V(\Phi), \label{eq:IHQCD3c}
\end{align}
where primes denote derivatives with respect to $z$. The holographic running coupling is defined as
\begin{equation}\label{eq:IHQCD4} 
\lambda=e^{\Phi}.
\end{equation}
The beta function of the dual gauge theory is then defined by
\begin{equation}\label{eq:IHQCD5} 
\beta(\lambda)
=
\frac{\lambda'}{b'/b}.
\end{equation}
The gravitational background is dual to a confining gauge theory provided that the infrared behavior of the beta function satisfies the condition~\cite{Gursoy:2007cb,Gursoy:2007er,Gursoy:2008bu}:
\begin{equation}\label{eq:IHQCD6}
\beta(\lambda)
=
-\frac32\lambda
\left[
1+\frac{3(\alpha-1)}{4\alpha\log\lambda}
+\mathcal{O}\!\left(\frac{1}{\log^2\lambda}\right)
\right],
\qquad
\alpha>1,
\end{equation}

We now consider the ansatz~\cite{Gursoy:2007cb,Gursoy:2007er,Kajantie:2011nx},
which we refer to as Model A throughout this work,
\begin{equation}\label{eq:IHQCD7}
    b(z)
    =
    \frac{R}{z}
    e^{-\frac{\Lambda^2z^2}{3}}.
\end{equation}
Here, $\Lambda$ sets the infrared scale of the model. This ansatz, together with
the Einstein equations~(\ref{eq:IHQCD3a})--(\ref{eq:IHQCD3c}),
determines the remaining background functions.

In particular, inserting Eq.~(\ref{eq:IHQCD7}) into
Eq.~(\ref{eq:IHQCD3a}), the dilaton equation becomes
\begin{equation}\label{eq:IHQCD8}
    \Phi'(z)
    =
    \frac{\lambda'(z)}{\lambda(z)}
    =
    \Lambda^2z
    \sqrt{1+\frac{9}{2\Lambda^2z^2}}.
\end{equation}
Upon integrating this equation and imposing the boundary condition
$\lambda(0)=\lambda_0$, we find
\begin{equation}\label{eq:IHQCD9}
    \frac{\lambda(z)}{\lambda_0}
    =
    e^{\frac12 \Lambda z
    \sqrt{\Lambda^2z^2+\frac92}}
    \left[\sqrt{\frac{2}{9}}\left((
    \Lambda z+\sqrt{\Lambda^2z^2+\frac{9}{2}}
   \right )\right]^{9/4}.
\end{equation}

The corresponding beta function can then be expressed as
\begin{equation}\label{eq:IHQCD10}
    \beta(\lambda)
    =
    -\lambda 
    \frac{\Lambda^2z^2\sqrt{1+\frac{9}{2\Lambda^2z^2}}}
    {1+\frac23\Lambda^2z^2}.
\end{equation}
In the infrared, $z\rightarrow\infty$, its asymptotic behavior is
\begin{equation}\label{eq:IHQCD11}
    \beta(\lambda)
    =
    -\frac32\lambda
    \left[
    1+\frac{3}{8\log\lambda}
    +\cdots
    \right].
\end{equation}
Thus, the infrared behavior corresponds to $\alpha=2$, satisfying the criterion for confinement \cite{Gursoy:2007cb,Gursoy:2007er,Kajantie:2011nx}.

The blackening function, $f(z)$, follows from Eq.~(\ref{eq:IHQCD3b}) after
imposing the boundary conditions
\begin{equation}\label{eq:IHQCD12}
    f(0)=1,
    \qquad
    f(z_h)=0,
\end{equation}
and is given by
\begin{equation}\label{eq:IHQCD13}
    f(z,z_h)
    =
    1-
    \frac{
        (\Lambda^2z^2-1)e^{\Lambda^2z^2}+1
    }{
        1+e^{\Lambda^2z_h^2}
        (\Lambda^2z_h^2-1)
    },
\end{equation}
where $z_h$ denotes the position of the horizon. Substituting this
solution into Eq.~(\ref{eq:IHQCD3c}) gives the dilaton potential,
\begin{equation}\label{eq:IHQCD14}
    V(z,z_h)
    =
    \frac{12}{R^2}
    e^{\frac23\Lambda^2z^2}
    \left[
        \left(
            \frac13\Lambda^4z^4
            +\frac56\Lambda^2z^2
            +1
        \right)f(z,z_h)
        -
        \left(
            \frac12+\frac13\Lambda^2z^2
        \right)
        z\frac{\partial f}{\partial z}
    \right].
\end{equation}
Finally, the temperature of the dual gauge theory is identified with
the Hawking temperature of the black brane. It is therefore determined
by the derivative of the blackening function at the horizon,
\begin{equation}\label{eq:IHQCD15}
    T
    =
    \frac{|f'(z_h)|}{4\pi}
    =
    \frac{\Lambda}{2\pi}
    \frac{y_h^3}
    {y_h^2-1+e^{-y_h^2}},
   \end{equation}
where $y_h=\Lambda z_h$.

We now turn to the pseudospectral analysis of the quasinormal modes associated with the tensor glueball sector in the IHQCD model. An important advantage of this channel is that the corresponding gravitational perturbation decouples from all other bulk fluctuations~\cite{Kiritsis:2006ua,Springer:2008js,Springer:2010mw,Kajantie:2011nx}, considerably simplifying the analysis.

Following the same procedure used in the soft-wall model, we first recast the linearized equation of motion as a standard eigenvalue problem suitable for pseudospectral analysis. To this end, we introduce ingoing Eddington--Finkelstein coordinates through the transformation in Eq.~(\ref{eqsec33}), which removes the coordinate singularity at the black-brane horizon. We also introduce the compact, dimensionless radial coordinate
\begin{equation}\label{eq:IHQCD16}
\rho=1-\frac{z}{z_h},
\end{equation}
such that the AdS boundary and the horizon are mapped to $\rho=1$ and $\rho=0$, respectively. As a result, the metric in Eq.~(\ref{eq:IHQCD2}) takes the form
\begin{equation}\label{eq:IHQCD17}
ds^2=b(\rho)^2
\left[
-f(\rho)d\mathfrak{t}^{\,2}
+d\vec{x}^{\,2}
-2\left(1-f(\rho)\right)d\mathfrak{t}\,d\rho
+\left(2-f(\rho)\right)d\rho^2
\right],
\end{equation}
where
\begin{equation}\label{eq:IHQCD18}
b(\rho)
=
\frac{R}{z_h(1-\rho)}
\exp\!\left[
-\frac{\bar{\Lambda}^2(1-\rho)^2}{3}
\right],
\end{equation}
and
\begin{equation}\label{eq:IHQCD19}
f(\rho)
=
1-
\frac{\left[\bar{\Lambda}^2(1-\rho)^2-1\right]
e^{\bar{\Lambda}^2(1-\rho)^2}+1}
{1+e^{\bar{\Lambda}^2}\left(\bar{\Lambda}^2-1\right)}.
\end{equation}
where $\bar{\Lambda}\equiv\Lambda z_h$. 

The tensor glueball perturbation dual to the operator $T_{12}$ is obtained by considering a fluctuation of the $12$ component of the background metric,
\begin{equation}\label{eq:IHQCD20}
g_{12}=b^2(z)h(t,z).
\end{equation}
Since this tensor fluctuation decouples from the other perturbation modes, its dynamics is governed by the following fluctuation equation:
\begin{eqnarray}\label{eq:IHQCD21}
   \mathfrak{w}^2h(\mathfrak{t},\rho)&=& \frac{i\mathfrak{w}}{f-2}\left[\frac{1}{b^3}\left(b^3(f-1)\right)' +2(f-1)\partial_\rho\right]h(\mathfrak{t},\rho)\cr \nonumber\\ && -\frac{1}{f-2}\left[-\frac{1}{b^3}\left(b^3f\right)'\partial_{\rho}-f\partial^2_\rho\right]h(\mathfrak{t},\rho)
\end{eqnarray}
Next, we introduce the auxiliary field
\begin{equation}\label{eq:IHQCD22}
\mathfrak{h}(\mathfrak{t},\rho)\equiv z_h,\partial_{\mathfrak{t}}h(\mathfrak{t},\rho),
\end{equation}
which allows us to recast the equation of motion as the following  eigenvalue problem:
\begin{equation}\label{eq:IHQCD23}
\mathfrak{w}
\begin{pmatrix}
h(\mathfrak{t},\rho)\\
\mathfrak{h}(\mathfrak{t},,\rho)
\end{pmatrix}
=
i
\begin{pmatrix}
0 & 1\\
\beta_{1}\!\left(\partial_{\rho}^{2},\partial_{\rho},\rho\right) &
\beta_{2}\!\left(\partial_{\rho},\rho\right)
\end{pmatrix}
\begin{pmatrix}
h(\mathfrak{t},\rho)\\
\mathfrak{h}(\mathfrak{t},\rho)
\end{pmatrix}.
\end{equation}
where the differential operators $\beta_{1} $ and $\beta_{2}$ are defined, respectively, as
\begin{align}\label{eq:IHQCD24}
\beta_{1} &= \frac{1}{f-2}\left[-\frac{1}{b^3}\left(b^3f\right)'\partial_{\rho}-f\partial^2_\rho\right],\\
\beta_{2} &= \frac{1}{f-2}\left[\frac{1}{b^3}\left(b^3(f-1)\right)' +2(f-1)\partial_\rho\right].
\end{align}
Therefore, we seek to study the spectral stability of the relevant operators,
which take the following form:
\begin{equation}\label{eq:IHQCD25}
\beta=
i
\begin{pmatrix}
0 & 1\\
\beta_1(\partial_\rho^2,\partial_\rho,\rho) &
\beta_2(\partial_\rho,\rho)
\end{pmatrix},
\end{equation}
Notice that these operators are non-normal, a feature associated with the
presence of a horizon: their adjoints are given by
\begin{equation}
\beta^{\dagger}=
i
\begin{pmatrix}
0 & 1\\
\beta_1(\partial_\rho^2,\partial_\rho,\rho) &
\beta_2(\partial_\rho,\rho)-2i\delta(\rho)\dfrac{f(\rho)-1}{f(\rho)-2}
\end{pmatrix},
\end{equation}
which differ from $\beta$ by a boundary term localized at the
horizon ($\rho=0$), reflecting the non-self-adjoint character of the operators in the presence of a horizon.

\subsubsection{Boundary Conditions}

Finally, we comment on the boundary conditions. Near the horizon, the
tensor field exhibits the same type of asymptotic behavior as the
scalar field in the soft-wall model, with the temperature now given by
Eq.~(\ref{eq:IHQCD15}). At the AdS boundary, the tensor field likewise
has an asymptotic behavior analogous to that of the scalar field in the
soft-wall model. Consequently, the corresponding horizon and boundary
conditions can be imposed following the same procedure used in the
scalar sector.

\subsubsection{Energy Norm}

To study the stability of the quasinormal modes, following Ref.~\cite{Kovtun:2005ev,Katz:2005ir}, we identify the tensor glueball perturbation with an effective scalar field. We then use the corresponding energy-momentum tensor to characterize the perturbation $h$:
\begin{equation}\label{eq:stress_tensorh}
T_{MN}
=
\nabla_{M}h\,\nabla_{N}h
-\frac12 g_{MN}
\left(
\nabla_{S}h \,\nabla^{S}h
\right),
\end{equation}
The energy inner product for a given field is constructed from the leading-order
contribution of that field's perturbation to the energy-momentum tensor. Here, this inner product takes the  form
\begin{equation}\label{eq:inner_productu}
\langle u_1,u_2\rangle_E
=
\int_{t=\mathrm{const}}
d\rho\, \,
\Psi_1^{\dagger}(t,\rho)\,
\mathcal{G}
\!\left(
\overleftarrow{\partial}_{\rho},
\overrightarrow{\partial}_{\rho},\rho
\right)
\,
\Psi_2(t,\rho),
\end{equation}
where $\Psi_i=(h_i,\mathfrak{h}_i)^{T}$, and $\mathcal{G}$ is a matrix-valued differential operator of the form
\begin{equation}\label{eq:Gmatrix_scalarh}
\mathcal{G}_h
=
\begin{pmatrix}
\overleftarrow{\partial}_{\rho}
fb^{3}
\overrightarrow{\partial}_{\rho}
&
0\\[0.4cm]
0 &
(2-f)b^{3}
\end{pmatrix}.
\end{equation}

\section{Numerical Implementation}\label{NU}

The spectral stability of the quasinormal modes is investigated using pseudospectral methods. We first discretize the radial
coordinate $\rho$ using a Chebyshev collocation scheme. Specifically,
we employ the Chebyshev--Lobatto points
\begin{equation}\label{Cheb}
    \rho_j =
    \frac{1}{2}
    \left(
    1-\cos\frac{j\pi}{N}
    \right),
    \qquad
    j=0,\ldots,N,
\end{equation}
where $N$ is the polynomial degree, corresponding to $N+1$ collocation
points. Throughout this section, we use $\mathfrak{A}$ as a generic
symbol for the differential operators defined in
Eqs.~\eqref{opes}--\eqref{opev} and~\eqref{eq:IHQCD25},
as well as for their adjoints $\mathfrak{A}^\dagger$. The specific operator and field content depend
on the sector under consideration and are given explicitly in the
previous sections. The differential operator $\mathfrak{A}$ is
represented on the Chebyshev--Lobatto grid by means of Chebyshev
differentiation matrices, yielding the finite-dimensional matrix
approximation $\mathfrak{A}^N$. In our case, the resulting
quasinormal-mode problem takes the standard matrix eigenvalue form
$\mathfrak{A}^N u_N = \lambda u_N$.

The boundary conditions are incorporated directly into the spectral
discretization. Since regularity in the bulk is automatically captured
by the spectral representation, the required asymptotic behavior at the
AdS boundary is imposed explicitly. For the rescaled fields considered
here, this amounts to homogeneous Dirichlet boundary conditions. These
conditions are implemented by removing the rows and columns associated
with the boundary collocation point from the discretized operator.

To define the pseudospectrum in a way that preserves the physical
inner product introduced in the previous section, we discretize the
energy norm following the prescription of Ref.~\cite{Jaramillo2021}.
For the scalar and vector sectors of the soft-wall model, we denote the
corresponding field vectors by
\begin{equation}
    u=(\hat{\varphi},\hat{\chi})^T,
    \qquad
    u=(\tilde{V},\tilde{\xi})^T,
\end{equation}
respectively. For the tensor sector of the IHQCD model, we instead
consider
\begin{equation}
    u=(\hat{h},\hat{\frak{h}})^T.
\end{equation}

For each sector, we then construct a positive-definite Gram matrix
$G_E$ such that
\begin{equation}
    \lim_{N\rightarrow\infty}
    u_N^\dagger G_E u_N
    =
    \langle u,u\rangle_E,
\end{equation}
where $u_N$ denotes the spectral representation of the corresponding
field vector $u$. The explicit construction of the matrix $G_E$ is
given in Appendix~\ref{app:chebyshev}. The boundary conditions imposed
on the differential operators are consistently incorporated into
$G_E$ by removing the corresponding rows and columns.

The matrix $G_E$ induces an inner product and an associated norm on the
finite-dimensional space $\mathbb{C}^n$. In particular, the adjoint of
a matrix $B\in M_n(\mathbb{C})$ with respect to this inner product is
defined as
\begin{equation}
    B^\dagger
    =
    G_E^{-1}B^*G_E,
\end{equation}
where $B^*$ denotes the Hermitian conjugate of $B$. For the
discretized operator, this gives
\begin{equation}
    (\mathfrak{A}^N)^\dagger
    =
    G_E^{-1}(\mathfrak{A}^N)^*G_E.
\end{equation}

The $\epsilon$-pseudospectrum of $\mathfrak{A}^N$ with respect to the
energy norm is defined by
\begin{equation}
    \sigma_\epsilon^E(\mathfrak{A}^N)
    =
    \left\{
    \lambda\in\mathbb{C}:
    s_{\min}^E\bigl(M(\lambda)\bigr)<\epsilon
    \right\},
\end{equation}
where
\begin{equation}
    M(\lambda)
    =
    \lambda I-\mathfrak{A}^N,
\end{equation}
and $I$ is the $n\times n$ identity matrix. The smallest singular value with respect to the energy inner product of $M$ is given by
\begin{equation}
    s_{\min}^E(M)
    =
    \min_{\mu\in\sigma(M^\dagger M)}
    \sqrt{\mu},
\end{equation}
where the adjoint of $M$ is taken with respect to the energy inner
product,
\begin{equation}
    M^\dagger
    =
    G_E^{-1}M^*G_E.
\end{equation}

This formulation provides a finite-dimensional representation of the
continuous problem that preserves the physical inner product and hence
the corresponding notion of perturbations. It allows us to compute
both the condition numbers of the quasinormal modes and their
pseudospectra using the matrix formulation described in
Sec.~\ref{Pseudo}. All numerical calculations are performed using the
\textit{Wolfram Language}.

\begin{figure}[H]
\centering
\includegraphics[width=0.45\linewidth]{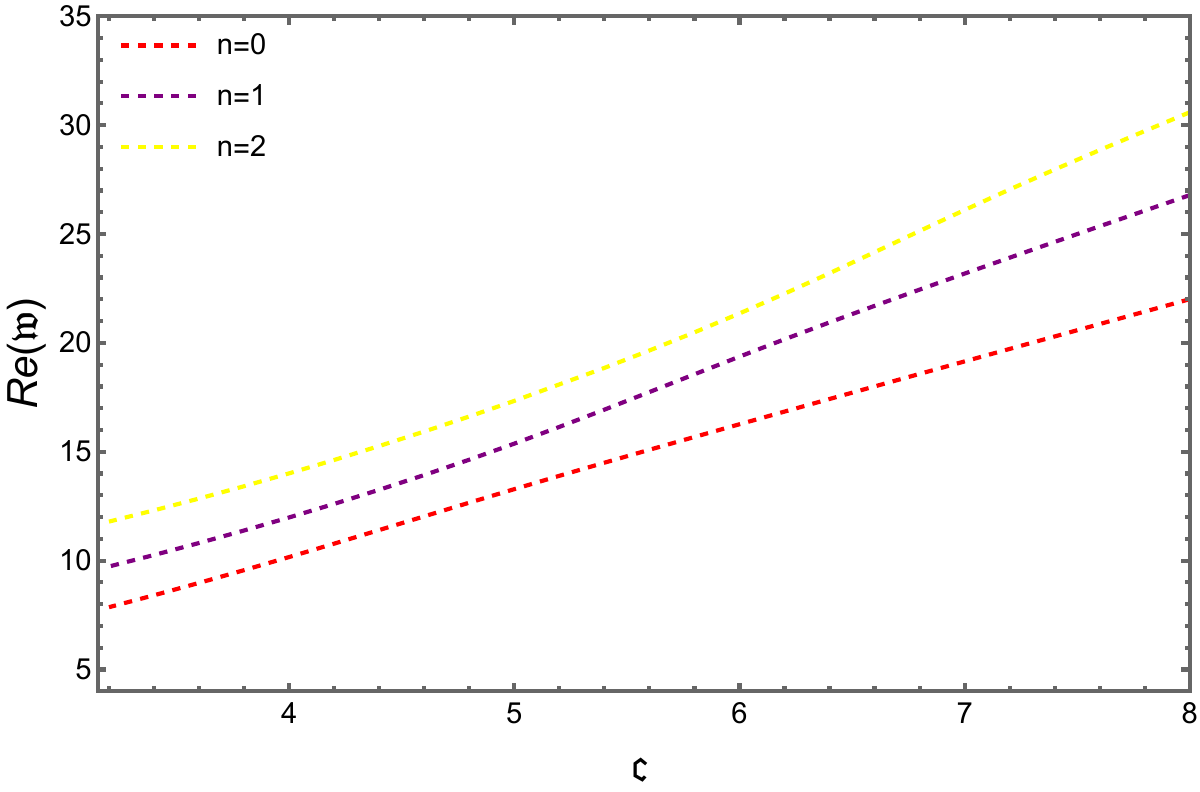}
\hfill
\includegraphics[width=0.45\linewidth]{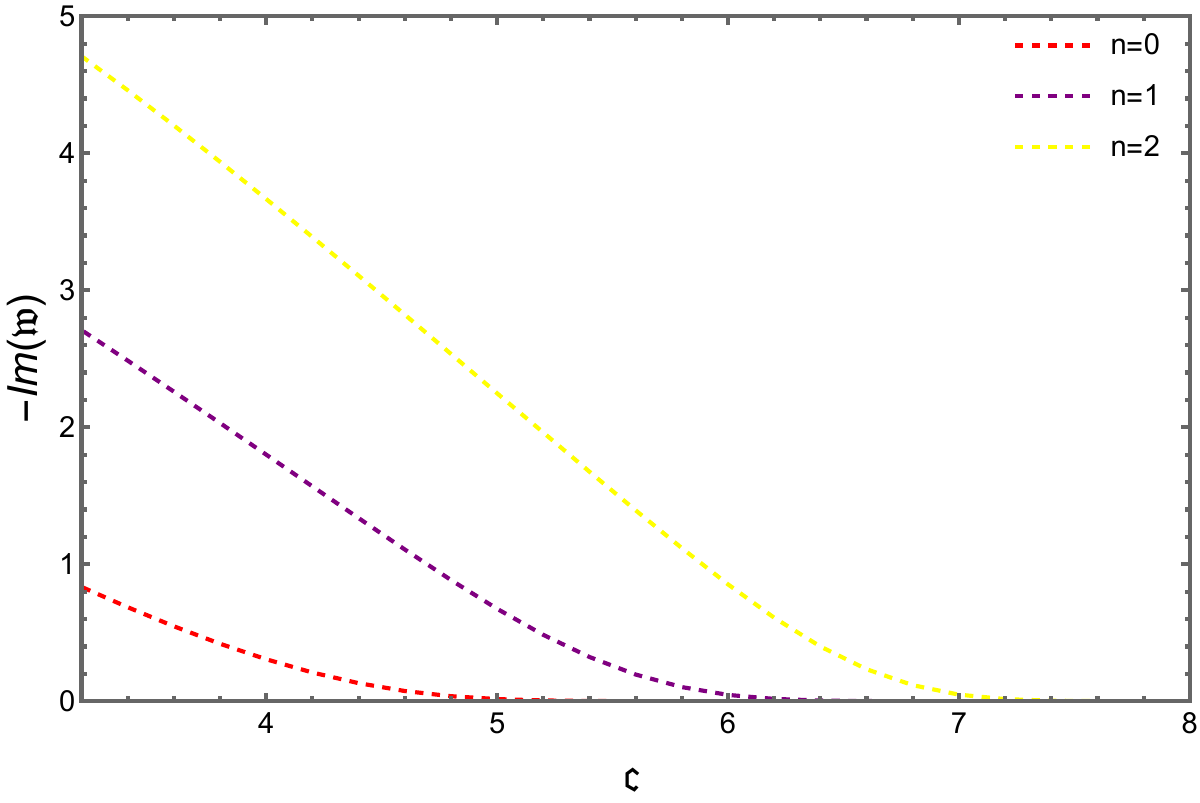}
\caption{Frequencies of the first three scalar glueball modes as
functions of $\mathfrak{c}$. The left panel shows the real parts of
the frequencies, while the right panel shows minus their imaginary parts.
}
\label{fig:qnm}
\end{figure}

\section{Results}\label{Re}

In this section, we present the numerical results for the pseudospectra of the quasinormal frequencies in the soft-wall and  Model A, following the procedure described above.

\subsection{Scalar Field}

The quasinormal mode spectra of the scalar field for different values of $\mathfrak{c}$ are shown in Fig.~\ref{fig:qnm}. The limit $\mathfrak{c}\rightarrow\infty$ corresponds to the low-temperature regime, whereas $\mathfrak{c}\rightarrow0$ corresponds to the high-temperature regime. As the temperature increases, the quasinormal mode spectrum gradually approaches that of the corresponding model without the dilaton, namely, the conformal AdS/CFT case. This behavior is expected, as the effects of the dilaton become progressively less significant at high temperatures.

\begin{figure}[H]
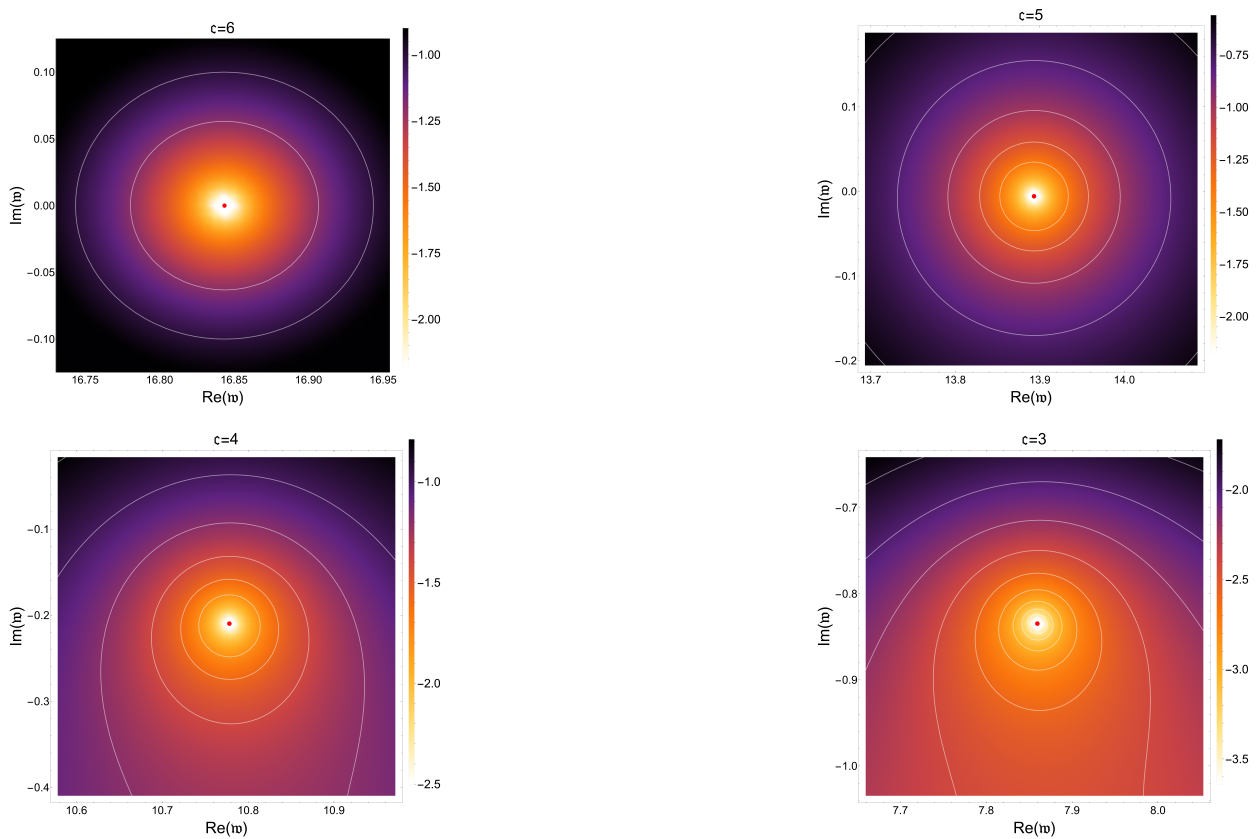

\centering

\includegraphics[width=0.35\linewidth]{Figure_3.png}
\hfill
\includegraphics[width=0.35\linewidth]{Figure_4.png}

\vspace{0.3cm}

\includegraphics[width=0.35\linewidth]{Figure_5.png}
\hfill
\includegraphics[width=0.35\linewidth]{Figure_6.png}

\caption{Close-up of the scalar glueball pseudospectrum in the energy norm around the ground-state quasinormal frequencies for different values of $\mathfrak{c}$. The red dot marks the corresponding quasinormal frequency, while the white curves delineate the boundaries of the full $\epsilon$-pseudospectra. The heatmap displays the base-10 logarithm of the inverse resolvent norm.}
\label{fig:panel}
\end{figure}

In Fig.~\ref{fig:panel}, we present the pseudospectra of the scalar-field ground state computed using the energy norm for different values of $\mathfrak{c}$. For $\mathfrak{c}=6$ and $\mathfrak{c}=5$, the pseudospectral contours remain relatively close to the
quasinormal frequency, indicating comparatively weak spectral instability. As $\mathfrak{c}$ decreases, the pseudospectral contours become increasingly extended, indicating greater instability of the quasinormal mode to perturbations. This behavior correlates with the thermal modification of the corresponding resonance and its eventual disappearance in the spectral function.

\begin{figure}[H]
    \centering
    \includegraphics[width=0.6\textwidth]{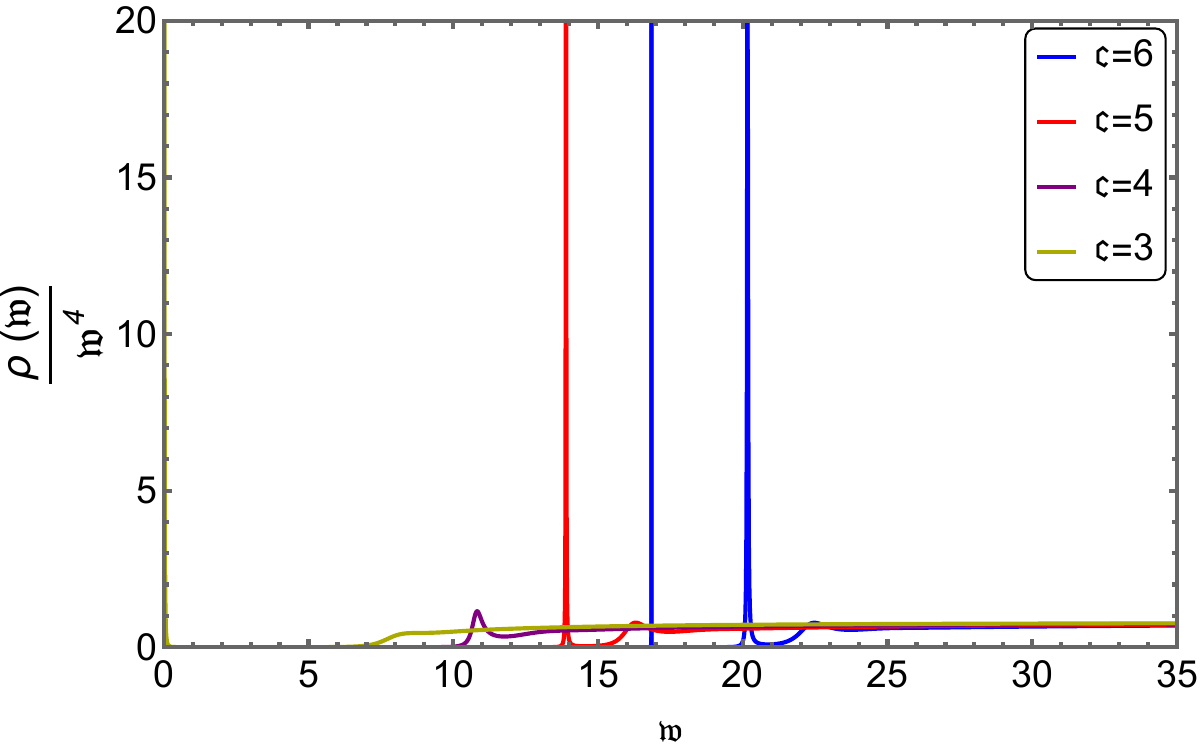}
    \caption{We show the spectral function of the scalar glueball, whose calculation is presented in  Appendix \ref{app:spectral_function}, for various values of $\mathfrak{c}$. The blue curves correspond to $\mathfrak{c}=6$, while $\mathfrak{c}=5$, $\mathfrak{c}=4$, and $\mathfrak{c}=3$ are represented by the red, purple, and dark-yellow curves, respectively. }
    \label{fig:figurex}
\end{figure}

To further investigate this behavior, we plot in
Fig.~\ref{fig:figurex} the spectral function of the scalar field for different values of $\mathfrak{c}$. The progressive broadening and eventual disappearance of the resonance peaks correlate with the enhanced spectral instability observed in the pseudospectral
analysis. In particular, for $\mathfrak{c}\lesssim 3$, the resonance becomes strongly broadened and eventually disappears from the spectral function, indicating the dissociation of the corresponding scalar state.

\begin{figure}[t]
    \centering
    \includegraphics[width=0.5\textwidth]{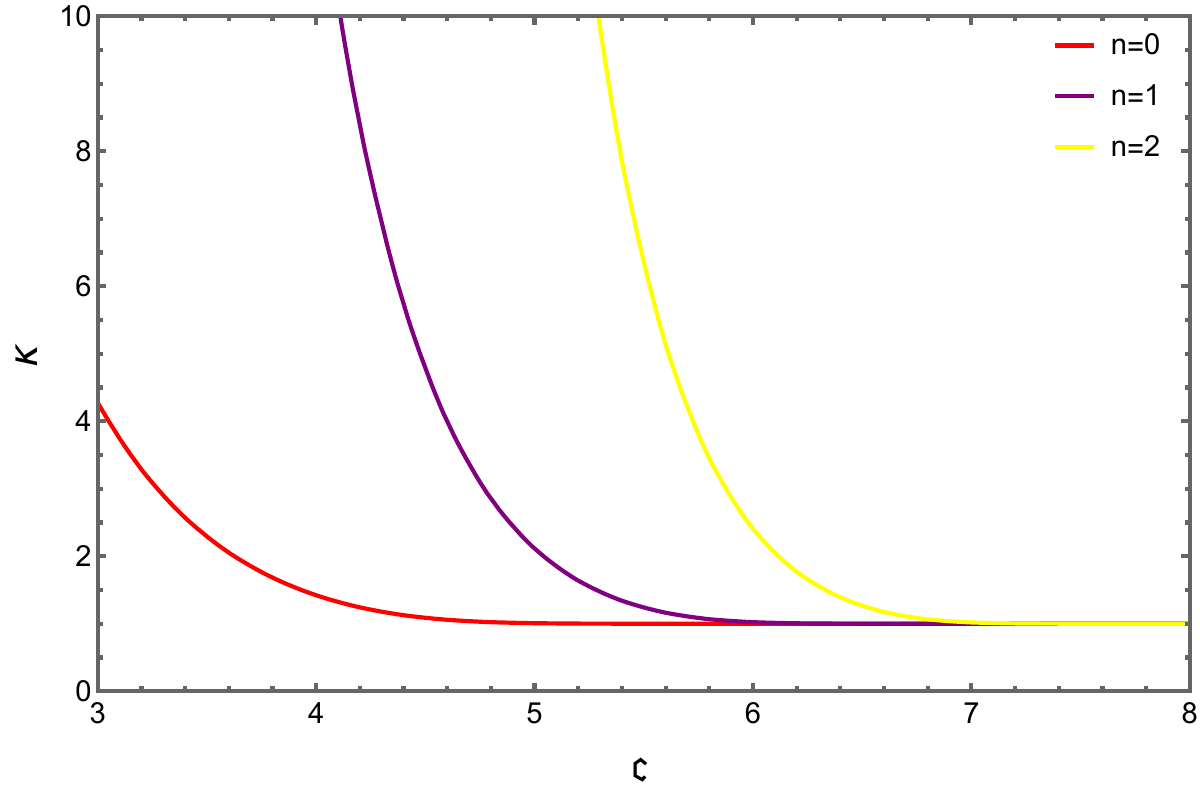}
    \caption{The condition number, $\kappa$, for the scalar glueball, computed
numerically, as a function of
$\mathfrak{c}$. The red, purple, and yellow curves correspond to the
ground, first excited, and second excited states, respectively.}
    \label{fig:figure4}
\end{figure}

To quantify the spectral instability, Fig.~\ref{fig:figure4} shows the condition number $\kappa$ as a function of $\mathfrak{c}$. Values of $\kappa\simeq1$ correspond to comparatively well-conditioned quasinormal modes with weak spectral
instability. As shown in the figure, $\kappa$ remains close to unity for sufficiently large values of $\mathfrak{c}$, corresponding to the low-temperature regime. As $\mathfrak{c}$ decreases toward the high-temperature regime, $\kappa$ increases rapidly, providing a quantitative measure of the enhanced spectral instability observed in the pseudospectra.
The simultaneous increase in the resonance width and its eventual disappearance in the spectral function therefore provide independent evidence for the thermal dissociation of the scalar state.

\subsection{Vector Field}

\begin{figure}[!ht]
\centering
\includegraphics[width=0.45\linewidth]{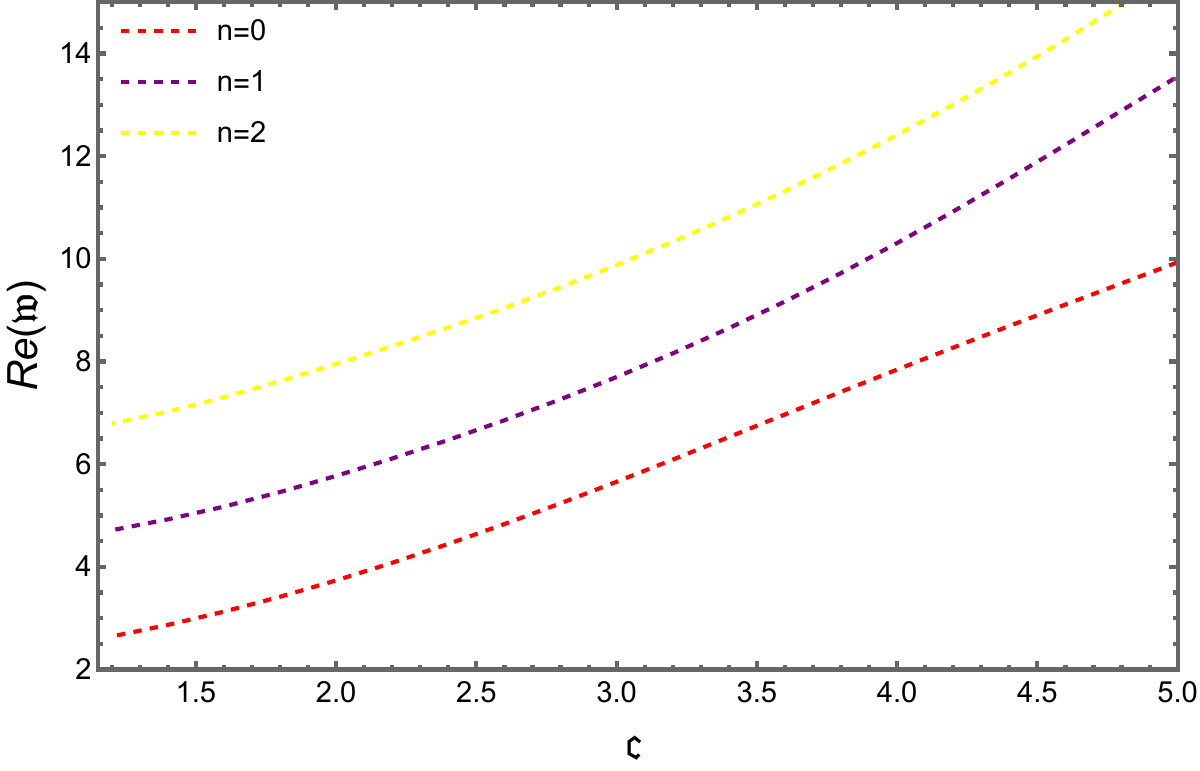}
\hfill
\includegraphics[width=0.45\linewidth]{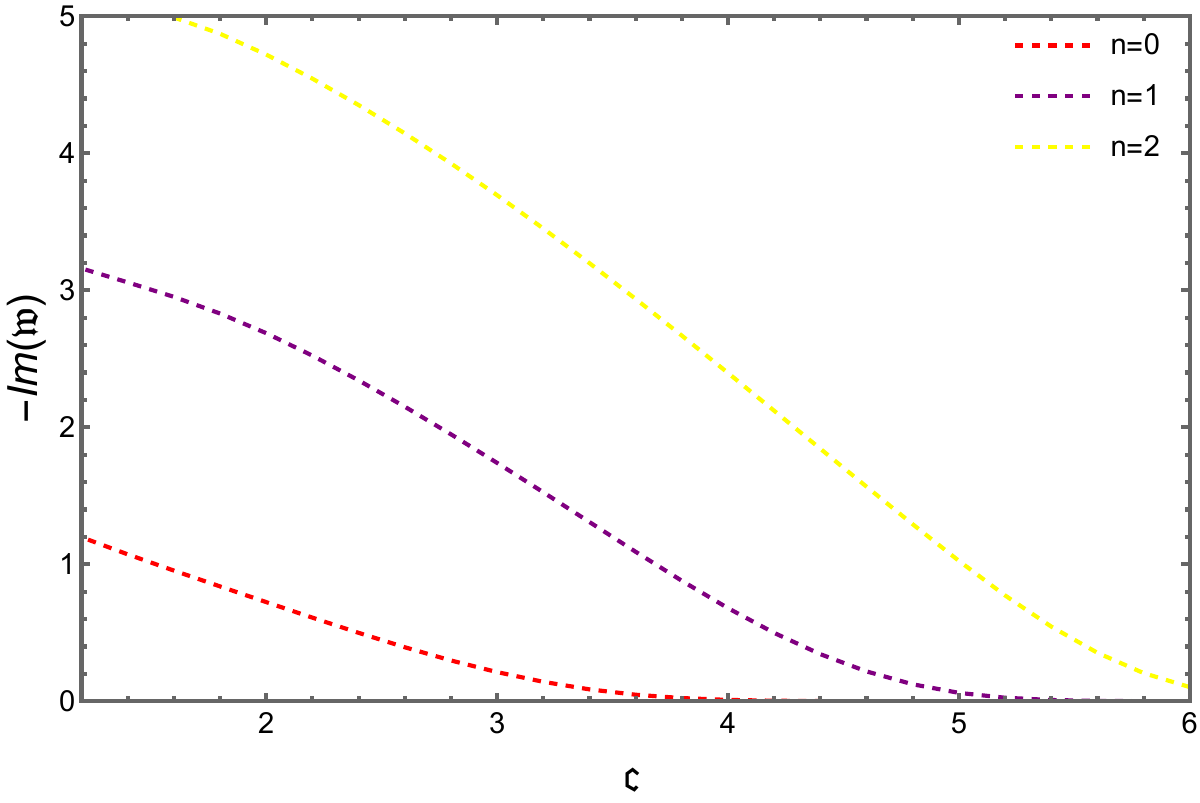}
\caption{The frequencies of the first three vector meson modes, computed as functions of $\mathfrak{c}$. The left panel shows the real parts of the frequencies, while the right panel shows minus their imaginary parts.}
\label{fig:qnmVec}
\end{figure}

We present in Fig.~\ref{fig:qnmVec} the quasinormal mode spectrum of the vector field for different values of the infrared energy scale $\mathfrak{c}$. As in the scalar case, decreasing $\mathfrak{c}$ corresponds to increasing temperature, whereas the low-temperature regime is recovered as $\mathfrak{c}\rightarrow\infty$.

\begin{figure}[t]
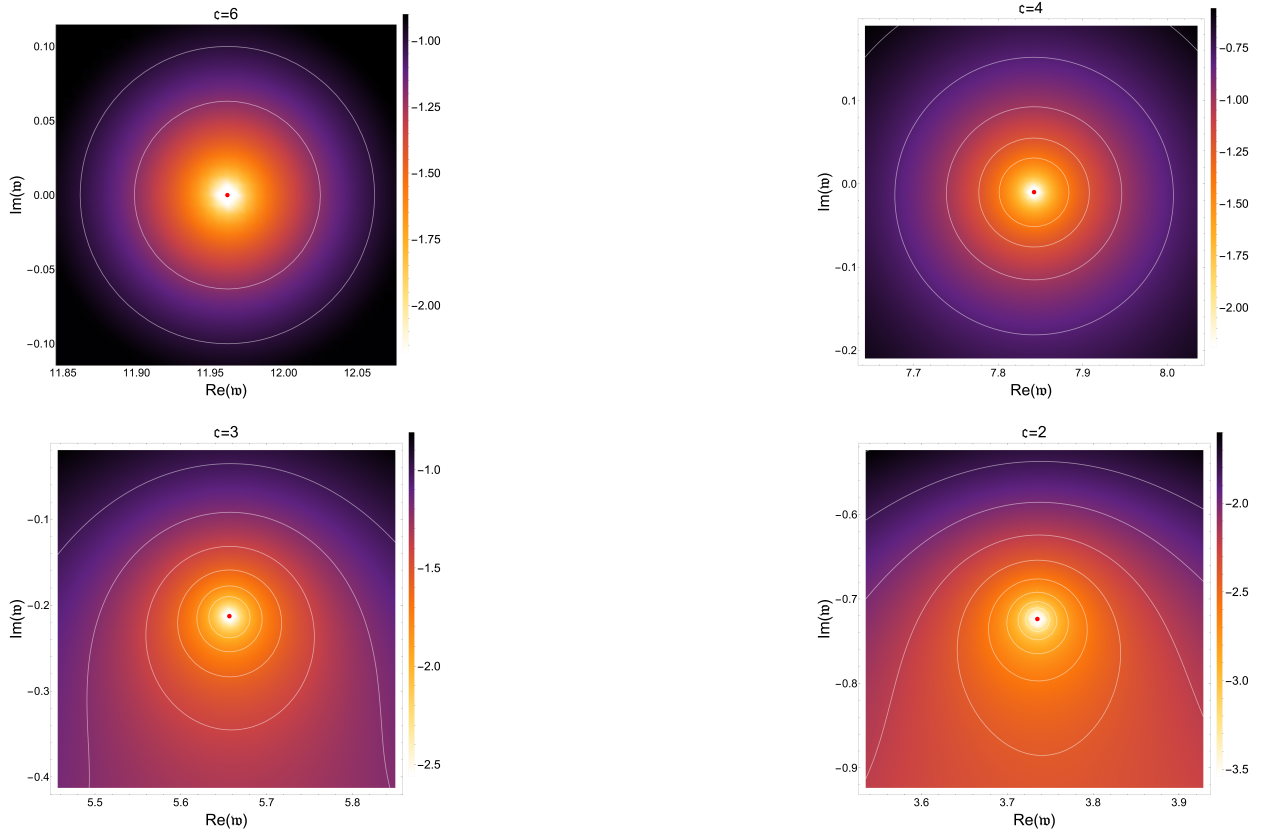

\centering

\includegraphics[width=0.35\linewidth]{Figure_11.png}
\hfill
\includegraphics[width=0.35\linewidth]{Figure_12.png}

\vspace{0.3cm}

\includegraphics[width=0.35\linewidth]{Figure_13.png}
\hfill
\includegraphics[width=0.35\linewidth]{Figure_14.png}

\caption{Close-up of the vector meson pseudospectrum in the energy norm around the ground-state quasinormal frequencies for different values of $\mathfrak{c}$. The red dot marks the corresponding quasinormal frequency, while the white curves delineate the boundaries of the full $\epsilon$-pseudospectra. The heatmap displays the base-10 logarithm of the inverse resolvent norm.}
\label{fig:panel2}
\end{figure}

The pseudospectra of the vector-field ground state, computed using the energy norm, are shown in  Fig.~\ref{fig:panel2}. For $\mathfrak{c}=6$ and $\mathfrak{c}=5$, the pseudospectral contours remain concentrated near the quasinormal frequency, reflecting the low instability of the mode to perturbations. As $\mathfrak{c}$ decreases, corresponding to increasing temperature, the pseudospectral contours progressively extend, signaling an increasing instability of the
quasinormal mode to perturbations. This behavior is illustrated by the cases $\mathfrak{c}=4$ and $\mathfrak{c}=3$, where the pseudospectral contours become significantly more extended.

\begin{figure}[!ht]
    \centering
    \includegraphics[width=0.5\textwidth]{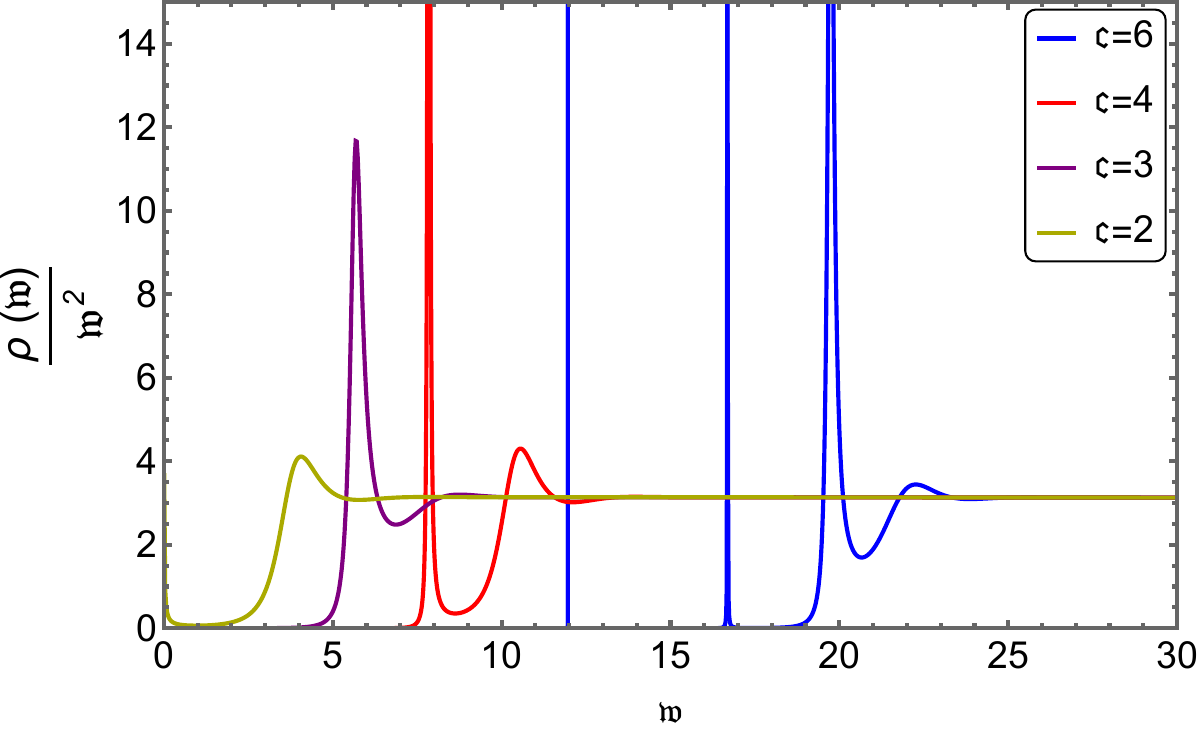}
    \caption{We show the spectral function of the vector case, whose calculation is presented in  Appendix \ref{app:spectral_function}, for various values of $\mathfrak{c}$. The blue curves correspond to $\mathfrak{c}=6$, while $\mathfrak{c}=4$, $\mathfrak{c}=3$, and $\mathfrak{c}=2$ are represented by the red, purple, and dark-yellow curves, respectively. Notice that the ground-state peak practically disappears for $\mathfrak{c}<3$.}
    \label{fig:figurex2}
\end{figure}

This increased spectral instability is correlated with the increasing widths of the vector-meson resonances. As shown in the spectral functions displayed in Fig.~\ref{fig:figurex2}, the resonance peaks
broaden and eventually disappear as the temperature increases, providing evidence for the thermal dissociation of the corresponding vector mesons. Consistent with the scalar case, the spectral-function analysis agrees with the pseudospectral results. In addition, Appendix~\ref{app:spectral_function} presents a comparison between the quasinormal modes and spectral-function analyses, showing good agreement between the peak position and width extracted from the spectral function and the real and imaginary parts of the corresponding quasinormal frequency for the cases considered.

\begin{figure}[!ht]
    \centering
    \includegraphics[width=0.45\textwidth]{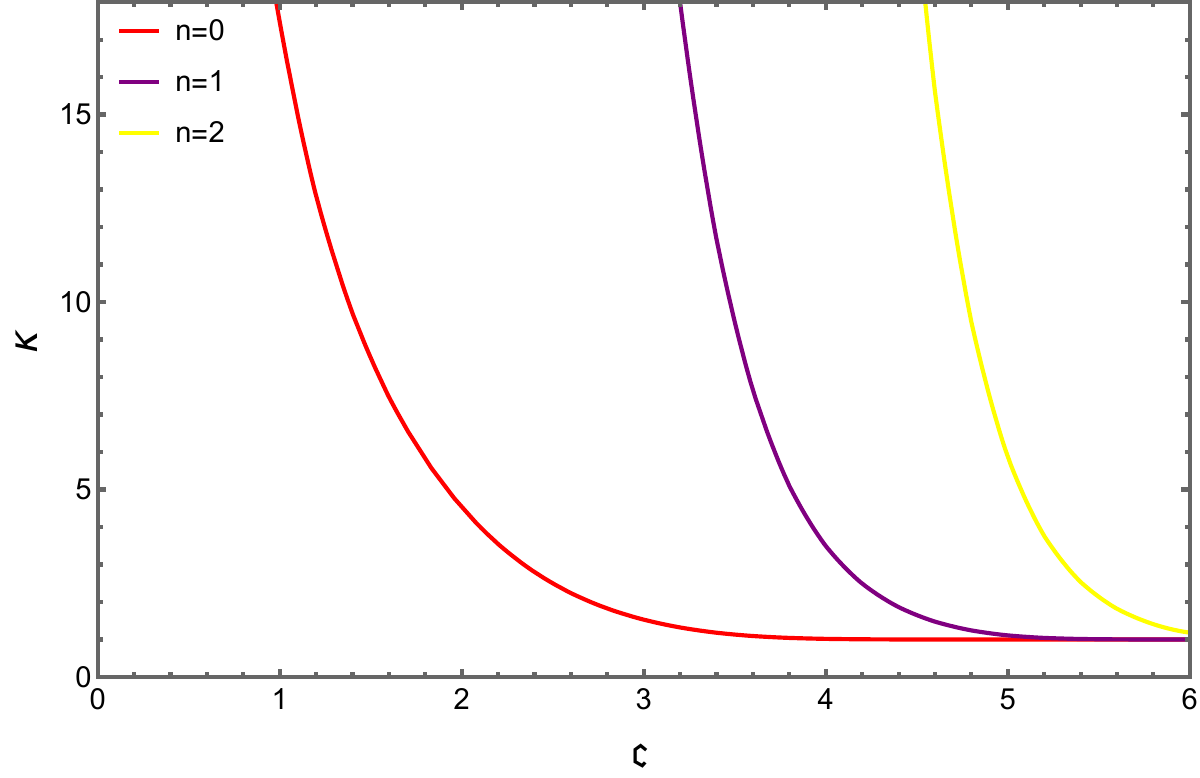}
    \caption{The condition number, $\kappa$, of the vector meson, computed
numerically, as a function of $\mathfrak{c}$. The red, purple, and
yellow curves correspond to the ground, first excited, and second
excited states, respectively.}
    \label{fig:figure44}
\end{figure}

Finally, Fig.~\ref{fig:figure44} shows the condition number $\kappa$ as a function of $\mathfrak{c}$. The behavior of $\kappa$ is consistent with the  pseudospectral analysis and provides a quantitative characterization of the spectral instability of the vector-field ground state. In particular, $\kappa$ remains close to unity in the low-temperature regime, indicating comparatively well-conditioned quasinormal modes with weak spectral instability. As $\mathfrak{c}$ decreases, $\kappa$ increases rapidly, providing a quantitative measure of the increasing instability of the quasinormal frequency to perturbations. This increase occurs in the same regime in which the spectral function exhibits a substantial increase in the resonance width and eventual disappearance, providing complementary evidence for the thermal dissociation of the vector meson.

\subsection{Model A}

\begin{figure}[t]
\centering
\includegraphics[width=0.45\linewidth]{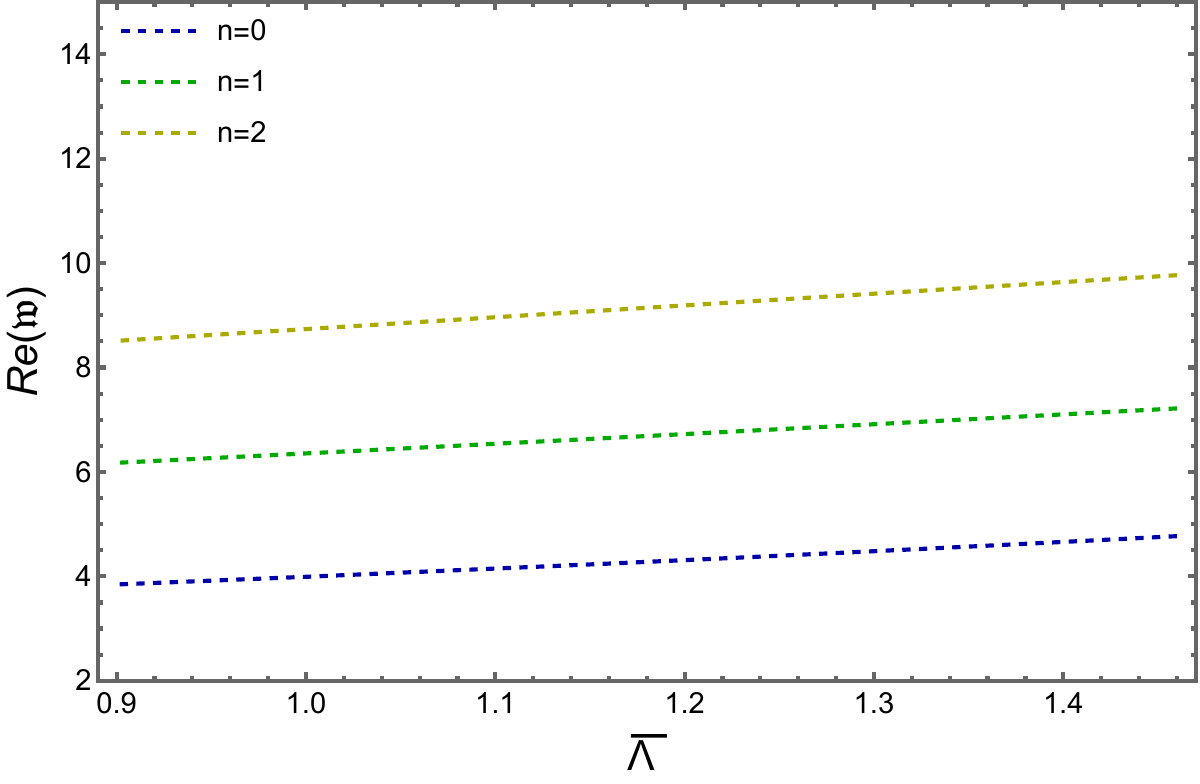}
\hfill
\includegraphics[width=0.459\linewidth]{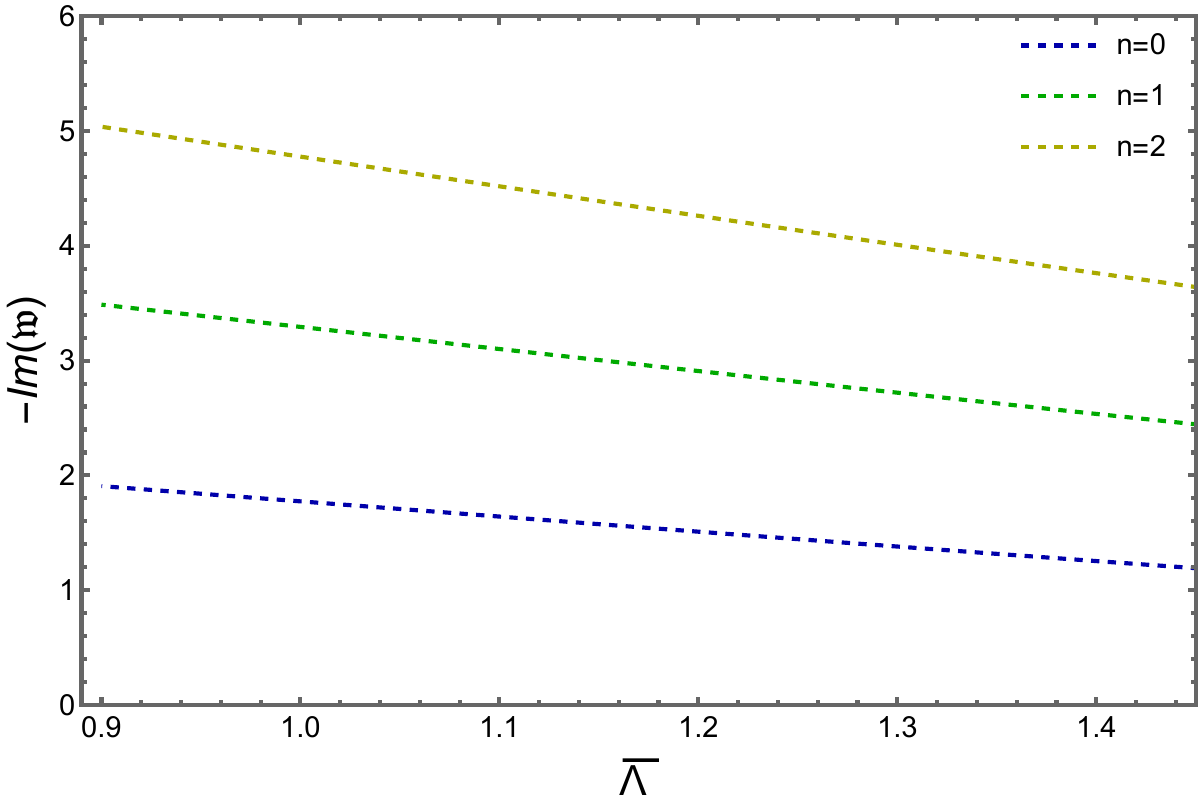}
\caption{The real and imaginary parts of the frequencies of the first three tensor glueball modes as functions of $\bar{\Lambda}$. The left panel shows the real parts of the frequencies, while the right panel shows minus their imaginary parts.}
\label{fig:qnmIHQCD}
\end{figure}

Let us now discuss the results for the holographic QCD model~\cite{
Gursoy:2007cb,Gursoy:2007er,Gursoy:2008bu,Kajantie:2011nx,Alanen:2011hh}. The quasinormal frequencies obtained for the tensor glueball are shown in Fig.~\ref{fig:qnmIHQCD} for values of $\bar{\Lambda}<\bar{\Lambda}_{\min}$. 
The results indicate that the tensor glueball is already in the dissociation regime, as suggested by the large damping rates of the corresponding quasinormal modes. We present our results for temperatures above the minimum temperature, corresponding to $\bar{\Lambda}_{\min}\approx1.46$. Values of $\bar{\Lambda}$ below $\bar{\Lambda}_{\min}$ correspond to the large-black-hole branch.

\begin{figure}[ht]
    \centering
    \includegraphics[width=0.5\textwidth]{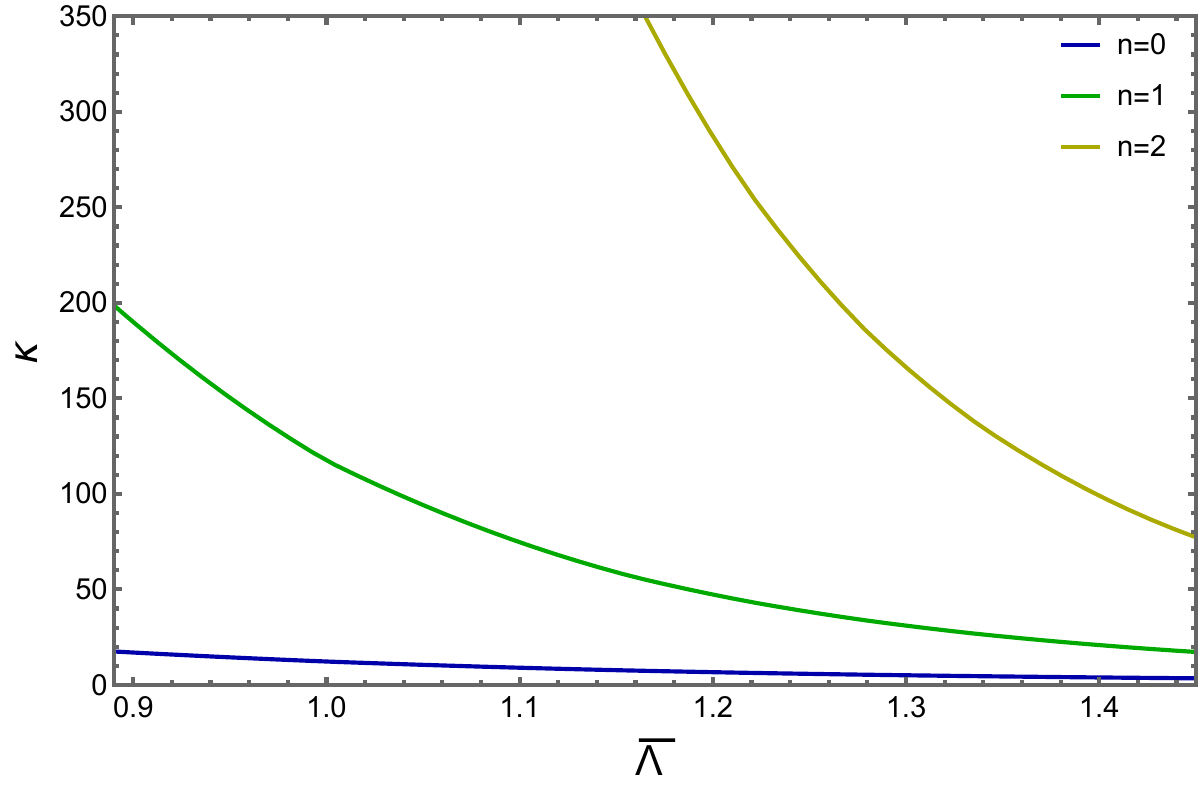}
    \caption{The condition number, $\kappa$, of the tensor glueball as a function of
$\bar{\Lambda}$, obtained numerically for Model A. The blue, green, and
yellow curves correspond to the ground, first excited, and second
excited states, respectively.}
    \label{fig:figureglueball}
\end{figure}

\begin{figure}[ht]
    \centering
    \includegraphics[width=0.5\textwidth]{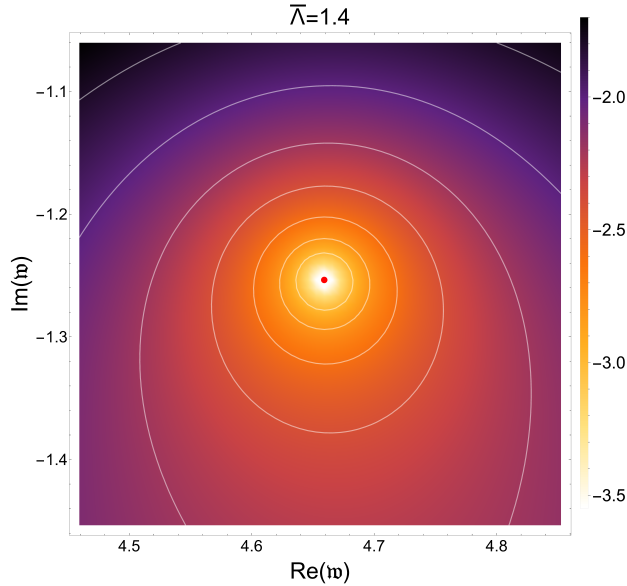}
    \caption{Close-up of the tensor glueball pseudospectrum in the energy norm around the ground-state quasinormal frequency for $\bar{\Lambda}=1.4$. The red
dot marks the corresponding quasinormal frequency, while the white curves delineate the
boundaries of the full $\epsilon$-pseudospectra. The heatmap shows the
base-10 logarithm of the inverse resolvent norm.}
    \label{fig:figureglueball2}
\end{figure}

Figure~\ref{fig:figureglueball} shows the condition numbers $\kappa$ as functions of $\bar{\Lambda}$ for the first three quasinormal modes. In
particular, the condition numbers reach very large values, indicating
that even small perturbations can produce significant changes in the
quasinormal frequencies. This strong spectral instability is consistent
with the rapid dissociation of the tensor glueball in the thermal
medium. To further support this interpretation, we present in
Fig.~\ref{fig:figureglueball2} the pseudospectra of the tensor glueball
ground state for $\bar{\Lambda}=1.4$. The pseudospectral analysis reveals the high sensitivity of the quasinormal modes to perturbations, manifested by the increasingly extended pseudospectral contours, indicating strong spectral instability.

\section{Conclusion and Discussion}\label{Con}

In this paper, we have analyzed the spectral stability of quasinormal modes in holographic QCD models. We first investigated the scalar and vector fields in the simplest soft-wall model and then extended the analysis to the quasinormal modes associated with the tensor glueball in an improved holographic QCD model.

The results obtained for the scalar field, which according to the gauge/gravity duality corresponds to the scalar glueball in the dual gauge theory, show that the quasinormal frequencies become increasingly spectrally stable as the temperature approaches zero. This behavior is consistent with the spectral theorem: in the zero-temperature limit, the horizon recedes toward the IR and the quasinormal-mode problem approaches the corresponding normal-mode problem governed by a self-adjoint operator. As the temperature increases, however, the spectral instability increases, and the quasinormal frequencies become progressively more sensitive to perturbations.

These results are also consistent with the usual holographic interpretation of quasinormal modes. As the temperature increases, the glueball state gradually dissociates in the thermal medium. At finite temperature, the bound states are replaced by quasibound states, whose frequencies are described by quasinormal modes. The increasing spectral instability of the quasinormal modes therefore
provides a complementary signature of the progressive dissociation
of the glueball in the thermal medium.

The results for the scalar sector are qualitatively similar to those obtained for the vector sector. The main difference is that, in the latter case, the corresponding state in the dual gauge theory is identified with a vector meson. The two sectors exhibit different degrees of spectral instability and different temperature dependence. In both cases, however, the pseudospectral analysis shows an increasing instability of the quasinormal frequencies as
$\mathfrak{c}$ decreases, in correlation with the attenuation and eventual disappearance of the corresponding resonance peaks in the spectral function. For the scalar field, the pseudospectral analysis indicates that the contours begin to become increasingly extended at $\mathfrak{c}\approx 4.2$, whereas for the vector case, a similar behavior is observed at $\mathfrak{c}\approx 3.5$. The corresponding condition numbers are $\kappa_{\varphi}\approx 1.2424$ and $\kappa_V\approx 1.13234$, respectively. These condition numbers provide a quantitative measure of the increased spectral instability. In Appendix \ref{Condition Number}, we present numerical values of the condition numbers as a function of the dimensionless parameter $\mathfrak{c}$ for the soft-wall model.

For holographic Model A, the quasinormal frequencies obtained in this work are restricted to temperatures above the minimum temperature of the model. This restriction arises from the existence of a minimum temperature below which the corresponding black-hole solutions cease to exist. Consequently, the low-temperature regime cannot be explored within this model. Moreover, above the minimum temperature, the tensor glueball is already in the dissociation regime. These observations are consistent with the pseudospectral analysis, which yields condition numbers significantly larger than unity, indicating a high degree of spectral instability throughout the accessible temperature range.

Finally, we remark that a natural extension of this work would be to investigate the pseudospectra of quasinormal modes in other holographic QCD models, such as those considered in Refs.~\cite{Alho:2012mh,Jena:2024cqs,Ballon-Bayona:2021ibm}. Furthermore, it would be interesting to study the effects of a magnetic field \cite{Braga:2018zlu}, a chemical potential \cite{Braga:2017oqw}, and rotation \cite{Chen:2020ath,Braga:2022yfe,Ferreira:2025iqe} on the pseudospectra. These extensions would provide a complementary investigation of quasinormal modes and offer further insights into the behavior of hadrons in a medium under different physical conditions, including  nonzero chemical potential, external magnetic fields, and rotation.

\noindent\textbf{Acknowledgments.}
L.F.F. is supported by the ANID Fondecyt Postdoctoral
Grant No.~3220304.

\appendix

\section{Discretization in the Chebyshev Grid}
\label{app:chebyshev}

In this appendix, we provide further details regarding the discretization procedure.

\subsection{Collocation Method}
\label{app:collocation}

Since we employ the Chebyshev grid~\eqref{Cheb}, consisting of $N+1$ points, the function $f(\rho)$ is approximated by a truncated expansion in Chebyshev polynomials
\begin{equation}
f(\rho) \approx \sum_{n=0}^{N} c_n T_n(2\rho-1),
\label{eq:cheb_expansion}
\end{equation}
where $T_n$ are the Chebyshev polynomials,
\begin{equation}
T_n(x) = \cos\left(n\arccos(x)\right).
\label{eq:cheb_polynomials}
\end{equation}

Following Ref.~\cite{Arean:2023ejh}, we work with the values of $f(\rho)$
at the collocation points of the Chebyshev grid~\eqref{Cheb}, rather than
with the expansion coefficients $\{c_n\}$ directly. The relation between
these two representations is established using the orthogonality relation
of the Chebyshev polynomials,
\begin{equation}
\int_0^1
\frac{d\rho}{\sqrt{1-(2\rho-1)^2}}\,
T_m(2\rho-1)T_n(2\rho-1)
=
\frac{\pi}{4}\delta_{mn}
\left(1+\delta_{n0}\right).
\label{eq:cheb_orthogonality}
\end{equation}
On the Chebyshev–Lobatto grid, the discrete orthogonality relation is approximated by
\begin{equation}
\frac{\pi}{N}
\sum_{j=0}^{N}
\frac{
T_m(2\rho_j-1)T_n(2\rho_j-1)
}{
1+\delta_{j0}+\delta_{jN}
}
\approx
\frac{\pi}{2}\delta_{mn}
\left(1+\delta_{n0}+\delta_{nN}\right).
\label{eq:discrete_orthogonality}
\end{equation}
Consequently, the expansion coefficients $c_n$ can be approximated in
terms of the values of $f$ at the grid points as
\begin{equation}
c_n =
\frac{2/N}{1+\delta_{n0}+\delta_{nN}}
\sum_{j=0}^{N}
\frac{
f(\rho_j)T_n(2\rho_j-1)
}{
1+\delta_{j0}+\delta_{jN}
}.
\label{eq:cheb_coefficients}
\end{equation}

\subsection{Construction of the $G_E$ Matrix}
\label{app:GE}

Once the functions on the grid have been expressed in terms of
Chebyshev polynomials, we can proceed to discretize a generic integral,
\begin{equation}
\int_0^1 d\rho\, f(\rho)g(\rho).
\label{eq:generic_integral}
\end{equation}

For the Chebyshev polynomials, we have
\begin{equation}
\int_0^1 d\rho\, T_n(2\rho-1)
=
\begin{cases}
0, & n \ \mathrm{odd}, \\[0.2cm]
\dfrac{1}{1-n^2}, & n \ \mathrm{even}.
\end{cases}
\label{eq:cheb_integral}
\end{equation}

Using Eq.~\eqref{eq:cheb_coefficients}, the integral in
Eq.~\eqref{eq:generic_integral} can be approximated as
\begin{align}
\int_0^1 d\rho\, f(\rho)g(\rho)
&\approx
\sum_{\substack{n=0 \\ n\,\mathrm{even}}}^{N}
\frac{1}{1-n^2}
\frac{2/N}{1+\delta_{n0}+\delta_{nN}}
\sum_{j=0}^{N}
\frac{
f(\rho_j)g(\rho_j)T_n(2\rho_j-1)
}{
1+\delta_{j0}+\delta_{jN}
}
\nonumber\\
&=
\sum_{j=0}^{N}
f(\rho_j)
\left[
\sum_{\substack{n=0 \\ n\,\mathrm{even}}}^{N}
\frac{2/N}{1-n^2}
\frac{
T_n(2\rho_j-1)
}{
\left(1+\delta_{n0}+\delta_{nN}\right)
\left(1+\delta_{j0}+\delta_{jN}\right)
}
\right]
g(\rho_j)
\nonumber\\
&=
\sum_{i=0}^{N}\sum_{j=0}^{N}
\mu_{ij}f(\rho_i)g(\rho_j),
\label{eq:discrete_integral}
\end{align}
where we have introduced the diagonal weight matrix $\mu$,
\begin{equation}
\mu_{ij}
=
\delta_{ij}
\sum_{\substack{n=0 \\ n\,\mathrm{even}}}^{N}
\frac{2/N}{1-n^2}
\frac{
T_n(2\rho_i-1)
}{
\left(1+\delta_{n0}+\delta_{nN}\right)
\left(1+\delta_{i0}+\delta_{iN}\right)
}.
\label{eq:mu_matrix}
\end{equation}
With this result, we can construct the matrix $G_E$. We first
factorize the operator $\mathcal{G}$, introduced in
Eqs.~\eqref{eq:Gmatrix_scalar}, \eqref{eq:Gmatrix_vector} and
\eqref{eq:Gmatrix_scalarh}, into a left-acting part and a right-acting part
\begin{equation}
\mathcal{G}\left(
\overleftarrow{\partial}_{\rho},
\overrightarrow{\partial}_{\rho};\rho
\right)
=
\begin{pmatrix}
\widetilde{\mathfrak{H}}_{11}
\left(\overleftarrow{\partial}_{\rho};\rho\right)
&
\widetilde{\mathfrak{H}}_{12}
\left(\overleftarrow{\partial}_{\rho};\rho\right)
\\[0.2cm]
\widetilde{\mathfrak{H}}_{21}[\rho]
&
\widetilde{\mathfrak{H}}_{22}[\rho]
\end{pmatrix}
\begin{pmatrix}
\widetilde{\mathfrak{H}}_{11}
\left(\overrightarrow{\partial}_{\rho};\rho\right)
&
\widetilde{\mathfrak{H}}_{12}[\rho]
\\[0.2cm]
\widetilde{\mathfrak{H}}_{21}
\left(\overrightarrow{\partial}_{\rho};\rho\right)
&
\widetilde{\mathfrak{H}}_{22}[\rho]
\end{pmatrix}.
\label{eq:G_factorization}
\end{equation}

Next, we discretize the differential operators
$\widetilde{\mathfrak{H}}_{ab}$ and $\mathfrak{H}_{ab}$, obtaining the corresponding
$(N+1)\times(N+1)$ matrices $\widetilde{\mathfrak{H}}_{ab}$ and $\mathfrak{H}_{ab}$.
The matrix $G_E$ is then constructed as
\begin{equation}
G_E =
\begin{pmatrix}
\widetilde{\mathfrak{H}}_{11} & \widetilde{\mathfrak{H}}_{12}\\
\widetilde{\mathfrak{H}}_{21} & \widetilde{\mathfrak{H}}_{22}
\end{pmatrix}
\begin{pmatrix}
\mu & 0\\
0 & \mu
\end{pmatrix}
\begin{pmatrix}
\mathfrak{H}_{11} & \mathfrak{H}_{12}\\
\mathfrak{H}_{21} & \mathfrak{H}_{22}
\end{pmatrix}.
\label{eq:GE_matrix}
\end{equation}

With this definition, as required, the operator $\mathcal{G}$ first acts on the
vectors, after which the integration over the radial coordinate is
performed through the matrix $\mu$.

\subsection{Interpolation Between Grids}
\label{app:interpolation}

When performing the integration as described in the previous subsection,
some information about the original functions is lost. This can be
understood from the fact that, although $2(N+1)$ coefficients are
required to represent the approximate functions $f(\rho)$ and $g(\rho)$,
the integral in Eq.~\eqref{eq:generic_integral} is evaluated on a grid
containing only $N+1$ points.

To minimize this effect in the construction of $G_E$, we first construct
the matrix on a grid with $M+1$ points and subsequently interpolate it
onto the original grid. To this end, we introduce the interpolation
matrix $\widetilde{I}$, which relates the two grids. Denoting the points
of the new grid by $\varrho_i$, the interpolation matrix is defined by
\begin{equation}
\sum_{j=0}^{N}
\widetilde{I}_{ij}f(\rho_j)
=
f(\varrho_i)
=
\sum_{n=0}^{N}
\frac{2/N}{1+\delta_{n0}+\delta_{nN}}
\sum_{j=0}^{N}
\frac{
f(\rho_j)T_n(2\rho_j-1)
}{
1+\delta_{j0}+\delta_{jN}
}
T_n(2\varrho_i-1).
\label{eq:interpolation_definition}
\end{equation}

It follows that
\begin{equation}
\widetilde{I}_{ij}
=
\sum_{n=0}^{N}
\frac{2/N}{1+\delta_{n0}+\delta_{nN}}
\frac{
T_n(2\varrho_i-1)T_n(2\rho_j-1)
}{
1+\delta_{j0}+\delta_{jN}
}.
\label{eq:interpolation_matrix}
\end{equation}

Finally, denoting by $G_E^{(M)}$ the matrix $G_E$ constructed on the
grid with $M+1$ points, the corresponding matrix on the original grid
with $N+1$ points is given by
\begin{equation}
G_E =
\begin{pmatrix}
\widetilde{I} & 0\\
0 & \widetilde{I}
\end{pmatrix}^{\!T}
G_E^{(M)}
\begin{pmatrix}
\widetilde{I} & 0\\
0 & \widetilde{I}
\end{pmatrix}.
\label{eq:GE_interpolation}
\end{equation}

\section{Spectral Function}
\label{app:spectral_function}

\subsection{Scalar Field}
Here, we present the derivation of the spectral function for a scalar field within the holographic framework, following the holographic prescription of Ref.~\cite{Son:2002sd}. The on-shell action for the scalar field reduces to a boundary term,
\begin{equation}\label{eqApC1}
S_{\mathrm{on-shell}}
=
\int d^4x\,
\sqrt{-g}\,e^{-\Phi}g^{zz}
\varphi\,\partial_z\varphi,
\end{equation}
where we have neglected the contribution from the horizon boundary.

After performing a Fourier transform of the scalar field, the on-shell
action can be written as
\begin{equation}\label{eqApC2}
S_{\mathrm{on-shell}}
=
\int \frac{d^4k}{(2\pi)^4}\,
\left.
\sqrt{-g}\,e^{-\Phi}g^{zz}
\bar{\varphi}(z,-k)\,
\partial_z\bar{\varphi}(z,k)
\right|_{z\rightarrow 0}.
\end{equation}

To proceed, we decompose the on-shell scalar field as
\begin{equation}\label{eqApC3}
\bar{\varphi}(z,k)
=
\varphi_k(z)\varphi_0(k),
\end{equation}
where $\varphi_k(z)$ is the bulk-to-boundary propagator, satisfying the
boundary condition
\begin{equation}\label{eqApC4}
\lim_{z\rightarrow 0}\varphi_k(z)=1.
\end{equation}
In addition, the bulk-to-boundary propagator must satisfy an infalling
boundary condition at the horizon, as required by the holographic
prescription for retarded correlation functions~\cite{Son:2002sd}.

Substituting Eq.~\eqref{eqApC3} into the on-shell action~\eqref{eqApC2}, we obtain
\begin{equation}\label{eqApC5}
S_{\mathrm{on-shell}}
=
\int \frac{d^4k}{(2\pi)^4}\,
\varphi_0(-k)\,
\mathcal{F}(k,z)\,
\varphi_0(k),
\end{equation}
where
\begin{equation}\label{eqApC6}
\mathcal{F}(k,z)
=
\left.
\sqrt{-g}\,g^{zz}e^{-\Phi}
\varphi_k^*(z)\,
\partial_z\varphi_k(z)
\right|_{z\rightarrow 0}.
\end{equation}

The retarded Green's function is then obtained from the on-shell action
according to the holographic prescription~\cite{Son:2002sd},
\begin{equation}\label{eqApC7}
G_R(k)
=
-2\mathcal{F}(k,z).
\end{equation}

Finally, the spectral function is defined as
\begin{equation}\label{eqApC8}
\rho(\omega)
=
-2\,\mathrm{Im}\,G_R(k)
=
4\,\mathrm{Im}\,\mathcal{F}(k,z).
\end{equation}

Thus, by numerically solving the scalar equation of motion,
\begin{equation}\label{eqz1A}
    \varphi''+\varphi'\left( \frac{f'}{f} - \frac{3}{z}-\Phi'\right)
    +\frac{\omega^{2}}{f^2}\varphi = 0,
\end{equation}
subject to the infalling boundary condition at the horizon,
\begin{eqnarray}
    \varphi(z\rightarrow z_h)
    =\left(1-\frac{z}{z_h}\right)^{-i\omega/4\pi T}
    \left[
    1+a_1\left(1-\frac{z}{z_h}\right)
    +a_2\left(1-\frac{z}{z_h}\right)^2+\cdots
    \right],
\end{eqnarray}
together with the prescribed boundary condition at the AdS boundary (\ref{eqApC4}), we obtain the retarded Green's function and, consequently, the spectral function through the relation~(\ref{eqApC8}).

\subsection{Vector Field}

\label{app:vector_spectral_function}

We now turn to the vector case. Following the holographic prescription of Ref.~\cite{Son:2002sd}, we start from the Maxwell action in the presence of the dilaton field:
\begin{equation}\label{ApeVec1}
S_V
=
-\frac{1}{4}
\int d^5x\,
\sqrt{-g}\,e^{-\Phi}
F_{MN}F^{MN},
\end{equation}
where
\begin{equation}\label{ApeVec2}
F_{MN}
=
\partial_M V_N-\partial_N V_M.
\end{equation}

Using the equations of motion and integrating by parts, the on-shell
action reduces to a boundary term,
\begin{equation}\label{ApeVec3}
S_{\mathrm{on-shell}}
=
-\frac{1}{2}
\int d^4x\,
\left.
\sqrt{-g}\,e^{-\Phi}
F^{z\mu}V_\mu
\right|_{z\rightarrow 0},
\end{equation}
where the contribution from the horizon has been neglected. For simplicity, we consider the vector field
\begin{equation}
V_M=(0,V_x,0,0,0).
\end{equation}
That is, we focus on a transverse vector perturbation, $V_x$. Performing a Fourier transform, we write
\begin{equation}\label{ApeVec4}
V_x(z,x)
=
\int\frac{d^4k}{(2\pi)^4}
e^{ik\cdot x}
\bar{V}_x(z,k),
\end{equation}
and substituting this expression into the on-shell action~\eqref{ApeVec3}, we obtain
\begin{equation}\label{ApeVec5}
S_{\mathrm{on-shell}}
=
-\frac{1}{2}
\int\frac{d^4k}{(2\pi)^4}
\left.
\sqrt{-g}\,e^{-\Phi}
g^{zz}g^{xx}
\bar{V}_x(z,-k)
\partial_z\bar{V}_x(z,k)
\right|_{z\rightarrow 0}.
\end{equation}

We decompose the on-shell field in terms of the bulk-to-boundary
propagator as
\begin{equation}\label{ApeVec6}
\bar{V}_x(z,k)
=
V_k(z)V_0(k),
\end{equation}
where $V_k(z)$ satisfies the boundary condition
\begin{equation}\label{ApeVec7}
\lim_{z\rightarrow 0}V_k(z)=1,
\end{equation}
together with the infalling boundary condition at the horizon required
for the retarded Green's function.

Substituting Eq.~\eqref{ApeVec6} into the on-shell action, we obtain
\begin{equation}\label{ApeVec8}
S_{\mathrm{on-shell}}
=
\int\frac{d^4k}{(2\pi)^4}
V_0(-k)
\mathcal{F}_V(k,z)
V_0(k),
\end{equation}
where
\begin{equation}\label{ApeVec10}
\mathcal{F}_V(k,z)
=
-\frac{1}{2}
\left.
\sqrt{-g}\,e^{-\Phi}
g^{zz}g^{xx}
V_k^*(z)
\partial_z V_k(z)
\right|_{z\rightarrow 0}.
\end{equation}

The retarded Green's function is obtained from the on-shell action
according to the holographic prescription \cite{Son:2002sd},
\begin{equation}\label{ApeVec11}
G_R^{V}(k)
=
-2\mathcal{F}_V(k,z).
\end{equation}

The corresponding spectral function is therefore given by
\begin{equation}\label{ApeVec12}
\rho_V(\omega)
=
-2\,\mathrm{Im}\,G_R^{V}(k)
=
4\,\mathrm{Im}\,\mathcal{F}_V(k,z).
\end{equation}

Similarly to the scalar case, we solve the equation of motion for the
vector field,
\begin{equation}\label{eqVector}
V'' + V'\left( \frac{f'}{f} - \frac{1}{z} - \Phi' \right)
+\frac{\omega^{2}}{f^2} V = 0,
\end{equation}
subject to the infalling boundary condition at the horizon,
\begin{eqnarray}
V(z\rightarrow z_h)
&=&
\left(1-\frac{z}{z_h}\right)^{-i\omega/4\pi T}
\left[
1+b_1\left(1-\frac{z}{z_h}\right)
+b_2\left(1-\frac{z}{z_h}\right)^2+\cdots
\right].
\end{eqnarray}
The solution is then normalized at the AdS boundary. The vector
spectral function is subsequently obtained numerically from the
retarded Green's function, following the same procedure as in the
scalar case.

To investigate the relation between the quasinormal modes and the spectral function, we use a Breit--Wigner distribution to characterize the peak position and width of the spectral function presented in Appendix~\ref{app:spectral_function}. The Breit--Wigner expression is given by~\cite{Miranda:2009uw}
\begin{equation}\label{BW}
R(\mathfrak{w})
=
\frac{a\mathfrak{w}^b}
{(\mathfrak{w}-\mathfrak{w}_R)^2+\mathfrak{w}_I^2},
\end{equation}
where $w_R=\operatorname{Re}w$ and $w_I=-\operatorname{Im}w>0$, so that
$w=w_R-iw_I$. The parameters $a$ and $b$ are adjustable constants determined from the fit and may depend on the temperature. The parameters $\mathfrak{w}_R$ and $\mathfrak{w}_I$ characterize the location and width of the resonance, respectively, while the peak position and width extracted from the spectral function provide estimates of the thermal mass and width. Table~\ref{TabBW} presents a comparison between the quasinormal frequencies and the results obtained from the Breit--Wigner fit for the ground states of the scalar and vector fields at two values of $\mathfrak{c}$.

\begin{table}[htbp]
  \centering
  \begin{tabular}{||c||c|c||c|c||c||}
    \hline
    \multicolumn{1}{||c||}{Vector Field} 
    & \multicolumn{2}{|c||}{Breit-Wigner} 
    & \multicolumn{2}{|c||}{QNMS} 
    \\
    \cline{2-5}
    \multicolumn{1}{||c||}{$\mathfrak{c}$} 
    & $\mathfrak{w}_R$  & $\mathfrak{w}_I$ 
    & $\mathfrak{w}_R$ & $\mathfrak{w}_I$   \\
    \hline
    $3$ & $5.650$ & $2.131 \times 10^{-1}$ & $5.656$ & $2.128 \times 10^{-1}$  \\
    $4$ & $7.842$ & $9.866 \times 10^{-3}$ & $7.842$ & $9.864 \times 10^{-3}$   \\
    \hline
    \multicolumn{1}{||c||}{Scalar Field} 
    & \multicolumn{2}{|c||}{Breit-Wigner} 
    & \multicolumn{2}{|c||}{QNMS} 
    \\
    \cline{2-5}
     \multicolumn{1}{||c||}{$\mathfrak{c}$} 
    & $\mathfrak{w}_R$  & $\mathfrak{w}_I$  
    & $\mathfrak{w}_R$  & $\mathfrak{w}_I$   \\
    \hline
    $4$ & $10.777$ & $2.10 \times 10^{-1}$ & $10.778$ & $2.10 \times 10^{-1}$  \\
    $5$ & $13.89$ & $5.734 \times 10^{-3}$ & $13.89$ & $5.734 \times 10^{-3}$   \\
    \hline
  \end{tabular}
  \caption{The ground-state frequencies are obtained from both the Breit--Wigner distribution (see Eq.~\eqref{BW}) and the quasinormal mode calculation.
}
  \label{TabBW}
\end{table}

\section{Condition Number for the soft-wall model}

\label{Condition Number}

In this Appendix, we present two tables showing the condition numbers of the scalar and vector fields, respectively, for different values of $\mathfrak{c}$. The results are presented in Tables~\ref{tab:kappascalar} and~\ref{tab:kappavector}.

\begin{table}[ht]
\caption{Condition number $\kappa$ of the scalar glueball as a function of $\mathfrak{c}$ for the ground state ($n=0$), first excited state ($n=1$), and second excited state ($n=2$).
}
\label{tab:kappascalar}
\begin{ruledtabular}
\begin{tabular}{cccc}
$\mathfrak{c}$ & \multicolumn{3}{c}{$\kappa$} \\
\cline{2-4}
 & $n=0$ & $n=1$ & $n=2$ \\
\hline
4.00  & 1.42136 & 12.374  & 201.626 \\
4.25  & 1.20898 & 7.62785 & 112.257 \\
4.50  & 1.08829 & 4.78531 & 62.265 \\
4.75  & 1.03012 & 3.09873 & 32.8023 \\
5.00  & 1.00795 & 2.11197 & 17.7463 \\
5.25  & 1.00181 & 1.55231 & 10.974 \\
5.50  & 1.00034 & 1.24166 & 6.36236 \\
5.75  & 1.00005 & 1.08611 & 3.80728 \\
6.00  & 1.00001 & 1.02280 & 2.40421 \\
6.25  & 1.00000 & 1.00486 & 1.65789 \\
6.50  & 1.00000 & 1.00081 & 1.26802 \\
6.75  & 1.00000 & 1.00010 & 1.08592 \\
7.00  & 1.00000 & 1.00001 & 1.01914 \\
7.25  & 1.00000 & 1.00000 & 1.00336 \\
7.50  & 1.00000 & 1.00000 & 1.00045 \\
7.75  & 1.00000 & 1.00000 & 1.00005 \\
8.00  & 1.00000 & 1.00000 & 1.00000 \\
\end{tabular}
\end{ruledtabular}
\end{table}

\begin{table}[ht]
\caption{Condition number $\kappa$ of the vector meson as a function of $\mathfrak{c}$ for the ground state ($n=0$), first excited state ($n=1$), and second excited state ($n=2$).}
\label{tab:kappavector}
\begin{ruledtabular}
\begin{tabular}{cccc}
$\mathfrak{c}$ & \multicolumn{3}{c}{$\kappa$} \\
\cline{2-4}
 & $n=0$ & $n=1$ & $n=2$ \\
\hline
2.00  & 4.5277 & 211.917  & 8846.33 \\
2.25  & 3.34075 & 133.389 & 5368.67 \\
2.50  & 2.49904 & 80.7228 & 3134.29 \\
2.75  & 1.91815 & 47.6612 & 1771.81 \\
3.00  & 1.52929 & 27.7385 & 972.416  \\
3.25  & 1.28126 & 16.1758 & 525.168 \\
3.50  & 1.13234 & 9.46776 &  277.647 \\
3.75  & 1.05291 & 5.64246 &  144.468  \\
4.00  & 1.0173 & 3.49311 & 74.4713  \\
4.25  & 1.00488 & 2.31166 &  38.8272 \\
4.50  & 1.00115 & 1.65132 & 20.3074\\
4.75  & 1.00022 & 1.29159 & 10.7606 \\
5.00  & 1.00003 & 1.10985 & 5.86696 \\
5.25  & 1.00001 & 1.0339 & 3.4068 \\
5.50  & 1.00000 & 1.00817 & 2.13821 \\
5.75  & 1.00000 & 1.00153 & 1.49418 \\
6.00  & 1.00000 & 1.00021 & 1.18246 \\
\end{tabular}
\end{ruledtabular}
\end{table}

\clearpage

\bibliographystyle{apsrev4-2}
\bibliography{refs.bib}

\begin{thebibliography}{69}%
\makeatletter
\providecommand \@ifxundefined [1]{%
 \@ifx{#1\undefined}
}%
\providecommand \@ifnum [1]{%
 \ifnum #1\expandafter \@firstoftwo
 \else \expandafter \@secondoftwo
 \fi
}%
\providecommand \@ifx [1]{%
 \ifx #1\expandafter \@firstoftwo
 \else \expandafter \@secondoftwo
 \fi
}%
\providecommand \natexlab [1]{#1}%
\providecommand \enquote  [1]{``#1''}%
\providecommand \bibnamefont  [1]{#1}%
\providecommand \bibfnamefont [1]{#1}%
\providecommand \citenamefont [1]{#1}%
\providecommand \href@noop [0]{\@secondoftwo}%
\providecommand \href [0]{\begingroup \@sanitize@url \@href}%
\providecommand \@href[1]{\@@startlink{#1}\@@href}%
\providecommand \@@href[1]{\endgroup#1\@@endlink}%
\providecommand \@sanitize@url [0]{\catcode `\\12\catcode `\$12\catcode `\&12\catcode `\#12\catcode `\^12\catcode `\_12\catcode `\%12\relax}%
\providecommand \@@startlink[1]{}%
\providecommand \@@endlink[0]{}%
\providecommand \url  [0]{\begingroup\@sanitize@url \@url }%
\providecommand \@url [1]{\endgroup\@href {#1}{\urlprefix }}%
\providecommand \urlprefix  [0]{URL }%
\providecommand \Eprint [0]{\href }%
\providecommand \doibase [0]{https://doi.org/}%
\providecommand \selectlanguage [0]{\@gobble}%
\providecommand \bibinfo  [0]{\@secondoftwo}%
\providecommand \bibfield  [0]{\@secondoftwo}%
\providecommand \translation [1]{[#1]}%
\providecommand \BibitemOpen [0]{}%
\providecommand \bibitemStop [0]{}%
\providecommand \bibitemNoStop [0]{.\EOS\space}%
\providecommand \EOS [0]{\spacefactor3000\relax}%
\providecommand \BibitemShut  [1]{\csname bibitem#1\endcsname}%
\let\auto@bib@innerbib\@empty
\bibitem [{\citenamefont {Trefethen}\ and\ \citenamefont {Embree}(2005)}]{Trefethen:2005}%
  \BibitemOpen
  \bibfield  {author} {\bibinfo {author} {\bibfnamefont {L.~N.}\ \bibnamefont {Trefethen}}\ and\ \bibinfo {author} {\bibfnamefont {M.}~\bibnamefont {Embree}},\ }\href {https://doi.org/10.1515/9780691213101} {\emph {\bibinfo {title} {Spectra and Pseudospectra: The Behavior of Non-Normal Matrices and Operators}}}\ (\bibinfo  {publisher} {Princeton University Press},\ \bibinfo {address} {Princeton},\ \bibinfo {year} {2005})\BibitemShut {NoStop}%
\bibitem [{\citenamefont {Sj{\"o}strand}(2019)}]{Sjöstrand:2019}%
  \BibitemOpen
  \bibfield  {author} {\bibinfo {author} {\bibfnamefont {J.}~\bibnamefont {Sj{\"o}strand}},\ }\href {https://doi.org/10.1007/978-3-030-10819-9} {\emph {\bibinfo {title} {Non-Self-Adjoint Differential Operators, Spectral Asymptotics and Random Perturbations}}}\ (\bibinfo {year} {2019})\BibitemShut {NoStop}%
\bibitem [{\citenamefont {Davies}(2007)}]{Davies:2007}%
  \BibitemOpen
  \bibfield  {author} {\bibinfo {author} {\bibfnamefont {E.~B.}\ \bibnamefont {Davies}},\ }\href {https://doi.org/10.1017/CBO9780511618864} {\emph {\bibinfo {title} {Linear Operators and their Spectra}}},\ Cambridge Studies in Advanced Mathematics\ (\bibinfo  {publisher} {Cambridge University Press},\ \bibinfo {year} {2007})\BibitemShut {NoStop}%
\bibitem [{\citenamefont {Jaramillo}\ \emph {et~al.}(2021{\natexlab{a}})\citenamefont {Jaramillo}, \citenamefont {Macedo},\ and\ \citenamefont {Al~Sheikh}}]{Jaramillo2021}%
  \BibitemOpen
  \bibfield  {author} {\bibinfo {author} {\bibfnamefont {J.~L.}\ \bibnamefont {Jaramillo}}, \bibinfo {author} {\bibfnamefont {R.~P.}\ \bibnamefont {Macedo}},\ and\ \bibinfo {author} {\bibfnamefont {L.}~\bibnamefont {Al~Sheikh}},\ }\href {https://doi.org/10.1103/PhysRevX.11.031003} {\bibfield  {journal} {\bibinfo  {journal} {Physical Review X}\ }\textbf {\bibinfo {volume} {11}},\ \bibinfo {pages} {031003} (\bibinfo {year} {2021}{\natexlab{a}})},\ \Eprint {https://arxiv.org/abs/2004.06434} {arXiv:2004.06434 [gr-qc]} \BibitemShut {NoStop}%
\bibitem [{\citenamefont {Destounis}\ \emph {et~al.}(2021)\citenamefont {Destounis}, \citenamefont {Macedo}, \citenamefont {Berti}, \citenamefont {Cardoso},\ and\ \citenamefont {Jaramillo}}]{Destounis:2021lum}%
  \BibitemOpen
  \bibfield  {author} {\bibinfo {author} {\bibfnamefont {K.}~\bibnamefont {Destounis}}, \bibinfo {author} {\bibfnamefont {R.~P.}\ \bibnamefont {Macedo}}, \bibinfo {author} {\bibfnamefont {E.}~\bibnamefont {Berti}}, \bibinfo {author} {\bibfnamefont {V.}~\bibnamefont {Cardoso}},\ and\ \bibinfo {author} {\bibfnamefont {J.~L.}\ \bibnamefont {Jaramillo}},\ }\href {https://doi.org/10.1103/PhysRevD.104.084091} {\bibfield  {journal} {\bibinfo  {journal} {Phys. Rev. D}\ }\textbf {\bibinfo {volume} {104}},\ \bibinfo {pages} {084091} (\bibinfo {year} {2021})},\ \Eprint {https://arxiv.org/abs/2107.09673} {arXiv:2107.09673 [gr-qc]} \BibitemShut {NoStop}%
\bibitem [{\citenamefont {Cheung}\ \emph {et~al.}(2022)\citenamefont {Cheung}, \citenamefont {Destounis}, \citenamefont {Macedo}, \citenamefont {Berti},\ and\ \citenamefont {Cardoso}}]{Cheung:2021bol}%
  \BibitemOpen
  \bibfield  {author} {\bibinfo {author} {\bibfnamefont {M.~H.-Y.}\ \bibnamefont {Cheung}}, \bibinfo {author} {\bibfnamefont {K.}~\bibnamefont {Destounis}}, \bibinfo {author} {\bibfnamefont {R.~P.}\ \bibnamefont {Macedo}}, \bibinfo {author} {\bibfnamefont {E.}~\bibnamefont {Berti}},\ and\ \bibinfo {author} {\bibfnamefont {V.}~\bibnamefont {Cardoso}},\ }\href {https://doi.org/10.1103/PhysRevLett.128.111103} {\bibfield  {journal} {\bibinfo  {journal} {Phys. Rev. Lett.}\ }\textbf {\bibinfo {volume} {128}},\ \bibinfo {pages} {111103} (\bibinfo {year} {2022})},\ \Eprint {https://arxiv.org/abs/2111.05415} {arXiv:2111.05415 [gr-qc]} \BibitemShut {NoStop}%
\bibitem [{\citenamefont {Jaramillo}(2022)}]{Jaramillo:2022kuv}%
  \BibitemOpen
  \bibfield  {author} {\bibinfo {author} {\bibfnamefont {J.~L.}\ \bibnamefont {Jaramillo}},\ }\href {https://doi.org/10.1088/1361-6382/ac8ddc} {\bibfield  {journal} {\bibinfo  {journal} {Class. Quant. Grav.}\ }\textbf {\bibinfo {volume} {39}},\ \bibinfo {pages} {217002} (\bibinfo {year} {2022})},\ \Eprint {https://arxiv.org/abs/2206.08025} {arXiv:2206.08025 [gr-qc]} \BibitemShut {NoStop}%
\bibitem [{\citenamefont {Konoplya}\ and\ \citenamefont {Zhidenko}(2024)}]{Konoplya:2022pbc}%
  \BibitemOpen
  \bibfield  {author} {\bibinfo {author} {\bibfnamefont {R.~A.}\ \bibnamefont {Konoplya}}\ and\ \bibinfo {author} {\bibfnamefont {A.}~\bibnamefont {Zhidenko}},\ }\href {https://doi.org/10.1016/j.jheap.2024.10.015} {\bibfield  {journal} {\bibinfo  {journal} {JHEAp}\ }\textbf {\bibinfo {volume} {44}},\ \bibinfo {pages} {419} (\bibinfo {year} {2024})},\ \Eprint {https://arxiv.org/abs/2209.00679} {arXiv:2209.00679 [gr-qc]} \BibitemShut {NoStop}%
\bibitem [{\citenamefont {Boyanov}\ \emph {et~al.}(2023)\citenamefont {Boyanov}, \citenamefont {Destounis}, \citenamefont {Panosso~Macedo}, \citenamefont {Cardoso},\ and\ \citenamefont {Jaramillo}}]{Boyanov:2022ark}%
  \BibitemOpen
  \bibfield  {author} {\bibinfo {author} {\bibfnamefont {V.}~\bibnamefont {Boyanov}}, \bibinfo {author} {\bibfnamefont {K.}~\bibnamefont {Destounis}}, \bibinfo {author} {\bibfnamefont {R.}~\bibnamefont {Panosso~Macedo}}, \bibinfo {author} {\bibfnamefont {V.}~\bibnamefont {Cardoso}},\ and\ \bibinfo {author} {\bibfnamefont {J.~L.}\ \bibnamefont {Jaramillo}},\ }\href {https://doi.org/10.1103/PhysRevD.107.064012} {\bibfield  {journal} {\bibinfo  {journal} {Phys. Rev. D}\ }\textbf {\bibinfo {volume} {107}},\ \bibinfo {pages} {064012} (\bibinfo {year} {2023})},\ \Eprint {https://arxiv.org/abs/2209.12950} {arXiv:2209.12950 [gr-qc]} \BibitemShut {NoStop}%
\bibitem [{\citenamefont {Sarkar}\ \emph {et~al.}(2023)\citenamefont {Sarkar}, \citenamefont {Rahman},\ and\ \citenamefont {Chakraborty}}]{Sarkar:2023rhp}%
  \BibitemOpen
  \bibfield  {author} {\bibinfo {author} {\bibfnamefont {S.}~\bibnamefont {Sarkar}}, \bibinfo {author} {\bibfnamefont {M.}~\bibnamefont {Rahman}},\ and\ \bibinfo {author} {\bibfnamefont {S.}~\bibnamefont {Chakraborty}},\ }\href {https://doi.org/10.1103/PhysRevD.108.104002} {\bibfield  {journal} {\bibinfo  {journal} {Phys. Rev. D}\ }\textbf {\bibinfo {volume} {108}},\ \bibinfo {pages} {104002} (\bibinfo {year} {2023})},\ \Eprint {https://arxiv.org/abs/2304.06829} {arXiv:2304.06829 [gr-qc]} \BibitemShut {NoStop}%
\bibitem [{\citenamefont {Courty}\ \emph {et~al.}(2023)\citenamefont {Courty}, \citenamefont {Destounis},\ and\ \citenamefont {Pani}}]{Courty:2023rxk}%
  \BibitemOpen
  \bibfield  {author} {\bibinfo {author} {\bibfnamefont {A.}~\bibnamefont {Courty}}, \bibinfo {author} {\bibfnamefont {K.}~\bibnamefont {Destounis}},\ and\ \bibinfo {author} {\bibfnamefont {P.}~\bibnamefont {Pani}},\ }\href {https://doi.org/10.1103/PhysRevD.108.104027} {\bibfield  {journal} {\bibinfo  {journal} {Phys. Rev. D}\ }\textbf {\bibinfo {volume} {108}},\ \bibinfo {pages} {104027} (\bibinfo {year} {2023})},\ \Eprint {https://arxiv.org/abs/2307.11155} {arXiv:2307.11155 [gr-qc]} \BibitemShut {NoStop}%
\bibitem [{\citenamefont {Destounis}\ and\ \citenamefont {Duque}(2023)}]{Destounis:2023ruj}%
  \BibitemOpen
  \bibfield  {author} {\bibinfo {author} {\bibfnamefont {K.}~\bibnamefont {Destounis}}\ and\ \bibinfo {author} {\bibfnamefont {F.}~\bibnamefont {Duque}}\ }(\bibinfo {year} {2023})\ \Eprint {https://arxiv.org/abs/2308.16227} {arXiv:2308.16227 [gr-qc]} \BibitemShut {NoStop}%
\bibitem [{\citenamefont {Destounis}\ \emph {et~al.}(2024)\citenamefont {Destounis}, \citenamefont {Boyanov},\ and\ \citenamefont {Panosso~Macedo}}]{Destounis:2023nmb}%
  \BibitemOpen
  \bibfield  {author} {\bibinfo {author} {\bibfnamefont {K.}~\bibnamefont {Destounis}}, \bibinfo {author} {\bibfnamefont {V.}~\bibnamefont {Boyanov}},\ and\ \bibinfo {author} {\bibfnamefont {R.}~\bibnamefont {Panosso~Macedo}},\ }\href {https://doi.org/10.1103/PhysRevD.109.044023} {\bibfield  {journal} {\bibinfo  {journal} {Phys. Rev. D}\ }\textbf {\bibinfo {volume} {109}},\ \bibinfo {pages} {044023} (\bibinfo {year} {2024})},\ \Eprint {https://arxiv.org/abs/2312.11630} {arXiv:2312.11630 [gr-qc]} \BibitemShut {NoStop}%
\bibitem [{\citenamefont {Cao}\ \emph {et~al.}(2024)\citenamefont {Cao}, \citenamefont {Chen}, \citenamefont {Wu}, \citenamefont {Xie},\ and\ \citenamefont {Zhou}}]{Cao:2024oud}%
  \BibitemOpen
  \bibfield  {author} {\bibinfo {author} {\bibfnamefont {L.-M.}\ \bibnamefont {Cao}}, \bibinfo {author} {\bibfnamefont {J.-N.}\ \bibnamefont {Chen}}, \bibinfo {author} {\bibfnamefont {L.-B.}\ \bibnamefont {Wu}}, \bibinfo {author} {\bibfnamefont {L.}~\bibnamefont {Xie}},\ and\ \bibinfo {author} {\bibfnamefont {Y.-S.}\ \bibnamefont {Zhou}},\ }\href {https://doi.org/10.1007/s11433-024-2435-5} {\bibfield  {journal} {\bibinfo  {journal} {Sci. China Phys. Mech. Astron.}\ }\textbf {\bibinfo {volume} {67}},\ \bibinfo {pages} {100412} (\bibinfo {year} {2024})},\ \Eprint {https://arxiv.org/abs/2401.09907} {arXiv:2401.09907 [gr-qc]} \BibitemShut {NoStop}%
\bibitem [{\citenamefont {Maldacena}(1998)}]{Maldacena:1997re}%
  \BibitemOpen
  \bibfield  {author} {\bibinfo {author} {\bibfnamefont {J.~M.}\ \bibnamefont {Maldacena}},\ }\href {https://doi.org/10.4310/ATMP.1998.v2.n2.a1} {\bibfield  {journal} {\bibinfo  {journal} {Adv. Theor. Math. Phys.}\ }\textbf {\bibinfo {volume} {2}},\ \bibinfo {pages} {231} (\bibinfo {year} {1998})},\ \Eprint {https://arxiv.org/abs/hep-th/9711200} {arXiv:hep-th/9711200} \BibitemShut {NoStop}%
\bibitem [{\citenamefont {Witten}(1998)}]{Witten:1998qj}%
  \BibitemOpen
  \bibfield  {author} {\bibinfo {author} {\bibfnamefont {E.}~\bibnamefont {Witten}},\ }\href {https://doi.org/10.4310/ATMP.1998.v2.n2.a2} {\bibfield  {journal} {\bibinfo  {journal} {Adv. Theor. Math. Phys.}\ }\textbf {\bibinfo {volume} {2}},\ \bibinfo {pages} {253} (\bibinfo {year} {1998})},\ \Eprint {https://arxiv.org/abs/hep-th/9802150} {arXiv:hep-th/9802150} \BibitemShut {NoStop}%
\bibitem [{\citenamefont {Gubser}\ \emph {et~al.}(1998)\citenamefont {Gubser}, \citenamefont {Klebanov},\ and\ \citenamefont {Polyakov}}]{Gubser:1998bc}%
  \BibitemOpen
  \bibfield  {author} {\bibinfo {author} {\bibfnamefont {S.~S.}\ \bibnamefont {Gubser}}, \bibinfo {author} {\bibfnamefont {I.~R.}\ \bibnamefont {Klebanov}},\ and\ \bibinfo {author} {\bibfnamefont {A.~M.}\ \bibnamefont {Polyakov}},\ }\href {https://doi.org/10.1016/S0370-2693(98)00377-3} {\bibfield  {journal} {\bibinfo  {journal} {Phys. Lett. B}\ }\textbf {\bibinfo {volume} {428}},\ \bibinfo {pages} {105} (\bibinfo {year} {1998})},\ \Eprint {https://arxiv.org/abs/hep-th/9802109} {arXiv:hep-th/9802109} \BibitemShut {NoStop}%
\bibitem [{\citenamefont {Are{\'a}n}\ \emph {et~al.}(2023)\citenamefont {Are{\'a}n}, \citenamefont {Fari{\~n}a},\ and\ \citenamefont {Landsteiner}}]{Arean:2023ejh}%
  \BibitemOpen
  \bibfield  {author} {\bibinfo {author} {\bibfnamefont {D.}~\bibnamefont {Are{\'a}n}}, \bibinfo {author} {\bibfnamefont {D.~G.}\ \bibnamefont {Fari{\~n}a}},\ and\ \bibinfo {author} {\bibfnamefont {K.}~\bibnamefont {Landsteiner}},\ }\href {https://doi.org/10.1007/JHEP12(2023)187} {\bibfield  {journal} {\bibinfo  {journal} {JHEP}\ }\textbf {\bibinfo {volume} {12}},\ \bibinfo {pages} {187}},\ \Eprint {https://arxiv.org/abs/2307.08751} {arXiv:2307.08751 [hep-th]} \BibitemShut {NoStop}%
\bibitem [{\citenamefont {Cownden}\ \emph {et~al.}(2024)\citenamefont {Cownden}, \citenamefont {Pantelidou},\ and\ \citenamefont {Zilh{\~a}o}}]{Cownden:2023dam}%
  \BibitemOpen
  \bibfield  {author} {\bibinfo {author} {\bibfnamefont {B.}~\bibnamefont {Cownden}}, \bibinfo {author} {\bibfnamefont {C.}~\bibnamefont {Pantelidou}},\ and\ \bibinfo {author} {\bibfnamefont {M.}~\bibnamefont {Zilh{\~a}o}},\ }\href {https://doi.org/10.1007/JHEP05(2024)202} {\bibfield  {journal} {\bibinfo  {journal} {JHEP}\ }\textbf {\bibinfo {volume} {05}},\ \bibinfo {pages} {202}},\ \Eprint {https://arxiv.org/abs/2312.08352} {arXiv:2312.08352 [gr-qc]} \BibitemShut {NoStop}%
\bibitem [{\citenamefont {Arean}\ \emph {et~al.}(2024)\citenamefont {Arean}, \citenamefont {Garcia-Fari{\~n}a},\ and\ \citenamefont {Landsteiner}}]{Arean:2024afl}%
  \BibitemOpen
  \bibfield  {author} {\bibinfo {author} {\bibfnamefont {D.}~\bibnamefont {Arean}}, \bibinfo {author} {\bibfnamefont {D.}~\bibnamefont {Garcia-Fari{\~n}a}},\ and\ \bibinfo {author} {\bibfnamefont {K.}~\bibnamefont {Landsteiner}},\ }\href {https://doi.org/10.3389/fphy.2024.1460268} {\bibfield  {journal} {\bibinfo  {journal} {Front. in Phys.}\ }\textbf {\bibinfo {volume} {12}},\ \bibinfo {pages} {1460268} (\bibinfo {year} {2024})},\ \Eprint {https://arxiv.org/abs/2407.04372} {arXiv:2407.04372 [hep-th]} \BibitemShut {NoStop}%
\bibitem [{\citenamefont {Garcia-Fari{\~n}a}\ \emph {et~al.}(2025)\citenamefont {Garcia-Fari{\~n}a}, \citenamefont {Landsteiner}, \citenamefont {Romeu},\ and\ \citenamefont {Saura-Bastida}}]{Garcia-Farina:2024pdd}%
  \BibitemOpen
  \bibfield  {author} {\bibinfo {author} {\bibfnamefont {D.}~\bibnamefont {Garcia-Fari{\~n}a}}, \bibinfo {author} {\bibfnamefont {K.}~\bibnamefont {Landsteiner}}, \bibinfo {author} {\bibfnamefont {P.~G.}\ \bibnamefont {Romeu}},\ and\ \bibinfo {author} {\bibfnamefont {P.}~\bibnamefont {Saura-Bastida}},\ }\href {https://doi.org/10.1007/JHEP01(2025)185} {\bibfield  {journal} {\bibinfo  {journal} {JHEP}\ }\textbf {\bibinfo {volume} {01}},\ \bibinfo {pages} {185}},\ \Eprint {https://arxiv.org/abs/2407.06104} {arXiv:2407.06104 [hep-th]} \BibitemShut {NoStop}%
\bibitem [{\citenamefont {Garcia-Fari{\~n}a}\ \emph {et~al.}(2026)\citenamefont {Garcia-Fari{\~n}a}, \citenamefont {Landsteiner}, \citenamefont {Romeu},\ and\ \citenamefont {Saura-Bastida}}]{Garcia-Farina:2026qug}%
  \BibitemOpen
  \bibfield  {author} {\bibinfo {author} {\bibfnamefont {D.}~\bibnamefont {Garcia-Fari{\~n}a}}, \bibinfo {author} {\bibfnamefont {K.}~\bibnamefont {Landsteiner}}, \bibinfo {author} {\bibfnamefont {P.~G.}\ \bibnamefont {Romeu}},\ and\ \bibinfo {author} {\bibfnamefont {P.}~\bibnamefont {Saura-Bastida}},\ }\href {https://doi.org/10.1007/JHEP08(2026)186} {\bibfield  {journal} {\bibinfo  {journal} {JHEP}\ }\textbf {\bibinfo {volume} {08}},\ \bibinfo {pages} {186}},\ \Eprint {https://arxiv.org/abs/2602.02659} {arXiv:2602.02659 [hep-th]} \BibitemShut {NoStop}%
\bibitem [{\citenamefont {Karch}\ \emph {et~al.}(2006)\citenamefont {Karch}, \citenamefont {Katz}, \citenamefont {Son},\ and\ \citenamefont {Stephanov}}]{Karch:2006pv}%
  \BibitemOpen
  \bibfield  {author} {\bibinfo {author} {\bibfnamefont {A.}~\bibnamefont {Karch}}, \bibinfo {author} {\bibfnamefont {E.}~\bibnamefont {Katz}}, \bibinfo {author} {\bibfnamefont {D.~T.}\ \bibnamefont {Son}},\ and\ \bibinfo {author} {\bibfnamefont {M.~A.}\ \bibnamefont {Stephanov}},\ }\href {https://doi.org/10.1103/PhysRevD.74.015005} {\bibfield  {journal} {\bibinfo  {journal} {Phys. Rev. D}\ }\textbf {\bibinfo {volume} {74}},\ \bibinfo {pages} {015005} (\bibinfo {year} {2006})},\ \Eprint {https://arxiv.org/abs/hep-ph/0602229} {arXiv:hep-ph/0602229} \BibitemShut {NoStop}%
\bibitem [{\citenamefont {Gursoy}\ and\ \citenamefont {Kiritsis}(2008)}]{Gursoy:2007cb}%
  \BibitemOpen
  \bibfield  {author} {\bibinfo {author} {\bibfnamefont {U.}~\bibnamefont {Gursoy}}\ and\ \bibinfo {author} {\bibfnamefont {E.}~\bibnamefont {Kiritsis}},\ }\href {https://doi.org/10.1088/1126-6708/2008/02/032} {\bibfield  {journal} {\bibinfo  {journal} {JHEP}\ }\textbf {\bibinfo {volume} {02}},\ \bibinfo {pages} {032}},\ \Eprint {https://arxiv.org/abs/0707.1324} {arXiv:0707.1324 [hep-th]} \BibitemShut {NoStop}%
\bibitem [{\citenamefont {Gursoy}\ \emph {et~al.}(2008{\natexlab{a}})\citenamefont {Gursoy}, \citenamefont {Kiritsis},\ and\ \citenamefont {Nitti}}]{Gursoy:2007er}%
  \BibitemOpen
  \bibfield  {author} {\bibinfo {author} {\bibfnamefont {U.}~\bibnamefont {Gursoy}}, \bibinfo {author} {\bibfnamefont {E.}~\bibnamefont {Kiritsis}},\ and\ \bibinfo {author} {\bibfnamefont {F.}~\bibnamefont {Nitti}},\ }\href {https://doi.org/10.1088/1126-6708/2008/02/019} {\bibfield  {journal} {\bibinfo  {journal} {JHEP}\ }\textbf {\bibinfo {volume} {02}},\ \bibinfo {pages} {019}},\ \Eprint {https://arxiv.org/abs/0707.1349} {arXiv:0707.1349 [hep-th]} \BibitemShut {NoStop}%
\bibitem [{\citenamefont {Colangelo}\ \emph {et~al.}(2007)\citenamefont {Colangelo}, \citenamefont {De~Fazio}, \citenamefont {Jugeau},\ and\ \citenamefont {Nicotri}}]{Colangelo:2007pt}%
  \BibitemOpen
  \bibfield  {author} {\bibinfo {author} {\bibfnamefont {P.}~\bibnamefont {Colangelo}}, \bibinfo {author} {\bibfnamefont {F.}~\bibnamefont {De~Fazio}}, \bibinfo {author} {\bibfnamefont {F.}~\bibnamefont {Jugeau}},\ and\ \bibinfo {author} {\bibfnamefont {S.}~\bibnamefont {Nicotri}},\ }\href {https://doi.org/10.1016/j.physletb.2007.06.072} {\bibfield  {journal} {\bibinfo  {journal} {Phys. Lett. B}\ }\textbf {\bibinfo {volume} {652}},\ \bibinfo {pages} {73} (\bibinfo {year} {2007})},\ \Eprint {https://arxiv.org/abs/hep-ph/0703316} {arXiv:hep-ph/0703316} \BibitemShut {NoStop}%
\bibitem [{\citenamefont {Colangelo}\ \emph {et~al.}(2008)\citenamefont {Colangelo}, \citenamefont {De~Fazio}, \citenamefont {Giannuzzi}, \citenamefont {Jugeau},\ and\ \citenamefont {Nicotri}}]{Colangelo:2008us}%
  \BibitemOpen
  \bibfield  {author} {\bibinfo {author} {\bibfnamefont {P.}~\bibnamefont {Colangelo}}, \bibinfo {author} {\bibfnamefont {F.}~\bibnamefont {De~Fazio}}, \bibinfo {author} {\bibfnamefont {F.}~\bibnamefont {Giannuzzi}}, \bibinfo {author} {\bibfnamefont {F.}~\bibnamefont {Jugeau}},\ and\ \bibinfo {author} {\bibfnamefont {S.}~\bibnamefont {Nicotri}},\ }\href {https://doi.org/10.1103/PhysRevD.78.055009} {\bibfield  {journal} {\bibinfo  {journal} {Phys. Rev. D}\ }\textbf {\bibinfo {volume} {78}},\ \bibinfo {pages} {055009} (\bibinfo {year} {2008})},\ \Eprint {https://arxiv.org/abs/0807.1054} {arXiv:0807.1054 [hep-ph]} \BibitemShut {NoStop}%
\bibitem [{\citenamefont {Gubser}\ and\ \citenamefont {Nellore}(2008)}]{Gubser:2008ny}%
  \BibitemOpen
  \bibfield  {author} {\bibinfo {author} {\bibfnamefont {S.~S.}\ \bibnamefont {Gubser}}\ and\ \bibinfo {author} {\bibfnamefont {A.}~\bibnamefont {Nellore}},\ }\href {https://doi.org/10.1103/PhysRevD.78.086007} {\bibfield  {journal} {\bibinfo  {journal} {Phys. Rev. D}\ }\textbf {\bibinfo {volume} {78}},\ \bibinfo {pages} {086007} (\bibinfo {year} {2008})},\ \Eprint {https://arxiv.org/abs/0804.0434} {arXiv:0804.0434 [hep-th]} \BibitemShut {NoStop}%
\bibitem [{\citenamefont {Gubser}\ \emph {et~al.}(2008)\citenamefont {Gubser}, \citenamefont {Pufu},\ and\ \citenamefont {Rocha}}]{Gubser:2008sz}%
  \BibitemOpen
  \bibfield  {author} {\bibinfo {author} {\bibfnamefont {S.~S.}\ \bibnamefont {Gubser}}, \bibinfo {author} {\bibfnamefont {S.~S.}\ \bibnamefont {Pufu}},\ and\ \bibinfo {author} {\bibfnamefont {F.~D.}\ \bibnamefont {Rocha}},\ }\href {https://doi.org/10.1088/1126-6708/2008/08/085} {\bibfield  {journal} {\bibinfo  {journal} {JHEP}\ }\textbf {\bibinfo {volume} {08}},\ \bibinfo {pages} {085}},\ \Eprint {https://arxiv.org/abs/0806.0407} {arXiv:0806.0407 [hep-th]} \BibitemShut {NoStop}%
\bibitem [{\citenamefont {Gursoy}\ \emph {et~al.}(2009)\citenamefont {Gursoy}, \citenamefont {Kiritsis}, \citenamefont {Mazzanti},\ and\ \citenamefont {Nitti}}]{Gursoy:2009jd}%
  \BibitemOpen
  \bibfield  {author} {\bibinfo {author} {\bibfnamefont {U.}~\bibnamefont {Gursoy}}, \bibinfo {author} {\bibfnamefont {E.}~\bibnamefont {Kiritsis}}, \bibinfo {author} {\bibfnamefont {L.}~\bibnamefont {Mazzanti}},\ and\ \bibinfo {author} {\bibfnamefont {F.}~\bibnamefont {Nitti}},\ }\href {https://doi.org/10.1016/j.nuclphysb.2009.05.017} {\bibfield  {journal} {\bibinfo  {journal} {Nucl. Phys. B}\ }\textbf {\bibinfo {volume} {820}},\ \bibinfo {pages} {148} (\bibinfo {year} {2009})},\ \Eprint {https://arxiv.org/abs/0903.2859} {arXiv:0903.2859 [hep-th]} \BibitemShut {NoStop}%
\bibitem [{\citenamefont {Colangelo}\ \emph {et~al.}(2009)\citenamefont {Colangelo}, \citenamefont {Giannuzzi},\ and\ \citenamefont {Nicotri}}]{Colangelo:2009ra}%
  \BibitemOpen
  \bibfield  {author} {\bibinfo {author} {\bibfnamefont {P.}~\bibnamefont {Colangelo}}, \bibinfo {author} {\bibfnamefont {F.}~\bibnamefont {Giannuzzi}},\ and\ \bibinfo {author} {\bibfnamefont {S.}~\bibnamefont {Nicotri}},\ }\href {https://doi.org/10.1103/PhysRevD.80.094019} {\bibfield  {journal} {\bibinfo  {journal} {Phys. Rev. D}\ }\textbf {\bibinfo {volume} {80}},\ \bibinfo {pages} {094019} (\bibinfo {year} {2009})},\ \Eprint {https://arxiv.org/abs/0909.1534} {arXiv:0909.1534 [hep-ph]} \BibitemShut {NoStop}%
\bibitem [{\citenamefont {Gursoy}\ \emph {et~al.}(2011)\citenamefont {Gursoy}, \citenamefont {Kiritsis}, \citenamefont {Mazzanti}, \citenamefont {Michalogiorgakis},\ and\ \citenamefont {Nitti}}]{Gursoy:2010fj}%
  \BibitemOpen
  \bibfield  {author} {\bibinfo {author} {\bibfnamefont {U.}~\bibnamefont {Gursoy}}, \bibinfo {author} {\bibfnamefont {E.}~\bibnamefont {Kiritsis}}, \bibinfo {author} {\bibfnamefont {L.}~\bibnamefont {Mazzanti}}, \bibinfo {author} {\bibfnamefont {G.}~\bibnamefont {Michalogiorgakis}},\ and\ \bibinfo {author} {\bibfnamefont {F.}~\bibnamefont {Nitti}},\ }\href {https://doi.org/10.1007/978-3-642-04864-7_4} {\bibfield  {journal} {\bibinfo  {journal} {Lect. Notes Phys.}\ }\textbf {\bibinfo {volume} {828}},\ \bibinfo {pages} {79} (\bibinfo {year} {2011})},\ \Eprint {https://arxiv.org/abs/1006.5461} {arXiv:1006.5461 [hep-th]} \BibitemShut {NoStop}%
\bibitem [{\citenamefont {Branz}\ \emph {et~al.}(2010)\citenamefont {Branz}, \citenamefont {Gutsche}, \citenamefont {Lyubovitskij}, \citenamefont {Schmidt},\ and\ \citenamefont {Vega}}]{Branz:2010ub}%
  \BibitemOpen
  \bibfield  {author} {\bibinfo {author} {\bibfnamefont {T.}~\bibnamefont {Branz}}, \bibinfo {author} {\bibfnamefont {T.}~\bibnamefont {Gutsche}}, \bibinfo {author} {\bibfnamefont {V.~E.}\ \bibnamefont {Lyubovitskij}}, \bibinfo {author} {\bibfnamefont {I.}~\bibnamefont {Schmidt}},\ and\ \bibinfo {author} {\bibfnamefont {A.}~\bibnamefont {Vega}},\ }\href {https://doi.org/10.1103/PhysRevD.82.074022} {\bibfield  {journal} {\bibinfo  {journal} {Phys. Rev. D}\ }\textbf {\bibinfo {volume} {82}},\ \bibinfo {pages} {074022} (\bibinfo {year} {2010})},\ \Eprint {https://arxiv.org/abs/1008.0268} {arXiv:1008.0268 [hep-ph]} \BibitemShut {NoStop}%
\bibitem [{\citenamefont {Li}\ \emph {et~al.}(2011)\citenamefont {Li}, \citenamefont {He}, \citenamefont {Huang},\ and\ \citenamefont {Yan}}]{Li:2011hp}%
  \BibitemOpen
  \bibfield  {author} {\bibinfo {author} {\bibfnamefont {D.}~\bibnamefont {Li}}, \bibinfo {author} {\bibfnamefont {S.}~\bibnamefont {He}}, \bibinfo {author} {\bibfnamefont {M.}~\bibnamefont {Huang}},\ and\ \bibinfo {author} {\bibfnamefont {Q.-S.}\ \bibnamefont {Yan}},\ }\href {https://doi.org/10.1007/JHEP09(2011)041} {\bibfield  {journal} {\bibinfo  {journal} {JHEP}\ }\textbf {\bibinfo {volume} {09}},\ \bibinfo {pages} {041}},\ \Eprint {https://arxiv.org/abs/1103.5389} {arXiv:1103.5389 [hep-th]} \BibitemShut {NoStop}%
\bibitem [{\citenamefont {Kajantie}\ \emph {et~al.}(2011)\citenamefont {Kajantie}, \citenamefont {Krssak}, \citenamefont {Vepsalainen},\ and\ \citenamefont {Vuorinen}}]{Kajantie:2011nx}%
  \BibitemOpen
  \bibfield  {author} {\bibinfo {author} {\bibfnamefont {K.}~\bibnamefont {Kajantie}}, \bibinfo {author} {\bibfnamefont {M.}~\bibnamefont {Krssak}}, \bibinfo {author} {\bibfnamefont {M.}~\bibnamefont {Vepsalainen}},\ and\ \bibinfo {author} {\bibfnamefont {A.}~\bibnamefont {Vuorinen}},\ }\href {https://doi.org/10.1103/PhysRevD.84.086004} {\bibfield  {journal} {\bibinfo  {journal} {Phys. Rev. D}\ }\textbf {\bibinfo {volume} {84}},\ \bibinfo {pages} {086004} (\bibinfo {year} {2011})},\ \Eprint {https://arxiv.org/abs/1104.5352} {arXiv:1104.5352 [hep-ph]} \BibitemShut {NoStop}%
\bibitem [{\citenamefont {Gutsche}\ \emph {et~al.}(2012)\citenamefont {Gutsche}, \citenamefont {Lyubovitskij}, \citenamefont {Schmidt},\ and\ \citenamefont {Vega}}]{Gutsche:2011vb}%
  \BibitemOpen
  \bibfield  {author} {\bibinfo {author} {\bibfnamefont {T.}~\bibnamefont {Gutsche}}, \bibinfo {author} {\bibfnamefont {V.~E.}\ \bibnamefont {Lyubovitskij}}, \bibinfo {author} {\bibfnamefont {I.}~\bibnamefont {Schmidt}},\ and\ \bibinfo {author} {\bibfnamefont {A.}~\bibnamefont {Vega}},\ }\href {https://doi.org/10.1103/PhysRevD.85.076003} {\bibfield  {journal} {\bibinfo  {journal} {Phys. Rev. D}\ }\textbf {\bibinfo {volume} {85}},\ \bibinfo {pages} {076003} (\bibinfo {year} {2012})},\ \Eprint {https://arxiv.org/abs/1108.0346} {arXiv:1108.0346 [hep-ph]} \BibitemShut {NoStop}%
\bibitem [{\citenamefont {Alho}\ \emph {et~al.}(2013)\citenamefont {Alho}, \citenamefont {J\"arvinen}, \citenamefont {Kajantie}, \citenamefont {Kiritsis},\ and\ \citenamefont {Tuominen}}]{Alho:2012mh}%
  \BibitemOpen
  \bibfield  {author} {\bibinfo {author} {\bibfnamefont {T.}~\bibnamefont {Alho}}, \bibinfo {author} {\bibfnamefont {M.}~\bibnamefont {J\"arvinen}}, \bibinfo {author} {\bibfnamefont {K.}~\bibnamefont {Kajantie}}, \bibinfo {author} {\bibfnamefont {E.}~\bibnamefont {Kiritsis}},\ and\ \bibinfo {author} {\bibfnamefont {K.}~\bibnamefont {Tuominen}},\ }\href {https://doi.org/10.1007/JHEP01(2013)093} {\bibfield  {journal} {\bibinfo  {journal} {JHEP}\ }\textbf {\bibinfo {volume} {01}},\ \bibinfo {pages} {093}},\ \Eprint {https://arxiv.org/abs/1210.4516} {arXiv:1210.4516 [hep-ph]} \BibitemShut {NoStop}%
\bibitem [{\citenamefont {Cai}\ \emph {et~al.}(2012)\citenamefont {Cai}, \citenamefont {He},\ and\ \citenamefont {Li}}]{Cai:2012xh}%
  \BibitemOpen
  \bibfield  {author} {\bibinfo {author} {\bibfnamefont {R.-G.}\ \bibnamefont {Cai}}, \bibinfo {author} {\bibfnamefont {S.}~\bibnamefont {He}},\ and\ \bibinfo {author} {\bibfnamefont {D.}~\bibnamefont {Li}},\ }\href {https://doi.org/10.1007/JHEP03(2012)033} {\bibfield  {journal} {\bibinfo  {journal} {JHEP}\ }\textbf {\bibinfo {volume} {03}},\ \bibinfo {pages} {033}},\ \Eprint {https://arxiv.org/abs/1201.0820} {arXiv:1201.0820 [hep-th]} \BibitemShut {NoStop}%
\bibitem [{\citenamefont {Li}\ and\ \citenamefont {Huang}(2013)}]{Li:2013oda}%
  \BibitemOpen
  \bibfield  {author} {\bibinfo {author} {\bibfnamefont {D.}~\bibnamefont {Li}}\ and\ \bibinfo {author} {\bibfnamefont {M.}~\bibnamefont {Huang}},\ }\href {https://doi.org/10.1007/JHEP11(2013)088} {\bibfield  {journal} {\bibinfo  {journal} {JHEP}\ }\textbf {\bibinfo {volume} {11}},\ \bibinfo {pages} {088}},\ \Eprint {https://arxiv.org/abs/1303.6929} {arXiv:1303.6929 [hep-ph]} \BibitemShut {NoStop}%
\bibitem [{\citenamefont {Dudal}\ and\ \citenamefont {Mertens}(2015)}]{Dudal:2014jfa}%
  \BibitemOpen
  \bibfield  {author} {\bibinfo {author} {\bibfnamefont {D.}~\bibnamefont {Dudal}}\ and\ \bibinfo {author} {\bibfnamefont {T.~G.}\ \bibnamefont {Mertens}},\ }\href {https://doi.org/10.1103/PhysRevD.91.086002} {\bibfield  {journal} {\bibinfo  {journal} {Phys. Rev. D}\ }\textbf {\bibinfo {volume} {91}},\ \bibinfo {pages} {086002} (\bibinfo {year} {2015})},\ \Eprint {https://arxiv.org/abs/1410.3297} {arXiv:1410.3297 [hep-th]} \BibitemShut {NoStop}%
\bibitem [{\citenamefont {Finazzo}\ and\ \citenamefont {Noronha}(2014)}]{Finazzo:2014zga}%
  \BibitemOpen
  \bibfield  {author} {\bibinfo {author} {\bibfnamefont {S.~I.}\ \bibnamefont {Finazzo}}\ and\ \bibinfo {author} {\bibfnamefont {J.}~\bibnamefont {Noronha}},\ }\href {https://doi.org/10.1103/PhysRevD.90.115028} {\bibfield  {journal} {\bibinfo  {journal} {Phys. Rev. D}\ }\textbf {\bibinfo {volume} {90}},\ \bibinfo {pages} {115028} (\bibinfo {year} {2014})},\ \Eprint {https://arxiv.org/abs/1411.4330} {arXiv:1411.4330 [hep-th]} \BibitemShut {NoStop}%
\bibitem [{\citenamefont {Braga}\ \emph {et~al.}(2016)\citenamefont {Braga}, \citenamefont {Martin~Contreras},\ and\ \citenamefont {Diles}}]{Braga:2016wkm}%
  \BibitemOpen
  \bibfield  {author} {\bibinfo {author} {\bibfnamefont {N.~R.~F.}\ \bibnamefont {Braga}}, \bibinfo {author} {\bibfnamefont {M.~A.}\ \bibnamefont {Martin~Contreras}},\ and\ \bibinfo {author} {\bibfnamefont {S.}~\bibnamefont {Diles}},\ }\href {https://doi.org/10.1140/epjc/s10052-016-4447-4} {\bibfield  {journal} {\bibinfo  {journal} {Eur. Phys. J. C}\ }\textbf {\bibinfo {volume} {76}},\ \bibinfo {pages} {598} (\bibinfo {year} {2016})},\ \Eprint {https://arxiv.org/abs/1604.08296} {arXiv:1604.08296 [hep-ph]} \BibitemShut {NoStop}%
\bibitem [{\citenamefont {Braga}\ \emph {et~al.}(2017)\citenamefont {Braga}, \citenamefont {Ferreira},\ and\ \citenamefont {Vega}}]{Braga:2017bml}%
  \BibitemOpen
  \bibfield  {author} {\bibinfo {author} {\bibfnamefont {N.~R.~F.}\ \bibnamefont {Braga}}, \bibinfo {author} {\bibfnamefont {L.~F.}\ \bibnamefont {Ferreira}},\ and\ \bibinfo {author} {\bibfnamefont {A.}~\bibnamefont {Vega}},\ }\href {https://doi.org/10.1016/j.physletb.2017.10.013} {\bibfield  {journal} {\bibinfo  {journal} {Phys. Lett. B}\ }\textbf {\bibinfo {volume} {774}},\ \bibinfo {pages} {476} (\bibinfo {year} {2017})},\ \Eprint {https://arxiv.org/abs/1709.05326} {arXiv:1709.05326 [hep-ph]} \BibitemShut {NoStop}%
\bibitem [{\citenamefont {Braga}\ and\ \citenamefont {Ferreira}(2018)}]{Braga:2018zlu}%
  \BibitemOpen
  \bibfield  {author} {\bibinfo {author} {\bibfnamefont {N.~R.~F.}\ \bibnamefont {Braga}}\ and\ \bibinfo {author} {\bibfnamefont {L.~F.}\ \bibnamefont {Ferreira}},\ }\href {https://doi.org/10.1016/j.physletb.2018.06.053} {\bibfield  {journal} {\bibinfo  {journal} {Phys. Lett. B}\ }\textbf {\bibinfo {volume} {783}},\ \bibinfo {pages} {186} (\bibinfo {year} {2018})},\ \Eprint {https://arxiv.org/abs/1802.02084} {arXiv:1802.02084 [hep-ph]} \BibitemShut {NoStop}%
\bibitem [{\citenamefont {Bohra}\ \emph {et~al.}(2021)\citenamefont {Bohra}, \citenamefont {Dudal}, \citenamefont {Hajilou},\ and\ \citenamefont {Mahapatra}}]{Bohra:2020qom}%
  \BibitemOpen
  \bibfield  {author} {\bibinfo {author} {\bibfnamefont {H.}~\bibnamefont {Bohra}}, \bibinfo {author} {\bibfnamefont {D.}~\bibnamefont {Dudal}}, \bibinfo {author} {\bibfnamefont {A.}~\bibnamefont {Hajilou}},\ and\ \bibinfo {author} {\bibfnamefont {S.}~\bibnamefont {Mahapatra}},\ }\href {https://doi.org/10.1103/PhysRevD.103.086021} {\bibfield  {journal} {\bibinfo  {journal} {Phys. Rev. D}\ }\textbf {\bibinfo {volume} {103}},\ \bibinfo {pages} {086021} (\bibinfo {year} {2021})},\ \Eprint {https://arxiv.org/abs/2010.04578} {arXiv:2010.04578 [hep-th]} \BibitemShut {NoStop}%
\bibitem [{\citenamefont {Ballon-Bayona}\ \emph {et~al.}(2021)\citenamefont {Ballon-Bayona}, \citenamefont {Mamani},\ and\ \citenamefont {Rodrigues}}]{Ballon-Bayona:2021ibm}%
  \BibitemOpen
  \bibfield  {author} {\bibinfo {author} {\bibfnamefont {A.}~\bibnamefont {Ballon-Bayona}}, \bibinfo {author} {\bibfnamefont {L.~A.~H.}\ \bibnamefont {Mamani}},\ and\ \bibinfo {author} {\bibfnamefont {D.~M.}\ \bibnamefont {Rodrigues}},\ }\href {https://doi.org/10.1103/PhysRevD.104.126029} {\bibfield  {journal} {\bibinfo  {journal} {Phys. Rev. D}\ }\textbf {\bibinfo {volume} {104}},\ \bibinfo {pages} {126029} (\bibinfo {year} {2021})},\ \Eprint {https://arxiv.org/abs/2107.10983} {arXiv:2107.10983 [hep-ph]} \BibitemShut {NoStop}%
\bibitem [{\citenamefont {Ballon-Bayona}\ \emph {et~al.}(2023)\citenamefont {Ballon-Bayona}, \citenamefont {Frederico}, \citenamefont {Mamani},\ and\ \citenamefont {de~Paula}}]{Ballon-Bayona:2023zal}%
  \BibitemOpen
  \bibfield  {author} {\bibinfo {author} {\bibfnamefont {A.}~\bibnamefont {Ballon-Bayona}}, \bibinfo {author} {\bibfnamefont {T.}~\bibnamefont {Frederico}}, \bibinfo {author} {\bibfnamefont {L.~A.~H.}\ \bibnamefont {Mamani}},\ and\ \bibinfo {author} {\bibfnamefont {W.}~\bibnamefont {de~Paula}},\ }\href {https://doi.org/10.1103/PhysRevD.108.106016} {\bibfield  {journal} {\bibinfo  {journal} {Phys. Rev. D}\ }\textbf {\bibinfo {volume} {108}},\ \bibinfo {pages} {106016} (\bibinfo {year} {2023})},\ \Eprint {https://arxiv.org/abs/2308.07503} {arXiv:2308.07503 [hep-ph]} \BibitemShut {NoStop}%
\bibitem [{\citenamefont {Aref'eva}\ \emph {et~al.}(2024)\citenamefont {Aref'eva}, \citenamefont {Hajilou}, \citenamefont {Nikolaev},\ and\ \citenamefont {Slepov}}]{Arefeva:2024xmg}%
  \BibitemOpen
  \bibfield  {author} {\bibinfo {author} {\bibfnamefont {I.~Y.}\ \bibnamefont {Aref'eva}}, \bibinfo {author} {\bibfnamefont {A.}~\bibnamefont {Hajilou}}, \bibinfo {author} {\bibfnamefont {A.}~\bibnamefont {Nikolaev}},\ and\ \bibinfo {author} {\bibfnamefont {P.}~\bibnamefont {Slepov}},\ }\href {https://doi.org/10.1103/PhysRevD.110.086021} {\bibfield  {journal} {\bibinfo  {journal} {Phys. Rev. D}\ }\textbf {\bibinfo {volume} {110}},\ \bibinfo {pages} {086021} (\bibinfo {year} {2024})},\ \Eprint {https://arxiv.org/abs/2407.11924} {arXiv:2407.11924 [hep-th]} \BibitemShut {NoStop}%
\bibitem [{\citenamefont {Ballon-Bayona}\ \emph {et~al.}(2025)\citenamefont {Ballon-Bayona}, \citenamefont {Bartz}, \citenamefont {Mamani},\ and\ \citenamefont {Rodrigues}}]{Ballon-Bayona:2024twa}%
  \BibitemOpen
  \bibfield  {author} {\bibinfo {author} {\bibfnamefont {A.}~\bibnamefont {Ballon-Bayona}}, \bibinfo {author} {\bibfnamefont {S.}~\bibnamefont {Bartz}}, \bibinfo {author} {\bibfnamefont {L.~A.~H.}\ \bibnamefont {Mamani}},\ and\ \bibinfo {author} {\bibfnamefont {D.~M.}\ \bibnamefont {Rodrigues}},\ }\href {https://doi.org/10.1103/PhysRevD.111.026011} {\bibfield  {journal} {\bibinfo  {journal} {Phys. Rev. D}\ }\textbf {\bibinfo {volume} {111}},\ \bibinfo {pages} {026011} (\bibinfo {year} {2025})},\ \Eprint {https://arxiv.org/abs/2410.23471} {arXiv:2410.23471 [hep-ph]} \BibitemShut {NoStop}%
\bibitem [{\citenamefont {Toniato}\ \emph {et~al.}(2025)\citenamefont {Toniato}, \citenamefont {Dudal}, \citenamefont {Mahapatra}, \citenamefont {da~Rocha},\ and\ \citenamefont {Jena}}]{Toniato:2025gts}%
  \BibitemOpen
  \bibfield  {author} {\bibinfo {author} {\bibfnamefont {B.}~\bibnamefont {Toniato}}, \bibinfo {author} {\bibfnamefont {D.}~\bibnamefont {Dudal}}, \bibinfo {author} {\bibfnamefont {S.}~\bibnamefont {Mahapatra}}, \bibinfo {author} {\bibfnamefont {R.}~\bibnamefont {da~Rocha}},\ and\ \bibinfo {author} {\bibfnamefont {S.~S.}\ \bibnamefont {Jena}},\ }\href {https://doi.org/10.1103/pn9x-ycnh} {\bibfield  {journal} {\bibinfo  {journal} {Phys. Rev. D}\ }\textbf {\bibinfo {volume} {111}},\ \bibinfo {pages} {126021} (\bibinfo {year} {2025})},\ \Eprint {https://arxiv.org/abs/2502.12694} {arXiv:2502.12694 [hep-th]} \BibitemShut {NoStop}%
\bibitem [{\citenamefont {Ferreira}(2025)}]{Ferreira:2025iqe}%
  \BibitemOpen
  \bibfield  {author} {\bibinfo {author} {\bibfnamefont {L.~F.}\ \bibnamefont {Ferreira}},\ }\href {https://doi.org/10.1103/32sr-nyrv} {\bibfield  {journal} {\bibinfo  {journal} {Phys. Rev. D}\ }\textbf {\bibinfo {volume} {112}},\ \bibinfo {pages} {074043} (\bibinfo {year} {2025})},\ \Eprint {https://arxiv.org/abs/2510.02647} {arXiv:2510.02647 [hep-ph]} \BibitemShut {NoStop}%
\bibitem [{\citenamefont {Courant}\ and\ \citenamefont {Hilbert}(1989)}]{Courant1989}%
  \BibitemOpen
  \bibfield  {author} {\bibinfo {author} {\bibfnamefont {R.}~\bibnamefont {Courant}}\ and\ \bibinfo {author} {\bibfnamefont {D.}~\bibnamefont {Hilbert}},\ }\href@noop {} {\emph {\bibinfo {title} {Methods of Mathematical Physics, Volume 1}}},\ \bibinfo {edition} {1st}\ ed.\ (\bibinfo  {publisher} {Wiley-VCH},\ \bibinfo {year} {1989})\BibitemShut {NoStop}%
\bibitem [{\citenamefont {Kato}(2013)}]{Kato2013}%
  \BibitemOpen
  \bibfield  {author} {\bibinfo {author} {\bibfnamefont {T.}~\bibnamefont {Kato}},\ }\href@noop {} {\emph {\bibinfo {title} {Perturbation Theory for Linear Operators}}},\ Vol.\ \bibinfo {volume} {132}\ (\bibinfo  {publisher} {Springer-Verlag},\ \bibinfo {address} {Berlin, Heidelberg},\ \bibinfo {year} {2013})\BibitemShut {NoStop}%
\bibitem [{\citenamefont {Warnick}(2015)}]{Warnick2015}%
  \BibitemOpen
  \bibfield  {author} {\bibinfo {author} {\bibfnamefont {C.~M.}\ \bibnamefont {Warnick}},\ }\href {https://doi.org/10.1007/s00220-014-2163-3} {\bibfield  {journal} {\bibinfo  {journal} {Communications in Mathematical Physics}\ }\textbf {\bibinfo {volume} {333}},\ \bibinfo {pages} {959} (\bibinfo {year} {2015})},\ \Eprint {https://arxiv.org/abs/1306.5760} {arXiv:1306.5760 [gr-qc]} \BibitemShut {NoStop}%
\bibitem [{\citenamefont {Jaramillo}\ \emph {et~al.}(2021{\natexlab{b}})\citenamefont {Jaramillo}, \citenamefont {Panosso~Macedo},\ and\ \citenamefont {Al~Sheikh}}]{Jaramillo:2020tuu}%
  \BibitemOpen
  \bibfield  {author} {\bibinfo {author} {\bibfnamefont {J.~L.}\ \bibnamefont {Jaramillo}}, \bibinfo {author} {\bibfnamefont {R.}~\bibnamefont {Panosso~Macedo}},\ and\ \bibinfo {author} {\bibfnamefont {L.}~\bibnamefont {Al~Sheikh}},\ }\href {https://doi.org/10.1103/PhysRevX.11.031003} {\bibfield  {journal} {\bibinfo  {journal} {Phys. Rev. X}\ }\textbf {\bibinfo {volume} {11}},\ \bibinfo {pages} {031003} (\bibinfo {year} {2021}{\natexlab{b}})},\ \Eprint {https://arxiv.org/abs/2004.06434} {arXiv:2004.06434 [gr-qc]} \BibitemShut {NoStop}%
\bibitem [{\citenamefont {Gasperin}\ and\ \citenamefont {Jaramillo}(2022)}]{Gasperin:2021kfv}%
  \BibitemOpen
  \bibfield  {author} {\bibinfo {author} {\bibfnamefont {E.}~\bibnamefont {Gasperin}}\ and\ \bibinfo {author} {\bibfnamefont {J.~L.}\ \bibnamefont {Jaramillo}},\ }\href {https://doi.org/10.1088/1361-6382/ac5054} {\bibfield  {journal} {\bibinfo  {journal} {Class. Quant. Grav.}\ }\textbf {\bibinfo {volume} {39}},\ \bibinfo {pages} {115010} (\bibinfo {year} {2022})},\ \Eprint {https://arxiv.org/abs/2107.12865} {arXiv:2107.12865 [gr-qc]} \BibitemShut {NoStop}%
\bibitem [{\citenamefont {Gursoy}\ \emph {et~al.}(2008{\natexlab{b}})\citenamefont {Gursoy}, \citenamefont {Kiritsis}, \citenamefont {Mazzanti},\ and\ \citenamefont {Nitti}}]{Gursoy:2008bu}%
  \BibitemOpen
  \bibfield  {author} {\bibinfo {author} {\bibfnamefont {U.}~\bibnamefont {Gursoy}}, \bibinfo {author} {\bibfnamefont {E.}~\bibnamefont {Kiritsis}}, \bibinfo {author} {\bibfnamefont {L.}~\bibnamefont {Mazzanti}},\ and\ \bibinfo {author} {\bibfnamefont {F.}~\bibnamefont {Nitti}},\ }\href {https://doi.org/10.1103/PhysRevLett.101.181601} {\bibfield  {journal} {\bibinfo  {journal} {Phys. Rev. Lett.}\ }\textbf {\bibinfo {volume} {101}},\ \bibinfo {pages} {181601} (\bibinfo {year} {2008}{\natexlab{b}})},\ \Eprint {https://arxiv.org/abs/0804.0899} {arXiv:0804.0899 [hep-th]} \BibitemShut {NoStop}%
\bibitem [{\citenamefont {Kiritsis}\ and\ \citenamefont {Nitti}(2007)}]{Kiritsis:2006ua}%
  \BibitemOpen
  \bibfield  {author} {\bibinfo {author} {\bibfnamefont {E.}~\bibnamefont {Kiritsis}}\ and\ \bibinfo {author} {\bibfnamefont {F.}~\bibnamefont {Nitti}},\ }\href {https://doi.org/10.1016/j.nuclphysb.2007.02.024} {\bibfield  {journal} {\bibinfo  {journal} {Nucl. Phys. B}\ }\textbf {\bibinfo {volume} {772}},\ \bibinfo {pages} {67} (\bibinfo {year} {2007})},\ \Eprint {https://arxiv.org/abs/hep-th/0611344} {arXiv:hep-th/0611344} \BibitemShut {NoStop}%
\bibitem [{\citenamefont {Springer}(2009)}]{Springer:2008js}%
  \BibitemOpen
  \bibfield  {author} {\bibinfo {author} {\bibfnamefont {T.}~\bibnamefont {Springer}},\ }\href {https://doi.org/10.1103/PhysRevD.79.046003} {\bibfield  {journal} {\bibinfo  {journal} {Phys. Rev. D}\ }\textbf {\bibinfo {volume} {79}},\ \bibinfo {pages} {046003} (\bibinfo {year} {2009})},\ \Eprint {https://arxiv.org/abs/0810.4354} {arXiv:0810.4354 [hep-th]} \BibitemShut {NoStop}%
\bibitem [{\citenamefont {Springer}\ \emph {et~al.}(2010)\citenamefont {Springer}, \citenamefont {Gale},\ and\ \citenamefont {Jeon}}]{Springer:2010mw}%
  \BibitemOpen
  \bibfield  {author} {\bibinfo {author} {\bibfnamefont {T.}~\bibnamefont {Springer}}, \bibinfo {author} {\bibfnamefont {C.}~\bibnamefont {Gale}},\ and\ \bibinfo {author} {\bibfnamefont {S.}~\bibnamefont {Jeon}},\ }\href {https://doi.org/10.1103/PhysRevD.82.126011} {\bibfield  {journal} {\bibinfo  {journal} {Phys. Rev. D}\ }\textbf {\bibinfo {volume} {82}},\ \bibinfo {pages} {126011} (\bibinfo {year} {2010})},\ \Eprint {https://arxiv.org/abs/1010.2760} {arXiv:1010.2760 [hep-th]} \BibitemShut {NoStop}%
\bibitem [{\citenamefont {Kovtun}\ and\ \citenamefont {Starinets}(2005)}]{Kovtun:2005ev}%
  \BibitemOpen
  \bibfield  {author} {\bibinfo {author} {\bibfnamefont {P.~K.}\ \bibnamefont {Kovtun}}\ and\ \bibinfo {author} {\bibfnamefont {A.~O.}\ \bibnamefont {Starinets}},\ }\href {https://doi.org/10.1103/PhysRevD.72.086009} {\bibfield  {journal} {\bibinfo  {journal} {Phys. Rev. D}\ }\textbf {\bibinfo {volume} {72}},\ \bibinfo {pages} {086009} (\bibinfo {year} {2005})},\ \Eprint {https://arxiv.org/abs/hep-th/0506184} {arXiv:hep-th/0506184} \BibitemShut {NoStop}%
\bibitem [{\citenamefont {Katz}\ \emph {et~al.}(2006)\citenamefont {Katz}, \citenamefont {Lewandowski},\ and\ \citenamefont {Schwartz}}]{Katz:2005ir}%
  \BibitemOpen
  \bibfield  {author} {\bibinfo {author} {\bibfnamefont {E.}~\bibnamefont {Katz}}, \bibinfo {author} {\bibfnamefont {A.}~\bibnamefont {Lewandowski}},\ and\ \bibinfo {author} {\bibfnamefont {M.~D.}\ \bibnamefont {Schwartz}},\ }\href {https://doi.org/10.1103/PhysRevD.74.086004} {\bibfield  {journal} {\bibinfo  {journal} {Phys. Rev. D}\ }\textbf {\bibinfo {volume} {74}},\ \bibinfo {pages} {086004} (\bibinfo {year} {2006})},\ \Eprint {https://arxiv.org/abs/hep-ph/0510388} {arXiv:hep-ph/0510388} \BibitemShut {NoStop}%
\bibitem [{\citenamefont {Alanen}\ \emph {et~al.}(2011)\citenamefont {Alanen}, \citenamefont {Alho}, \citenamefont {Kajantie},\ and\ \citenamefont {Tuominen}}]{Alanen:2011hh}%
  \BibitemOpen
  \bibfield  {author} {\bibinfo {author} {\bibfnamefont {J.}~\bibnamefont {Alanen}}, \bibinfo {author} {\bibfnamefont {T.}~\bibnamefont {Alho}}, \bibinfo {author} {\bibfnamefont {K.}~\bibnamefont {Kajantie}},\ and\ \bibinfo {author} {\bibfnamefont {K.}~\bibnamefont {Tuominen}},\ }\href {https://doi.org/10.1103/PhysRevD.84.086007} {\bibfield  {journal} {\bibinfo  {journal} {Phys. Rev. D}\ }\textbf {\bibinfo {volume} {84}},\ \bibinfo {pages} {086007} (\bibinfo {year} {2011})},\ \Eprint {https://arxiv.org/abs/1107.3362} {arXiv:1107.3362 [hep-th]} \BibitemShut {NoStop}%
\bibitem [{\citenamefont {Jena}\ \emph {et~al.}(2024)\citenamefont {Jena}, \citenamefont {Barman}, \citenamefont {Toniato}, \citenamefont {Dudal},\ and\ \citenamefont {Mahapatra}}]{Jena:2024cqs}%
  \BibitemOpen
  \bibfield  {author} {\bibinfo {author} {\bibfnamefont {S.~S.}\ \bibnamefont {Jena}}, \bibinfo {author} {\bibfnamefont {J.}~\bibnamefont {Barman}}, \bibinfo {author} {\bibfnamefont {B.}~\bibnamefont {Toniato}}, \bibinfo {author} {\bibfnamefont {D.}~\bibnamefont {Dudal}},\ and\ \bibinfo {author} {\bibfnamefont {S.}~\bibnamefont {Mahapatra}},\ }\href {https://doi.org/10.1007/JHEP12(2024)096} {\bibfield  {journal} {\bibinfo  {journal} {JHEP}\ }\textbf {\bibinfo {volume} {12}},\ \bibinfo {pages} {096}},\ \Eprint {https://arxiv.org/abs/2408.14813} {arXiv:2408.14813 [hep-th]} \BibitemShut {NoStop}%
\bibitem [{\citenamefont {Braga}\ and\ \citenamefont {Ferreira}(2017)}]{Braga:2017oqw}%
  \BibitemOpen
  \bibfield  {author} {\bibinfo {author} {\bibfnamefont {N.~R.~F.}\ \bibnamefont {Braga}}\ and\ \bibinfo {author} {\bibfnamefont {L.~F.}\ \bibnamefont {Ferreira}},\ }\href {https://doi.org/10.1016/j.physletb.2017.08.037} {\bibfield  {journal} {\bibinfo  {journal} {Phys. Lett. B}\ }\textbf {\bibinfo {volume} {773}},\ \bibinfo {pages} {313} (\bibinfo {year} {2017})},\ \Eprint {https://arxiv.org/abs/1704.05038} {arXiv:1704.05038 [hep-ph]} \BibitemShut {NoStop}%
\bibitem [{\citenamefont {Chen}\ \emph {et~al.}(2021)\citenamefont {Chen}, \citenamefont {Zhang}, \citenamefont {Li}, \citenamefont {Hou},\ and\ \citenamefont {Huang}}]{Chen:2020ath}%
  \BibitemOpen
  \bibfield  {author} {\bibinfo {author} {\bibfnamefont {X.}~\bibnamefont {Chen}}, \bibinfo {author} {\bibfnamefont {L.}~\bibnamefont {Zhang}}, \bibinfo {author} {\bibfnamefont {D.}~\bibnamefont {Li}}, \bibinfo {author} {\bibfnamefont {D.}~\bibnamefont {Hou}},\ and\ \bibinfo {author} {\bibfnamefont {M.}~\bibnamefont {Huang}},\ }\href {https://doi.org/10.1007/JHEP07(2021)132} {\bibfield  {journal} {\bibinfo  {journal} {JHEP}\ }\textbf {\bibinfo {volume} {07}},\ \bibinfo {pages} {132}},\ \Eprint {https://arxiv.org/abs/2010.14478} {arXiv:2010.14478 [hep-ph]} \BibitemShut {NoStop}%
\bibitem [{\citenamefont {Braga}\ \emph {et~al.}(2022)\citenamefont {Braga}, \citenamefont {Faulhaber},\ and\ \citenamefont {Junqueira}}]{Braga:2022yfe}%
  \BibitemOpen
  \bibfield  {author} {\bibinfo {author} {\bibfnamefont {N.~R.~F.}\ \bibnamefont {Braga}}, \bibinfo {author} {\bibfnamefont {L.~F.}\ \bibnamefont {Faulhaber}},\ and\ \bibinfo {author} {\bibfnamefont {O.~C.}\ \bibnamefont {Junqueira}},\ }\href {https://doi.org/10.1103/PhysRevD.105.106003} {\bibfield  {journal} {\bibinfo  {journal} {Phys. Rev. D}\ }\textbf {\bibinfo {volume} {105}},\ \bibinfo {pages} {106003} (\bibinfo {year} {2022})},\ \Eprint {https://arxiv.org/abs/2201.05581} {arXiv:2201.05581 [hep-th]} \BibitemShut {NoStop}%
\bibitem [{\citenamefont {Son}\ and\ \citenamefont {Starinets}(2002)}]{Son:2002sd}%
  \BibitemOpen
  \bibfield  {author} {\bibinfo {author} {\bibfnamefont {D.~T.}\ \bibnamefont {Son}}\ and\ \bibinfo {author} {\bibfnamefont {A.~O.}\ \bibnamefont {Starinets}},\ }\href {https://doi.org/10.1088/1126-6708/2002/09/042} {\bibfield  {journal} {\bibinfo  {journal} {JHEP}\ }\textbf {\bibinfo {volume} {09}},\ \bibinfo {pages} {042}},\ \Eprint {https://arxiv.org/abs/hep-th/0205051} {arXiv:hep-th/0205051} \BibitemShut {NoStop}%
\bibitem [{\citenamefont {Miranda}\ \emph {et~al.}(2009)\citenamefont {Miranda}, \citenamefont {Ballon~Bayona}, \citenamefont {Boschi-Filho},\ and\ \citenamefont {Braga}}]{Miranda:2009uw}%
  \BibitemOpen
  \bibfield  {author} {\bibinfo {author} {\bibfnamefont {A.~S.}\ \bibnamefont {Miranda}}, \bibinfo {author} {\bibfnamefont {C.~A.}\ \bibnamefont {Ballon~Bayona}}, \bibinfo {author} {\bibfnamefont {H.}~\bibnamefont {Boschi-Filho}},\ and\ \bibinfo {author} {\bibfnamefont {N.~R.~F.}\ \bibnamefont {Braga}},\ }\href {https://doi.org/10.1088/1126-6708/2009/11/119} {\bibfield  {journal} {\bibinfo  {journal} {JHEP}\ }\textbf {\bibinfo {volume} {2009}}\bibfield  {number} {\bibinfo  {number} { (11)},\ \bibinfo {pages} {119}},\ }\Eprint {https://arxiv.org/abs/0909.1790} {arXiv:0909.1790 [hep-th]} \BibitemShut {NoStop}%
\end{thebibliography}%

\end{document}